\documentclass[longauth]{aa}
\usepackage{graphicx,natbib,times,amssymb,amsmath}
\usepackage[varg]{txfonts}

\bibpunct{(}{)}{;}{a}{}{,} 

\def \kms {\,km\,s$^{-1}$}

\def \hda {H$\delta_{\rm A}$}
\def \hga {H$\gamma_{\rm A}$}

\def \ha {H$\alpha$}
\def \hb {H$\beta$}
\def \hd {H$\delta$}
\def \hg {H$\gamma$}

\def \hii {[H~{\sc ii}]}
\def \HIIr {H~{\sc ii}}

\def \oi {[O~{\sc i}]}
\def \oiii {[O~{\sc iii}]}
\def \oii {[O~{\sc ii}]}

\def \nii {[N~{\sc ii}]}
\def \heii {He~{\sc ii}}
\def \neii {[Ne~{\sc ii}]}

\def \sii {[S~{\sc ii}]}
\def \neiii {[Ne~{\sc iii}]}
\def \ariv {[Ar~{\sc iv}]}

\def \hi {H~{\sc i}}

\def \Msol {M$_{\odot}$}

\def \msol {M$_{\odot}$}
\def \lsol {L$_{\odot}$}

\def \msolyr {M$_{\odot}$yr$^{-1}$}

\def \mum {\,$\mu$m}
\def \ergflux {erg\,s$^{-1}$\,cm$^{-2}$}
\def \ergs {erg\,s$^{-1}$}
\def \sfrsd {\msol\,yr$^{-1}$\,kpc$^{-2}$}
\def \kms {km\,s$^{-1}$}
\def \cmcube {cm$^{-3}$}

\begin{document}
\title{The Mice at play in the CALIFA survey}

\subtitle{A case study of a gas-rich major merger between first
  passage and coalescence}
\author{
Vivienne Wild \inst{\ref{sta},\ref{roe}}\fnmsep\thanks{Send correspondence to vw8@st-andrews.ac.uk}
  \and
  Fabian Rosales-Ortega\inst{\ref{madrid},\ref{mex1}}
  \and
  Jesus Falc\'on-Barroso\inst{\ref{iac},\ref{ull}}
  \and
  Rub\'en Garc\'ia-Benito\inst{\ref{csic}}
  \and
  Anna Gallazzi\inst{\ref{inaf},\ref{dark}}
  \and
  Rosa M. Gonz\'alez Delgado\inst{\ref{csic}}
  \and
  Simona Bekerait\.{e} \inst{\ref{aip}}
  \and
  Anna Pasquali\inst{\ref{heidel}}
  \and
  Peter H. Johansson\inst{\ref{helsin}}
  \and
  Bego\~{n}a Garc\'ia Lorenzo\inst{\ref{iac},\ref{ull}}
  \and 
  Glenn van de Ven\inst{\ref{mpia}}
  \and
  Milena Pawlik\inst{\ref{sta}}
  \and
  Enrique Per\'ez\inst{\ref{csic}}
  \and
  Ana Monreal-Ibero\inst{\ref{aip},\ref{gepi}}
  \and
  Mariya Lyubenova\inst{\ref{mpia}}
  \and
  Roberto Cid Fernandes\inst{\ref{brazil1}}
  \and
  Jairo M\'endez-Abreu\inst{\ref{sta}}
  \and
  Jorge Barrera-Ballesteros\inst{\ref{iac},\ref{ull}}
  \and
  Carolina Kehrig\inst{\ref{csic}}
  \and
  Jorge Iglesias-P\'{a}ramo\inst{\ref{csic},\ref{caha}}
  \and
  Dominik J. Bomans\inst{\ref{boc1},\ref{boc2}}
  \and
  Isabel M\'arquez\inst{\ref{csic}}
  \and
  Benjamin D. Johnson\inst{\ref{iap}}
  \and
  Robert C. Kennicutt\inst{\ref{ioa}}
  \and
   Bernd Husemann\inst{\ref{eso},\ref{aip}}
  \and
  Damian Mast\inst{\ref{brazil2}}
 \and 
 Sebastian F. S\'anchez\inst{\ref{csic},\ref{caha},\ref{unam}} 
 \and
 C. Jakob Walcher\inst{\ref{aip}}
  \and
 Jo\~ao Alves\inst{\ref{vienna}}
 \and
 Alfonso L. Aguerri\inst{\ref{iac},\ref{ull}}
  \and
  Almudena Alonso Herrero\inst{\ref{santan}}
  \and
  Joss Bland-Hawthorn\inst{\ref{sydney}}
  \and
  Cristina Catal\'an-Torrecilla\inst{\ref{madrid2}}
 \and
  Estrella Florido\inst{\ref{granada}}  
  \and
  Jean Michel Gomes\inst{\ref{porto}}
  \and
  Knud Jahnke\inst{\ref{mpia}}
  \and
  \'A.R. L\'opez-S\'anchez\inst{\ref{aao},\ref{macq}}
  \and
  Adriana de Lorenzo-C\'aceres\inst{\ref{sta}}
  \and
  Raffaella A. Marino\inst{\ref{madrid3}}
  \and
  Esther M\'armol-Queralt\'o\inst{\ref{roe}}
  \and 
  Patrick Olden\inst{\ref{sta}}
  \and
  Ascensi\'on del Olmo\inst{\ref{csic}}
  \and
  Polychronis Papaderos\inst{\ref{porto}}
  \and
  Andreas Quirrenbach\inst{\ref{zah}}
  \and
  Jose M.~V\'ilchez\inst{\ref{csic}}
  \and
  Bodo Ziegler\inst{\ref{vienna}}
}

\institute{School of Physics and Astronomy, University of St Andrews, North
  Haugh, St Andrews, KY16 9SS, U.K. (SUPA) \label{sta}
  \and 
  Institute for Astronomy, University of Edinburgh, Royal
  Observatory, Blackford Hill, Edinburgh, EH9 3HJ, U.K. (SUPA)\label{roe}
  \and
  Departamento de F{\'i}sica Te{\'o}rica, Universidad Aut{\'o}noma de
  Madrid, 28049 Madrid, Spain \label{madrid}
  \and
  Instituto Nacional de Astrof{\'i}sica, {\'O}ptica y Electr{\'o}nica, Luis E. Erro 1, 72840 Tonantzintla, Puebla, Mexico \label{mex1}
  \and
  Instituto de Astrof\'isica de Canarias (IAC), E-38205 La Laguna, Tenerife, Spain \label{iac}
  \and
  Depto. Astrof\'isica, Universidad de La Laguna (ULL), E-38206 La Laguna, Tenerife, Spain \label{ull}
  \and
  Instituto de Astrof\'{\i}sica de Andaluc\'{\i}a (CSIC), C/Camino Bajo de Hu\'etor, 50, 18008 Granada, Spain \label{csic}
  \and
  Leibniz-Institut f\"ur Astrophysik Potsdam (AIP), An der Sternwarte
  16, D-14482 Potsdam, Germany \label{aip}
  \and
  European Southern Observatory (ESO), Karl-Schwarzschild-Str. 2,
  D-85748 Garching b. Muenchen, Germany \label{eso} \\
  \and
  Astronomisches Rechen-Institut, Zentrum f\"ur Astronomie der Universit\"at  Heidelberg, M\"onchhofstr. 12 - 14, D-69120 Heidelberg, Germany \label{heidel}
  \and
  INAF – Osservatorio Astrofisico di Arcetri, Largo Enrico Fermi 5, 50125 Firenze, Italy \label{inaf}
  \and 
  Dark Cosmology Centre, Niels Bohr Institute, University of Copenhagen, Juliane Mariesvej 30, 2100 Copenhagen, Denmark \label{dark}
  \and  
  Department of Physics, University of Helsinki, Gustaf H\"allstr\"omin katu 2a, FI-00014 Helsinki, Finland\label{helsin} 
  \and
   Max Planck Institute for Astronomy, K\"onigstuhl 17, 69117 Heidelberg, Germany \label{mpia}
   \and
   GEPI Observatoire de Paris, CNRS, Université Paris Diderot, Place Jules Janssen, 92190 Meudon, France \label{gepi}
   \and
   Departamento de Fisica, Universidade Federal de Santa Catarina, Brazil \label{brazil1}
    \and
   Centro Astron\'{o}mico Hispano Alem\'{a}n, C/ Jes\'{u}s Durb\'{a}n Rem\'{o}n 2-2, 04004 Almer\'{\i}a, Spain \label{caha}
   \and
   Astronomical Institute of the Ruhr-University Bochum, Universit\"atsstr. 150, 44580 Bochum, Germany \label{boc1}
   \and
   RUB Research Department 'Plasmas with Complex Interactions', Universit\"atsstr. 150, 44580 Bochum, Germany  \label{boc2}
   \and
   Institut d'Astrophysique de Paris, CNRS UMR7095, Université Pierre  et Marie Curie, 98bis Bd Arago, 75014 Paris, France \label{iap}
   \and
   Institute of Astronomy, University of Cambridge, Madingly Road, Cambridge, CB3 0HA, UK \label{ioa}
  \and
  \label{unam} Instituto de Astronom\'\i a,Universidad Nacional Auton\'oma de Mexico, A.P. 70-264, 04510, M\'exico,D.F. 
  \and
  Instituto de Cosmologia, Relatividade e Astrof\'{i}sica – ICRA, Centro Brasileiro de Pesquisas F\'{i}sicas, Rua Dr.Xavier Sigaud 150, CEP 22290-180, Rio de Janeiro, RJ, Brazil \label{brazil2}
  \and 
  University of Vienna, Türkenschanzstrasse 17, 1180, Vienna, Austria \label{vienna}
  \and
  Instituto de Fisica de Cantabria, CSIC-UC, Avenida de los Castros s/n, 39006 Santander, Spain \label{santan}
  \and
  Sydney Institute for Astronomy, School of Physics A28, The University of Sydney, NSW 2006, Australia \label{sydney}
  \and
  Departamento de Astrofísica y CC. de la Atm\'{o}sfera, Universidad Complutense de Madrid,E-28040, Madrid, Spain \label{madrid2}
  \and
  Departamento de Física Teórica y del Cosmos, Universidad de Granada, Spain \label{granada} 
  \and
  Centro de Astrof{\'i}sica and Faculdade de Ci\^{e}ncias, Universidade do Porto, Rua das Estrelas, 4150-762 Porto, Portugal\label{porto} 
  \and
  CEI Campus Moncloa, UCM-UPM, Departamento de Astrof\'{i}sica y CC$.$ de la Atm\'{o}sfera, Facultad de CC$.$ F\'{i}sicas, Universidad Complutense de Madrid, Avda.\,Complutense s/n, 28040 Madrid, Spain \label{madrid3}
 \and
  Landessternwarte, Zentrum für Astronomie der Universit\"{a}t Heidelberg (ZAH), K\"{o}nigstuhl 12, D-69117 Heidelberg, Germany \label{zah}
  \and
  Australian Astronomical Observatory, PO Box 915, North Ryde, NSW 1670, Australia \label{aao}
  \and
  Department of Physics and Astronomy, Macquarie University, NSW 2109, Australia \label{macq}
}

\date{Received xx/xx/xxxx; accepted xx/xx/xxxx} 

\abstract{
  We present optical integral field spectroscopy (IFS) observations of
  the Mice, a major merger between two massive ($\gtrsim10^{11}$\msol)
  gas-rich spirals NGC~4676A and B, observed between first passage and
  final coalescence.  The spectra provide stellar and gas kinematics,
  ionised gas properties and stellar population diagnostics, over the
  full optical extent of both galaxies with $\sim$1.6\,kpc spatial
  resolution. The Mice galaxies provide a perfect case study
  highlighting the importance of IFS data for improving our
  understanding of local galaxies.
  The impact of first passage on the kinematics of the stars and gas
  has been significant, with strong bars likely induced in both
  galaxies. The barred spiral NGC~4676B exhibits a strong twist in
  both its stellar and ionised gas disk. The edge-on disk galaxy
  NGC~4676A appears to be bulge free, with a strong bar causing its
  ``boxy'' light profile.
  On the other hand, the impact of the merger on the stellar
  populations has been minimal thus far.  By combining the IFS data
  with archival multiwavelength observations we show that star
  formation induced by the recent close passage has not contributed
  significantly to the global star formation rate or stellar mass of the
  galaxies.
  Both galaxies show bicones of high ionisation gas extending along
  their minor axes. In NGC~4676A the high gas velocity dispersion and
  Seyfert-like line ratios at large scaleheight indicate a powerful
  outflow.  Fast shocks ($v_s\sim$350\,km\,s$^{-1}$) extend to
  $\sim6.6$\,kpc above the disk plane.  The measured ram pressure
  ($P/k = 4.8\times10^{6}$K\,\cmcube) and mass outflow rate
  ($\sim8-20$\msolyr) are similar to superwinds from local
  ultra-luminous infrared galaxies, although NGC~4676A has only a
  moderate infrared luminosity of $3\times10^{10}$\lsol.  Energy
  beyond that provided by the mechanical energy of the starburst
  appears to be required to drive the outflow.
  Finally, we compare the observations to mock kinematic and stellar
  population maps extracted from a hydrodynamical merger
  simulation. The models show little enhancement in star formation
  during and following first passage, in agreement with the
  observations. We highlight areas where IFS data could help further
  constrain the models. 
} 


\maketitle

\section{Introduction}\label{sec:intro}

\begin{figure}[t]
 \centering
\includegraphics[scale=0.4]{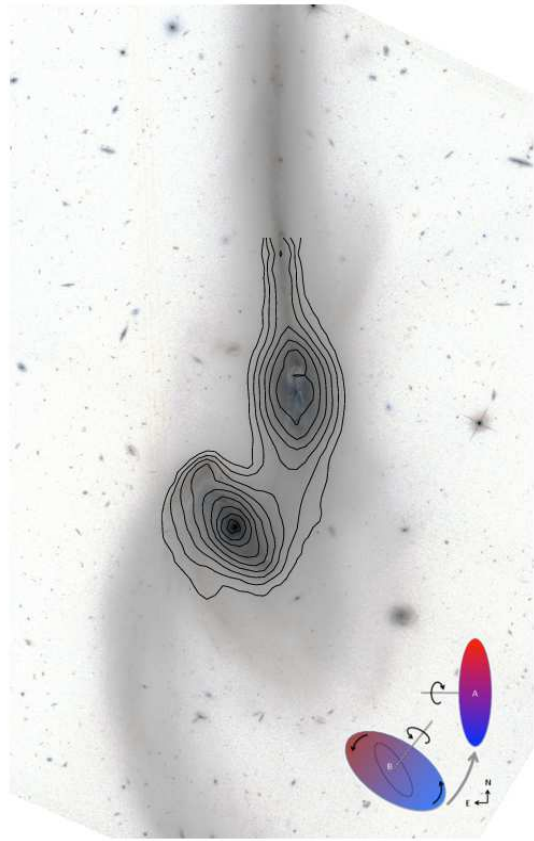}
\caption{False colour image 2.4\arcmin$\times$3.5\arcmin\ in size in
  filters F606W ($V$, green channel) and F814W ($I$, red channel) for
  \object{NGC~4676} using the HST-ACS/WFC (ACS Early Release
  Observations, Proposal ID 8992, P.I. Ford).  Overlaid are V-band
  contours constructed from the {\sc CALIFA} integral field datacubes,
  with the outermost contour at 23\,mag/arcsec$^2$ and contours spaced
  by 0.45\,mag/arcsec$^2$. The inlaid schematic shows the main
  kinematical properties of the Mice, as determined from long-slit
  spectroscopic observations \citep{Burbidge:1961p7316,
    Stockton:1974p7470}, the CALIFA IFS data (Section \ref{sec:kin}),
  HI maps and comparisons with N-body simulations
  \citep{Barnes:2004p6826}. Black arrows indicate the direction of
  rotation of the disks, defined such that the disk of NGC~4676B (SE)
  is inclined away from the line-of-sight. NGC~4676A is viewed
  almost exactly edge-on.  The thick grey arrow indicates the
  approximate track that NGC~4676A (NW) has taken relative to
  NGC~4676B. The pair are observed close to apocentre, with NGC~4676A
  receding from NGC~4676B at $\sim$160\,\kms\ (Section
  \ref{sec:kin}).}\label{fig:hst}
\end{figure}

NGC~4676A and B are members of the original \citet{Toomre:1977p7267}
sequence of merging galaxies, otherwise known as ``the playing mice''
\citep{VorontsovVelIaminov:1958p7858}. They are a classic example of a
major gas-rich prograde merger, where the roughly equal mass of the
progenitors and coincidence of the sense of rotation and orbital
motion leads to lengthy tidal tails. NGC~4676A and B (hereafter
referred to as the Mice) are one of the earliest stage gas-rich major
mergers visible in the nearby Universe, observed close to their
apocentre, with N-body simulations and observations agreeing that
first passage must have occurred around 170\,Myr ago
\citep{Barnes:2004p6826, Chien:2007p7317}.
The system is an outlying member of the Coma cluster, located about
$4^{\circ}$ or 1.7 virial radii from the centre
\citep{Kubo:2007p8229}, with a velocity about 350\,\kms\ from the mean
cluster velocity \citep{Burbidge:1961p7316}.  Figure \ref{fig:hst} shows an image of the Mice taken with
the Hubble Space Telescope (HST) Advanced Camera for Surveys (ACS). 

Massive gas-rich galaxy mergers, of which the Mice are a classic
example, may play a key role in the evolution of the galaxy population
and in explaining the galaxy demographics in the present day Universe.
Within the currently favoured cosmological model of a cold dark matter
dominated Universe, structure formation is hierarchical, with small
overdensities forming early on and subsequently merging to form larger
structures. Galaxies form and evolve within these overdensities, or
dark matter halos. When the dark matter halos merge, the galaxies are
thought to behave likewise, forming a single more massive
system. In order to obtain a complete understanding of how
galaxies formed and evolved within this gravitational framework, we
need to disentangle the relative importance of the many processes
affecting the baryonic material.  

By combining high quality spatially resolved optical spectroscopy from
the {\sc CALIFA} survey with archival multiwavelength observations,
the aim of this paper is to present a detailed picture of the physical
processes induced by the first passage of two massive gas-rich
galaxies. In Figure \ref{fig:hst} the contours of the reconstructed
{\sc CALIFA} $V-$band flux are superimposed on the HST image, showing
how the large field-of-view of the integral-field unit used by the
{\sc CALIFA} survey allows coverage of the full extent of the bodies
of the Mice, including the inter-galaxy region and some of the tidal
tails. While the Mice are only one example of an early-stage gas-rich
merger, and initial conditions of mergers (galaxy properties, impact
parameter, orbital motions etc.)  are expected to lead to a range of
final outcomes, this study aims to serve as a reference point for
statistical studies of close galaxy pairs, of particular use when
combined with other similar case studies (e.g. Engel et al. 2010
[NGC~6240]; Alonso-Herrero et al. 2012
[NGC~7771$+$7770])\nocite{Engel:2010p8436,AlonsoHerrero:2012p8640}.

In the following section we collate published observed and derived
properties of the Mice galaxies. Details of the observations taken as
part of the {\sc CALIFA} survey are given in Section \ref{sec:califa}.
We study the morphology, gas and stellar kinematics of the Mice
galaxies in Section \ref{sec:kin}, including a new image decomposition
of the archival HST images.  In Section \ref{sec:sfh} we use the
stellar continuum to constrain the star formation history of the
galaxies.  We present maps of emission line strengths and ratios in
Section \ref{sec:emission}. In Section \ref{sec:MWsfr} we collate
multiwavelength observations from the literature to obtain an accurate
estimate of the ongoing star formation in the Mice galaxies. In
Sections \ref{sec:outflowA} and \ref{sec:biconeB} we discuss the
origin of the high ionisation bicones in each of the galaxies.  In Section
\ref{sec:simn} we present a mock integral field spectroscopy (IFS)
datacube of the Mice merger, created from an hydrodynamical simulation, which
we analyse using the same codes as the real data cube. We collate
the results in Section \ref{sec:disc}, to reveal the extent to which the
interaction has affected the properties of the progenitor galaxies.

Throughout the paper we assume a flat cosmology with $\Omega_{\rm
  M}=0.272$, $\Omega_\Lambda=0.728$ and
$H_0=70.4$km\,s$^{-1}$Mpc$^{-1}$
(WMAP-7)\footnote{http://lambda.gsfc.nasa.gov/product/map/current/best\_params.cfm}. Masses
assume a Salpeter initial mass function unless otherwise stated.  From
the redshift of the Mice we obtain a distance of 95.5\,Mpc, assuming
negligible peculiar velocity contribution. The proximity of the Mice
to the Coma cluster leads to some uncertainty on their distance;
unfortunately no redshift-independent distance measures are available.
At a distance of 95.5\,Mpc, 1\arcsec\ corresponds to 0.44\,kpc and the
effective spatial resolution of the {\sc CALIFA} observations
(3.7\arcsec) is about 1.6\,kpc.

\section{Summary of previous observations of the Mice pre-merger}

\begin{table*}
    \caption{Properties of the Mice galaxies culled from the
      literature.  \label{tab:basic}}
    \centering
    \begin{tabular}{ccc}\hline\hline
      Parameter      & NGC~4676A (NW) / IC~819  & NGC~4676B (SE) / IC~820\\\hline
      RA (EquJ2000)  & 12h46m10.110s   & 12h46m11.237s \\
      Dec (EquJ2000) & +30d43m54.9s    & +30d43m21.87s\\
      CALIFA ID & 577 & 939\\
      SDSS objid    & 587739721900163101 & 587739721900163099 \\
      Redshift\tablefootmark{a}   & 0.02206       & 0.02204 \\
      $r$-band magnitude\tablefootmark{b} & 13.22 & 13.03\\
      M(\hi) [$10^9$M$_\odot$]\tablefootmark{c}  & 3.6 & 4.0 \\
      M(H$_2$) [$10^9$M$_\odot$]\tablefootmark{d} & 5.7  & 3.6 \\
      M$_{dyn}$ ($10^9$M$_\odot$\tablefootmark{e} & 74 & 129 \\
      L$_{\rm FIR}$ [$10^{10}$L$_{\odot}$]\tablefootmark{f}  & 3.3 & 0.9\\
      L$_{X,0.5-2keV}$ [$10^{40}$\ergs]\tablefootmark{g} & 0.6  & 1.2 \\
      L$_{X,2-10keV}$ [$10^{40}$\ergs]\tablefootmark{g} & 0.9  & 1.5 \\
      \hline
    \end{tabular}
    \tablefoot{  Physical parameters have been converted to the same distance and Hubble
    parameter used in this paper (95.5\,Mpc; $H_0=70.4$km\,s$^{-1}$Mpc$^{-1}$). The
    unresolved pair is also known as Arp 242 and IRAS
    12437+3059.
    \tablefoottext{a}{From \hi\ 21cm \citep{deVaucouleurs:1991p7177}.}
    \tablefoottext{b}{From a growth curve analysis of the SDSS $r$-band image (Walcher
      et~al. in prep.). The SDSS catalogue magnitude for
      NGC~4676B is incorrect (see also Section \ref{sec:mass}).}
    \tablefoottext{c}{\citet{Hibbard:1996p7306}.}
    \tablefoottext{d}{\citet{Yun:2001p7305} for NGC~4676A and \citet{Casoli:1991p7310}
      for NGC~4676B.}
    \tablefoottext{e}{From optical extent and \hi\ line width \citep{Hibbard:1996p7306}.}
    \tablefoottext{f}{$L_{\rm FIR}$ as defined by \citet{Helou:1988p1394}. The Mice
      are not individually resolved by IRAS; these estimates from
      \citet{Yun:2001p7305} are from the total FIR flux and 1.4GHz
      radio continuum ratio.}
    \tablefoottext{g}{\citet{GonzalezMartin:2009p6907}.}
  }
\end{table*}

As one member of the original \citet{Toomre:1977p7267} merger
sequence, the Mice have been observed at most wavelengths available to
astronomers. The basic kinematics of the Mice were first studied using
long-slit spectroscopy covering the \ha+\nii\ lines
\citep{Burbidge:1961p7316}, who found the northern sides of both hulks
to be receding. \citet{Stockton:1974p7470} found \ha\ emitting gas in
the northern tail to be receding from the body of NGC~4676A with a
velocity of 250\,\kms\ at a distance of 90\arcsec\
(40\,kpc). \citet{Barnes:2004p6826} presented a hydrodynamic
simulation of the Mice, where a reasonable match to the observed
hydrogen gas dynamics was obtained. In the remainder of this Section
we give a brief summary of the known physical properties of the two
galaxies, based on previous observations. These properties are
collated in Table \ref{tab:basic}. We focus on each galaxy in turn.

The northern NGC~4676A is a massive, gas-rich, actively star-forming
disk galaxy, viewed almost edge-on. It was classified as an S0 galaxy
by \citet{deVaucouleurs:1991p7177}, although we suggest a revised
classification below. 
From neutral Hydrogen and optical observations, it is estimated to
have a dynamical mass  within the optical disk of 7.4$\times10^{10}$\msol\
\citep{Hibbard:1996p7306}. Molecular gas is primarily located in a
central disk with scalelength of 2\,kpc and thickness of 270\,pc and this 
disk has a large molecular-to-dynamical mass ratio of 20\%
\citep{Yun:2001p7305}. The two galaxies were not individually resolved
by IRAS, but recent Spitzer Space Telescope observations find that
NGC~4676A accounts for 83\% of the total 24\mum\ flux of the system
\citep[][ and Section \ref{sec:MWsfr}]{Smith:2007p8201}. Previous
estimates of its star formation rate (SFR) range from
$\sim$1\,\msolyr\ from a narrow band \ha\ image where corrections for
\nii\ emission, stellar absorption, dust attenuation or non-stellar
emission were not possible \citep{Mihos:1993p7313}, to 10\,\msolyr\
using the unresolved IRAS flux and resolved radio continuum
\citep{Hibbard:1996p7306,Yun:2001p7305}. We update this estimate in
Section \ref{sec:MWsfr} using the full multi-wavelength information
available to us today. The steep mid-IR continuum of NGC~4676A
suggests a high global heating intensity compared to normal
star-forming galaxies, perhaps caused by a nuclear starburst
\citep{Dale:2000p7291, Haan:2011p7283}.

NGC~4676A has LINER-like optical emission line ratios
\citep{Keel:1985p7462}, but has a diffuse X-ray morphology, leading
\citet{GonzalezMartin:2009p6907} to conclude that NGC~4676A does not
harbour an AGN.  Plumes of \ha\ extending along the minor-axis of
NGC~4676A were identified by \citet{Hibbard:1996p7306}. These have
been linked to outflowing gas from the nucleus, coincident with
diffuse soft X-ray emission\citep{Read:2003p7304}. The unusually high
ionised-to-neutral (7.7/11.3\mum) PAH ratio found to the east of the
nucleus by \citet{Haan:2011p7283} may also be associated with this
outflow, or its driving source.

The south-east NGC~4676B is a massive, strongly barred spiral
galaxy. Its faintness at FIR wavelengths indicates a lower overall
star formation rate than its partner, and the mid-IR continuum shape
suggests a star formation intensity consistent with normal
star-forming galaxies \citep{Haan:2011p7283,Dale:2000p7291}.
The dynamical mass  within the optical disk is measured to be 1.29$\times10^{11}$\msol\
\citep{Hibbard:1996p7306}, about 50\% larger than that of
NGC~4676A. Concentrations of molecular gas are located at either end
of the strong bar \citep{Yun:2001p7305}.  Single-dish CO observations
by \citet{Casoli:1991p7310} result in a CO line flux that is a factor
of 2.5 greater than the interferometric measurement of
\citet{Yun:2001p7305}, which suggests that a large fraction of the
molecular gas in NGC~4676B is either low surface brightness or
extended over scales larger than those probed by the interferometric
observations ($\theta>45$\arcsec, 20\,kpc). The total Hydrogen mass of
NGC~4676B is a little lower than that of NGC~4676A
\citep{Hibbard:1996p7306}; given its larger dynamical mass, NGC~4676B
has a significantly lower gas mass fraction than NGC~4676A.

With a LINER-like emission line spectrum \citep{Keel:1985p7462},
compact X-ray emission and a detection in hard X-rays
($L_{2-10}=1.48\times10^{40}$\ergs), NGC~4676B is classified as a
``candidate'' AGN by \citet{GonzalezMartin:2009p6907}.  Additional
evidence for the presence of an AGN comes from: (i) the unusually high
ratio of excited H$_2$ emission to PAH emission
\citep{Haan:2011p7283}, consistent with additional excitation of H$_2$
by hard X-rays from a central AGN \citep{Roussel:2007p7296}; (ii) the
hard X-ray emission is offset from the nuclear H$\alpha$ emission
\citep{Masegosa:2011p6966}, showing that the hard X-rays are not
related to a nuclear starburst.


In summary, previous observations have shown that NGC~4676A is an
edge-on disk, rich in dust, molecular and atomic gas, while NGC~4676B
is a strongly barred, inclined disk, with a lower gas mass
fraction. There is some evidence for a low luminosity AGN in NGC~4676B, but
no evidence for an AGN in NGC~4676A.  A bipolar outflow is
found in NGC~4676A. Table \ref{tab:basic} summarises the
properties of the Mice galaxies that can be found in the literature.

\section{The CALIFA Data}\label{sec:califa}

\begin{figure*}
\centering
\includegraphics[width=17cm]{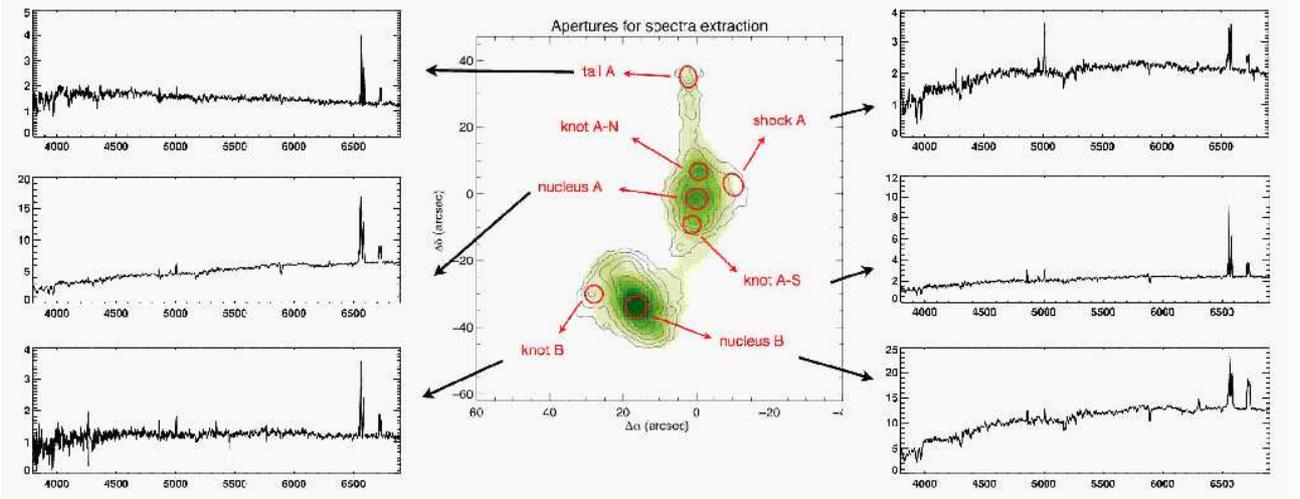}
\caption{The \ha\ map of the Mice galaxies with examples of V500
  CALIFA spectra, extracted from key regions.  For the nucleus and
  knots of both galaxies the spectra are extracted using a circular
  aperture of 2\arcsec\ radius, for the tail and outflow regions in
  NGC~4676A an elliptical aperture of 4\arcsec$\times$3\arcsec\ was
  extracted.}\label{fig:regions}
\end{figure*}

The Calar Alto Legacy Integral Field Area ({\sc CALIFA}) survey,
currently in progress on the 3.5m Calar Alto telescope in Spain, aims
to obtain spatially resolved spectra of 600 local galaxies spanning
the full colour-magnitude diagram, with the PPAK integral field unit
(IFU) of the PMAS instrument \citep{Roth:2005p7546}. The galaxies are
diameter selected to be volume correctable from the seventh data
release (DR7) of the Sloan Digital Sky Survey (SDSS).  Observations are
performed using two overlapping grating setups (V500 and V1200), with
resolutions of 6.3\AA\ and 2.3\AA\ (FWHM) and wavelength ranges of
3745--7500\AA\ and 3650--4840\AA\ respectively (velocity resolution,
$\sigma\sim$100--210\,\kms\ and 60--80\,\kms). The spectrophotometric
calibration accuracy is close to that achieved by the SDSS DR7. The
large field-of-view of PPAK ($1.3\square\arcmin$) allows coverage of
the full optical extent of the galaxies. The datacubes reach a
$3\sigma$ limiting surface brightness of 23.0\,mag/arcsec$^2$ for the
V500 grating data and 22.8\,mag/arcsec$^2$ for V1200.  The survey is
described in detail in the presentation article
\citep{Sanchez:2012p7464} and first Data Release article
\citep[DR1,][]{Husemann:2013p8375}.

Observations of the Mice galaxies were taken during the nights of
2011-05-04 (V500), 2011-05-05 (V1200) and 2011-06-29 (V1200). Three
dithered pointings were made, centred around each of the two galaxy
nuclei, with exposure times of 900\,s per pointing for V500 and
$2\times900$\,s for V1200.  
The PPAK IFU data of the Mice were reduced with the dedicated {\sc
  CALIFA} data reduction pipeline and are available to the public as
part of DR1. The basic outline of the pipeline is given in
\citet{Sanchez:2012p7464}, with significant improvements made for the
DR1 release and described in \citet{Husemann:2013p8375}. In addition
to the data available in the public release, a combined cube with
observations from both gratings was created to improve the vignetting
of the blue and red end of the V500 and V1200 data
respectively. Wavelengths shorter (longer) than 4500\AA\ originate
from the V1200 (V500) data, with the V1200 data cube degraded to the
spectral resolution of the V500 data. The V500 and degraded V1200
cubes were then spatially co-registered during the process of
differential atmospheric refraction (DAR)
correction and their relative spectrophotometry matched in the
overlapping spectral regions before being combined into a single cube.

Finally, for the purposes of presentation in this paper, the cubes for
NGC~4676A and NGC~4676B were cropped diagonally between the galaxies
and joined to form a single cube, with the relative offset between the
two nuclei determined from their SDSS centroids. This procedure is
only astrometrically accurate to within the size of the spaxels,
i.e. 1\arcsec, which was deemed sufficient given the spatial
resolution of the data of $\sim$3.7\arcsec. Voronoi binning was
performed on the individual cubes, in order to retain independence of
the two sets of observations.

Figure \ref{fig:regions} presents the \ha\ emission line intensity map of the
Mice galaxies, measured from the V500 data, and provides examples of
spectra extracted from key regions of interest. 

\section{Morphology and kinematics}\label{sec:kin}

We begin by combining archival HST imaging and {\sc CALIFA}
spectroscopy to build a global picture of the morphology and
kinematics of the merging galaxies. 

\subsection{Image decomposition}\label{sec:galfit}
\begin{table*}
  \caption{ Results of a GALFIT 3.0 decomposition of the HST ACS F814W
     image. \label{tab:galfit}}
 \centering
 \begin{tabular}{lll} \hline\hline
   Component           & magnitude   & parameters \\\hline
   A : Edge-on Disk  & 14.7:\tablefootmark{a} & edgedisk: $\mu_0$=21.8  ;  h$_s$=3.5  ; $r_s$=29:  ;  PA= 5\\
   A : Bar                   & 14.1\tablefootmark{a} & ferrer2: $\mu_{\rm  FWHM}$=19.7 ;   R$_{\rm trunc}$=19;  $\alpha$=2; $\beta$=1; b/a=0.6; PA= 4\\
   B : Bulge                & 14.2   & S\'ersic: R$_e$=6.0: ;   n=4.3: ; b/a=0.75 ; PA=-26 \\ 
   B : Disk+Bar          & 13.8   & expdisk: R$_s$=6.2 ; b/a=0.4 ; PA= 36\\
   B : Point source    & 20.9:  & PSF \\
      \hline
    \end{tabular}
\tablefoot{  Parameters are given as described in the GALFIT
  documentation. Length scales are in arcseconds, position angles
  are from N to E (anti-clockwise in the figures) and surface
  brightnesses are in mag/arcsec$^2$. Parameter values
  should be taken as indicative only, and those marked with ``:'' are
  particularly uncertain due to the complex morphology of the
   system.
   \tablefoottext{a}{Measured from the model image.}
}
\end{table*}

 To measure the main morphological components of the Mice galaxies
  we use GALFIT 3.0 \citep{Peng:2010p7175} on the archival HST ACS
  F814W image. The instrumental point-spread-function (PSF) was
synthesised using TinyTim; Galactic stars, tidal arms and prominent
dust lanes in NGC~4676A were masked using SExtractor. The parameters
of the fit are presented in Table \ref{tab:galfit}.

We find that NGC~4676A is dominated by an edge-on disk and/or bar. No evidence
could be found for either a nuclear point source or bulge, although
the central dust lanes may still hide these components. The boxy shape
apparent in the HST image could arise from vertical motions of stars
in a bar, rather than a bulge
\citep[e.g.][]{Kuijken:1995,Bureau:1999p9430,Bureau:2005p8704,
  MartinezValpuesta:2006p7585,Williams:2011,Lang:2014p9453}, and the model fit
improves upon inclusion of a Ferrer bar component. Our
results show that the previous classification of NGC~4676A as an S0
galaxy is likely incorrect \citep{deVaucouleurs:1991p7177}, and SBd is
more appropriate. This revised classification is also consistent with
the high dust and gas content of the galaxy.

The inclination of NGC~4676B allows the identification of a
strong bar, although our simple image decomposition is unable to
separate the bar from the disk component.  By ellipse fitting the HST
image isophotes using the method presented in
\citet{Aguerri:2009p7278}, we find the radius of the bar in NGC~4676B
to be 10\arcsec\ (4.4\,kpc) and position angle (PA)
$\sim$20$^\circ$. This is consistent with the peaks of CO, separated
by $\sim16$\arcsec\ and thought to lie at either end of the bar.  A
significant bulge is required to obtain a good fit. Including a
nuclear point-source component also improves the fit slightly and
decreases the S\'ersic index of the bulge as expected
\citep{Gadotti:2008p7586}.  The best fit model has a bulge-to-total
flux ratio of $\sim0.5$ i.e. an S0/a galaxy \citep{Simien:1986}. 

\subsection{Stellar kinematics}

\begin{figure*}
\centering
\includegraphics[scale=0.6]{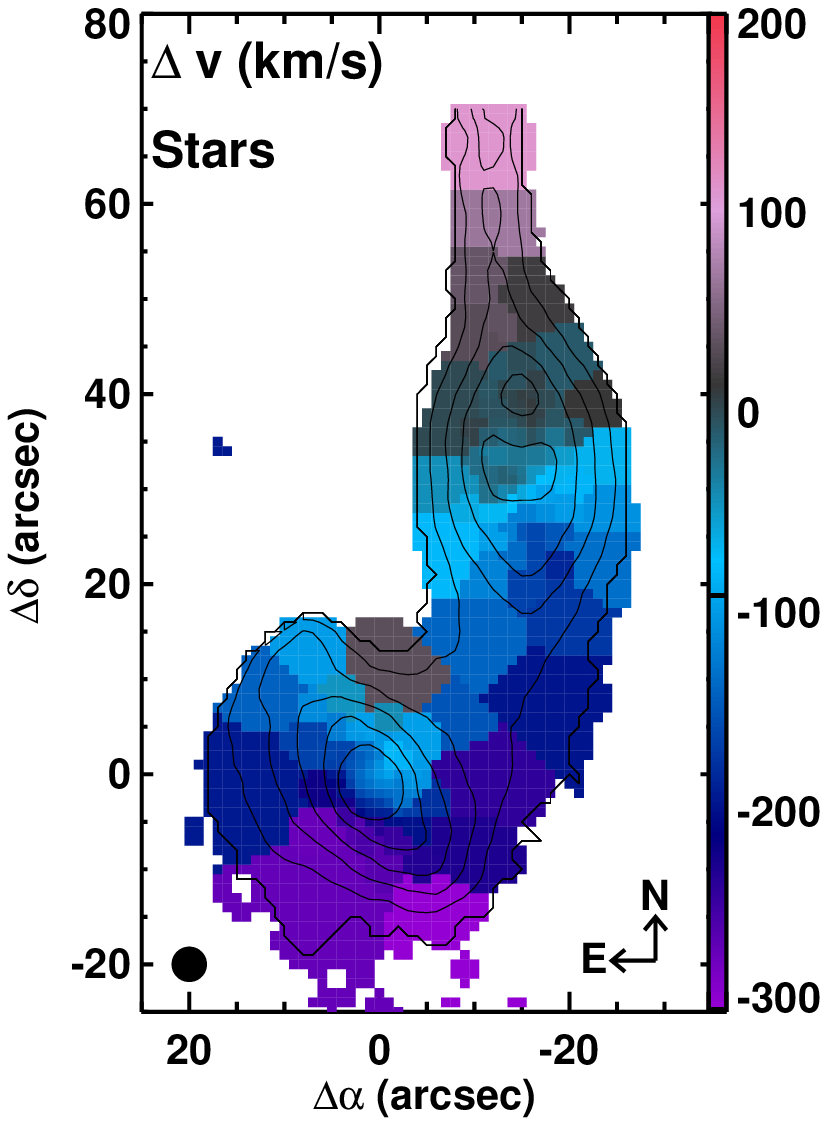}
\includegraphics[scale=0.6]{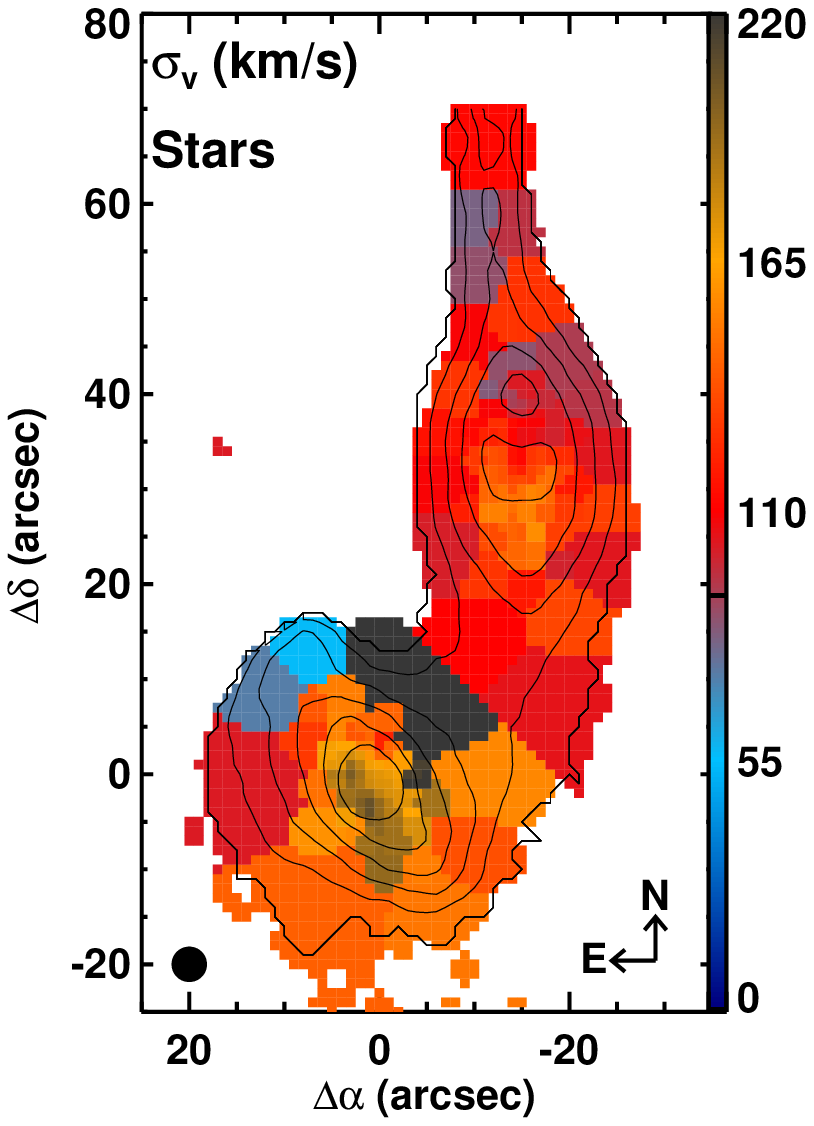}\\
\includegraphics[scale=0.6]{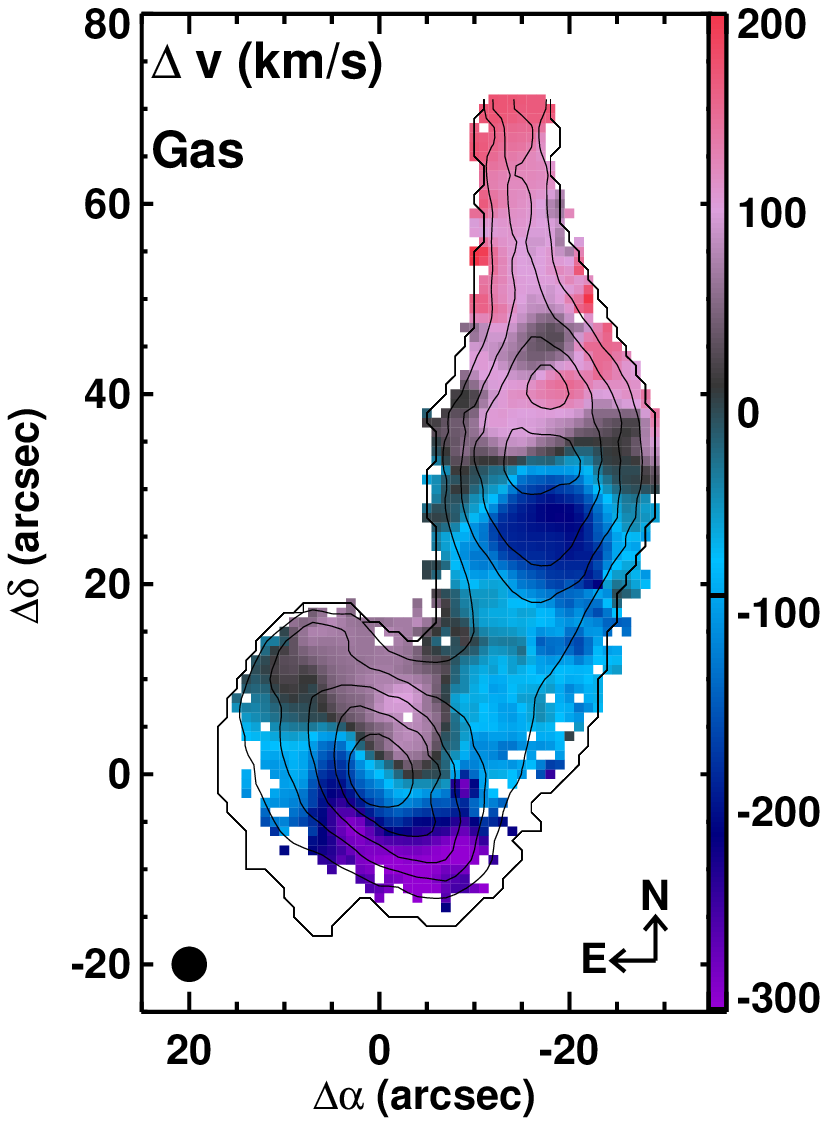}
\includegraphics[scale=0.6]{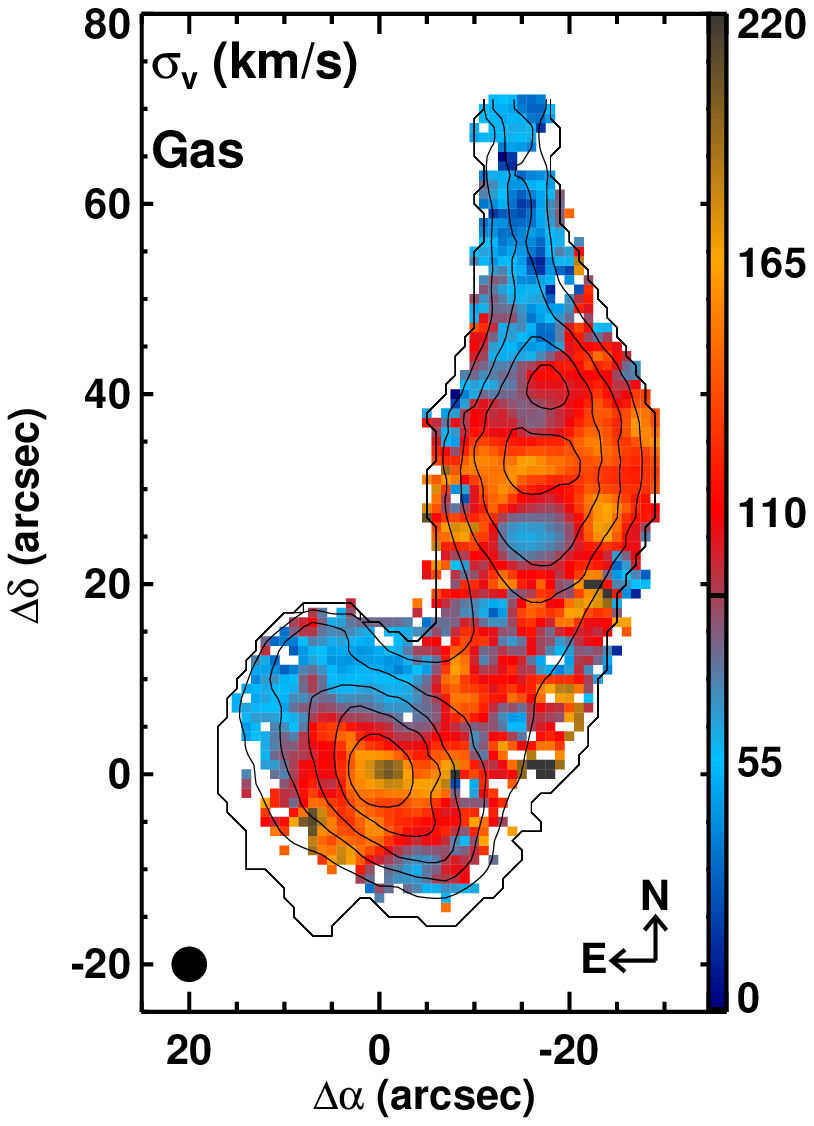}
\caption{Stellar (top) and ionised gas (bottom) velocity field (left)
  and velocity dispersion (right). The ionised gas maps are measured
  from the \ha\ emission line in the V500 grating. Velocities are
  relative to 6652\,\kms\ which corresponds to the heliocentric
  velocity of stars measured within the central 5\arcsec\ of
  NGC~4676A. The velocity dispersion maps have been corrected for
  instrumental resolution; velocity dispersions below $\sim$35\kms\
  for stars and $\sim$58\kms\ for the gas are unresolved at the
  instrumental resolution. Typical errors are 5-10\kms\ for the
  stellar velocities and 15\kms\ for the ionised gas. In these maps
  and all the following maps, the black contours indicate $V$-band
  isophotes with the outermost contour at 23\,mag/arcsec$^2$ and
  contours spaced by 0.6\,mag/arcsec$^2$. The black circle shows
  the effective spatial resolution of the CALIFA observations before
  Voronoi binning of the data.
}\label{fig:kin}
\end{figure*}

The stellar kinematics of the Mice galaxies are measured from the
V1200 grating datacube using the penalised pixel-fitting (pPXF) method
of \citet{Cappellari:2004p7594}. Full details of the method will be
given in Falc\'on-Barroso et al. (in prep.). Briefly, a non-negative
linear combination of a subset of 328 stellar templates from the
Indo-US library \citep{Valdes:2004p8280} covering a full range of
stellar parameters (T$_{eff}$, $\log$(g), [Fe/H]) is fit to the
spectra, following a Voronoi binning of spaxels
to achieve a minimum signal-to-noise (SNR) of
20 \citep{Cappellari:2003p7606}. These Voronoi binned spaxels are termed ``voxels'', and the
effective spatial resolution of Voronoi binned maps is naturally
degraded below the CALIFA effective resolution of 3.7\arcsec in the outer
regions. In the computation of the SNR we have taken into account the
correlation among the different spectra in the datacube as explained
in \citet{Husemann:2013p8375}. Spaxels in the original V1200 datacube
with per pixel SNR$<3$
are deemed unreliable and not included in the analysis. Emission lines
in the covered wavelength range are masked during the fitting
procedure (i.e. \oii, \neiii, \hd, \hg, \oiii, \heii, and
\ariv). Error estimates are determined via Monte Carlo simulations,
and are typically 5 (10)\kms\ for velocities and dispersions in the
inner (outer) voxels. Velocity dispersions are corrected for
instrumental resolution during the fitting process, velocity
dispersions below $\sim$35\kms\ are unresolved at the resolution of
the data.

The stellar velocity and velocity dispersion maps are shown in the top
panels of Figure \ref{fig:kin}, where the systemic redshift of the
system is taken as the velocity of the stars in the central 5\arcsec\
of NGC~4676A. The nucleus of NGC~4676A is receding at $\sim$160\,\kms\
relative to the nucleus of NGC~4676B. The stellar kinematics of the
main body of NGC~4676A show a rotating edge-on disk, with rotation
axis coincident with the minor disk axis. This rotation continues into
the northern tidal tail, which is receding at 130\,\kms\ relative to
the nucleus of NGC~4676A in the northernmost voxel (30-35\arcsec\ from
the nucleus). No evidence for a classical bulge is seen in this
galaxy, with low nuclear velocity dispersion and constant rotation
with height above the major axis. This supports the results from the
image analysis above. However, given the disturbed nature of the
system and the high dust attenuation close to the nucleus, high
spatial resolution longer wavelength observations would be required to
give a robust upper limit on bulge size. We note that we are unable to
identify ``pseudo'' bulges at the spatial resolution of the CALIFA
data.

The stellar velocity field of NGC~4676B shows a twisted inclined disk
(Z or S-shaped isovelocity contours), with the inner rotation axis
offset from the minor axis of the disk.  A direct analysis of the
kinematic maps provides a quantitative measurement of the kinematic
centres and PAs of the galaxies \citep[see][ for details of the
method]{BarreraBallesteros:2014p9452}. For a radius internal to
10\arcsec\ we measure receding and approaching stellar kinematic PAs
of 165 and 150 degrees respectively (anti-clockwise from north),
compared to a morphological PA of 33 degrees, confirming that the
dominant rotation in the inner regions is around the major axis of the
galaxy. The classical bulge in NGC~4676B is clearly visible from the
significant increase in velocity dispersion in the centre of the
galaxy, confirming the results from the morphological decomposition.

\subsection{Ionised gas kinematics}\label{sec:gaskin}

We measure the ionised gas kinematics from the \ha\ line in the V500
datacube, favouring the increased SNR provided by the stronger \ha\ line, over
the higher velocity resolution afforded
by the weaker lines in the V1200 cube.  We verified that our results
are consistent with those derived from \oii\ in the higher spectral
resolution V1200 datacube in the inner high surface brightness
regions. We use standard IRAF routines to fit simultaneously three
Gaussian line profiles to \ha\ and the two \nii\ lines, with both
amplitude and width free to vary independently, in every spaxel that
is included in the \ha\ emission line quality mask described in
Section \ref{sec:emission}.  The instrumental resolution was
subtracted from the measured line widths in quadrature
($\sigma=$116\,\kms\ at the wavelength of \ha). Typical errors on the
line velocities were estimated to be $\sim$15\,\kms\ from fitting
different species. Line widths below $\sim$58\kms\ are unresolved
at the CALIFA resolution.

The ionised gas kinematics derived from the \ha\ emission line are
shown in the bottom panels of Figure \ref{fig:kin}. The same global
kinematics are seen as in the stellar kinematics: NGC~4676A is
dominated by a rotating edge-on disk and NGC~4676B by a twisted disk
inclined to the line-of-sight. Evidence of a dynamically hot bulge is
seen in NGC~4676B, but not in NGC~4676A. The receding and approaching
kinematic PAs of the ionised gas disk in NGC~4676B are 350 and 161
degrees respectively, consistent with those of the stars i.e. the
twist in the disk is as strong in both the stellar and ionised gas
kinematic fields. The ionised gas in the N-E spiral arm of NGC~4676B and
northern tidal tail of NGC~4676A are the dynamically coldest regions of the system
with observed line widths close to the resolution of the CALIFA data,
implying velocity dispersions below $\sim$58\kms. The ionised gas in
the tail of NGC~4676A is receding at $\sim$180\,\kms, relative to the
body of NGC~4676A, at a distance of $\sim$35\arcsec. This is in
agreement with long-slit observations by \citet{Stockton:1974p7470}.

The zero-velocity curve of the ionised gas in NGC~4676A shows a
V-shape along the minor axis, indicating lower line-of-sight
velocities above the plane of the disk than within the disk. This
extra-planar gas also shows enhanced velocity dispersion, in streamers
extending radially outwards from the nucleus in the same direction as
the soft X-ray emission \citep{Read:2003p7304}.  These are dynamical
signatures of outflowing gas, i.e. a galactic superwind, which we will
return to discuss in more detail in Section
\ref{sec:outflowA}. Additional regions of high velocity dispersion are
observed close to the nucleus of NGC~4676B. In this case, the velocity
gradient is high and this effect might be equivalent to ``beam
smearing'' seen in \hi\
surveys, where multiple components are observed in a single resolution
element (3.7\arcsec\ in the case of {\sc CALIFA}).

We detect a blueshift of 30\kms\ of the gas relative to stars in the
nucleus of NGC~4676A, and an 88\kms\ redshift of the gas relative to
the stars in the nucleus of NGC~4676B.  In the case of NGC~4676A this
is a clear case for an outflow with some transverse velocity
component. The cause of the offset for NGC~4676B is less obvious, but
may be due to a tidal component passing in front of the
bulge and contributing significantly to the \ha\ luminosity. 

The spectral resolution of CALIFA is not sufficient to identify
dynamically distinct components in the emission lines, however
measurements of line asymmetries indicate where such multiple
components may exist.  Following the cross-correlation method of
\citet{GarciaLorenzo:2012p8279} for the \oiii\ emission line we find a
mixture of blue- and redshifted asymmetries in the western bicone of
NGC~4676A, again consistent with gas outflowing in bicones
perpendicular to the disk.

\section{Stellar populations}\label{sec:sfh}


\begin{figure*}
\centering
\includegraphics[scale=0.5]{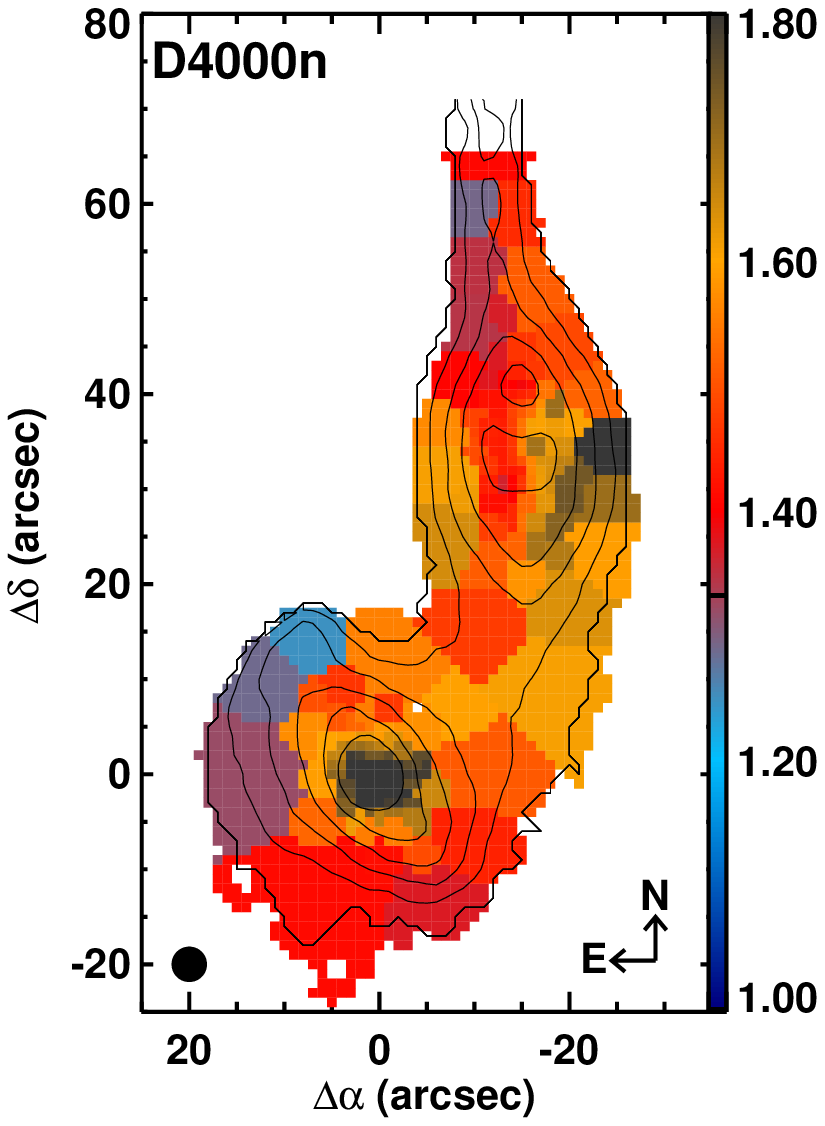}
\includegraphics[scale=0.5]{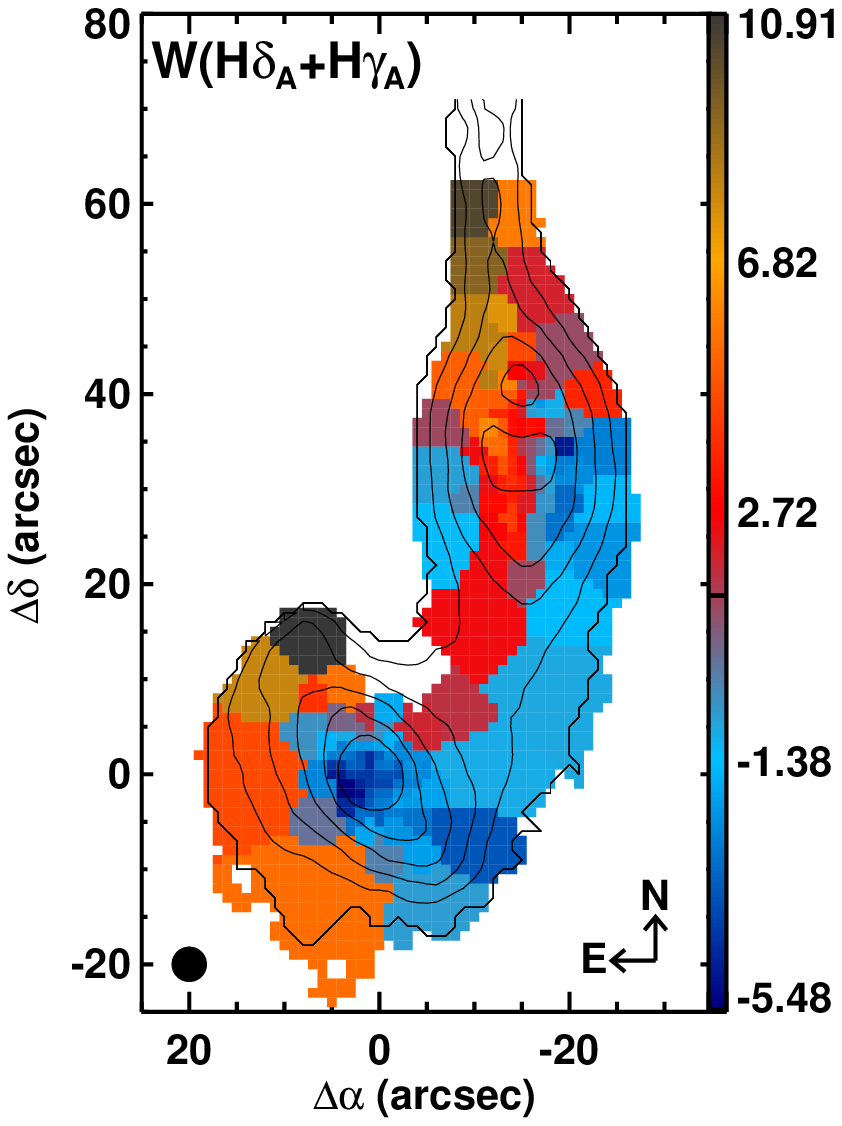}
\includegraphics[scale=0.5]{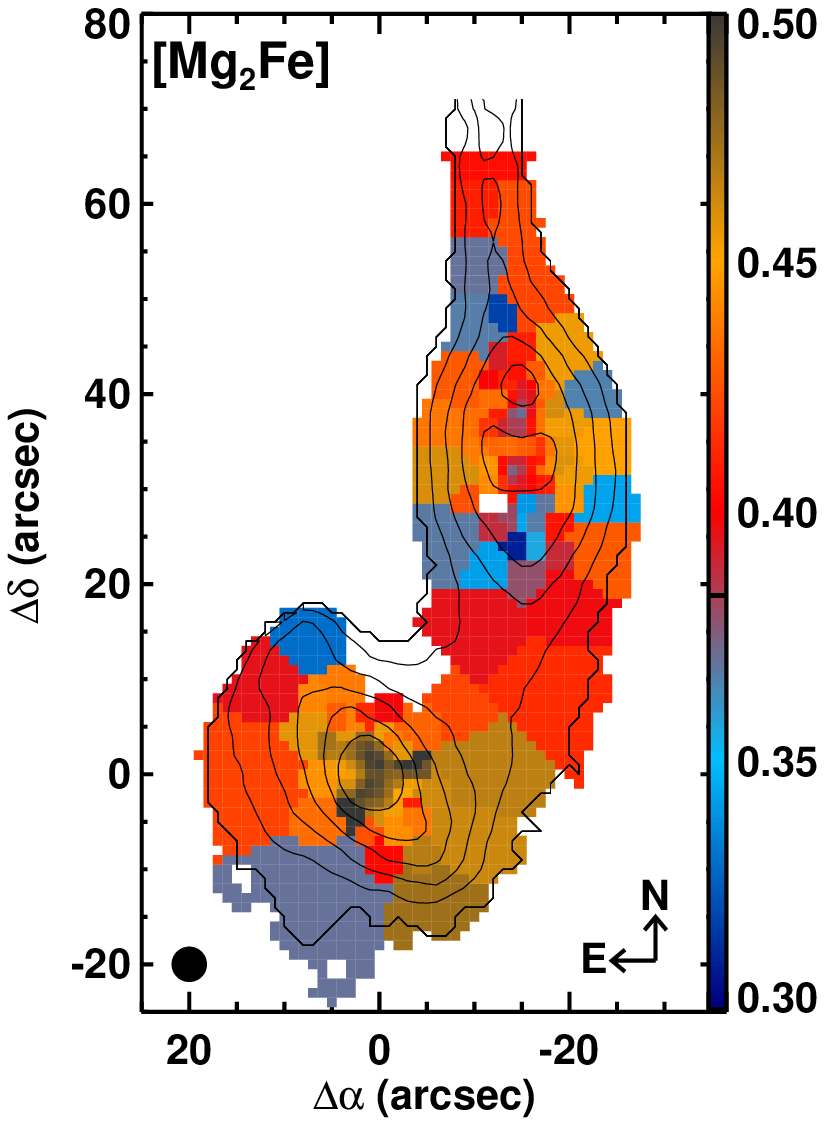}\\
\includegraphics[scale=0.5]{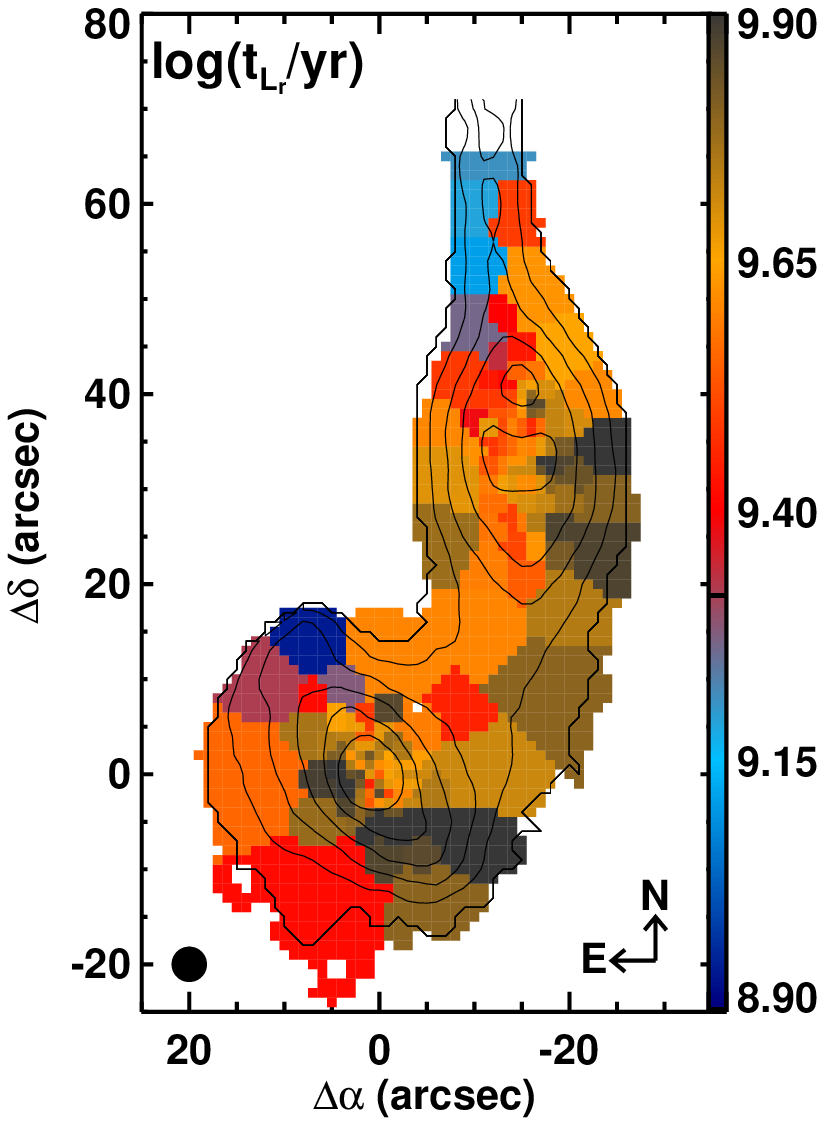}
\includegraphics[scale=0.5]{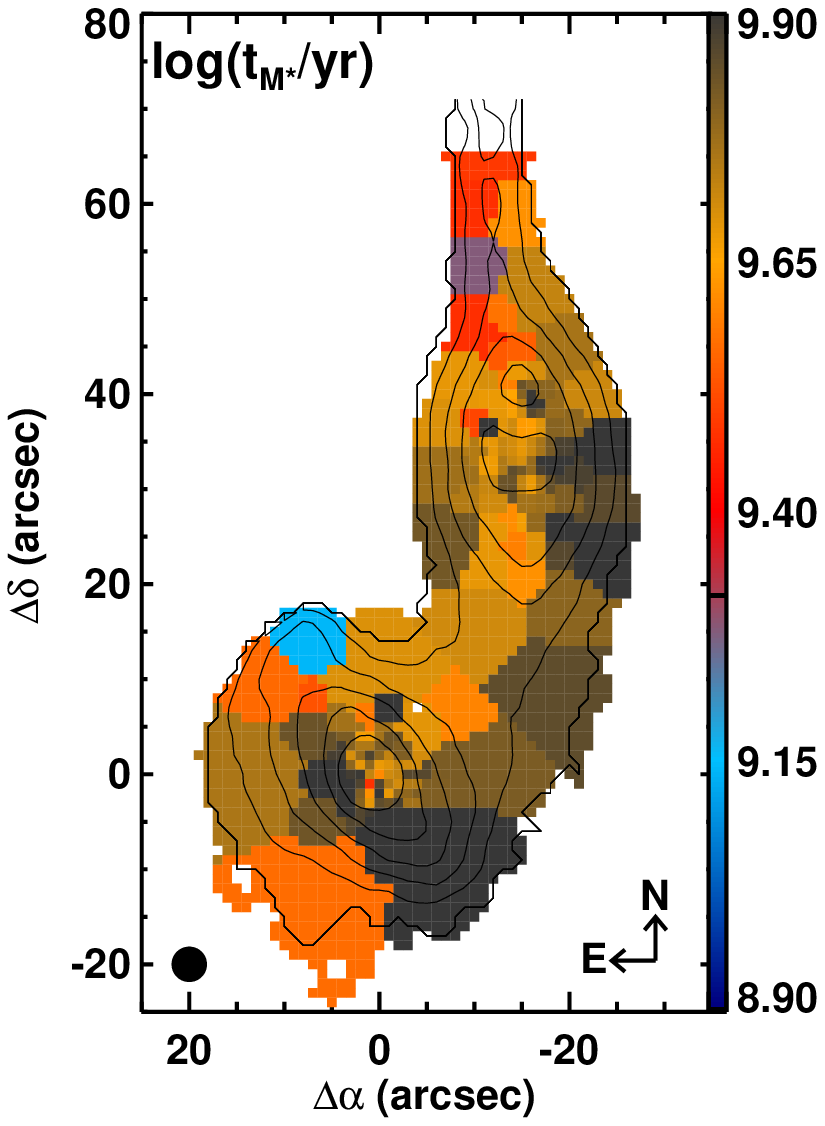}
\includegraphics[scale=0.5]{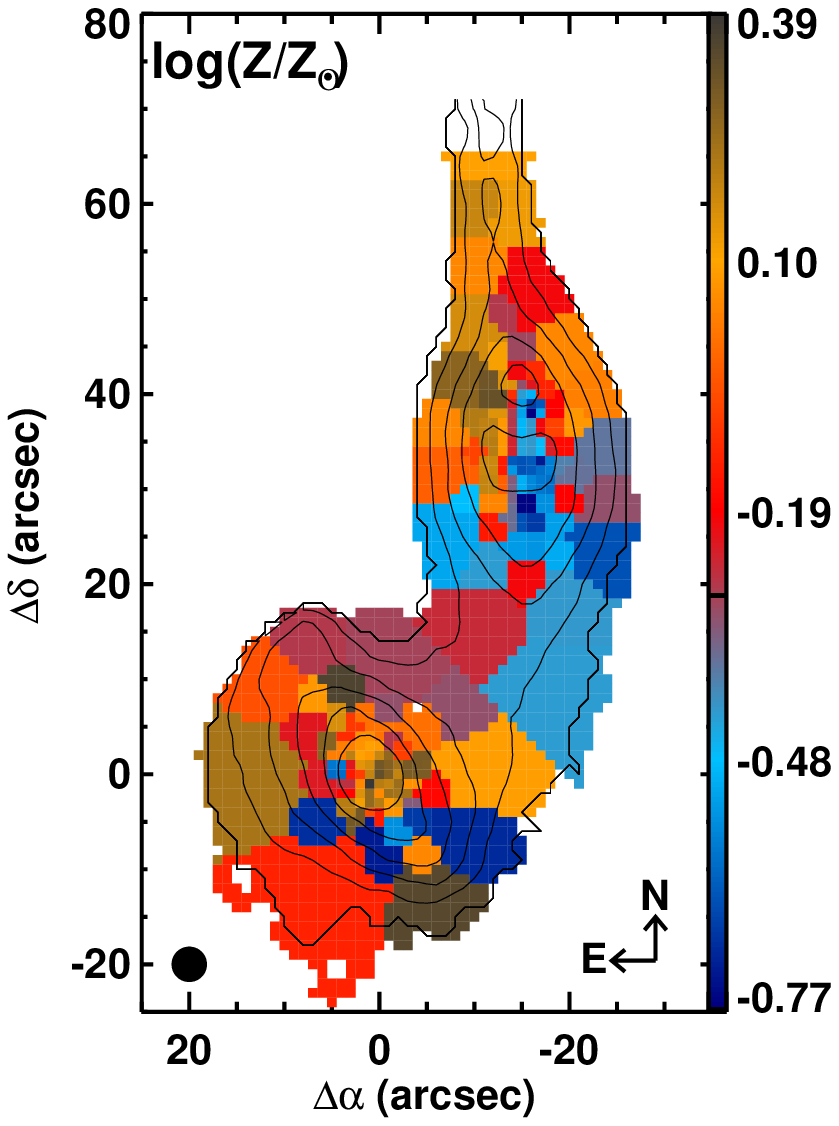}
\caption{\emph{Top:} Maps of three commonly used spectral line
  indices: the 4000\AA\ break strength (D$_n$4000); the combined
  equivalent width of the higher-order Balmer absorption lines \hda\
  and \hga (in \AA); the Lick index [Mg$_2$Fe]. \emph{Bottom:} Mean
  light-weighted age in the $r$-band, stellar mass-weighted age, and
  stellar metallicity. These have been measured from comparison of the
  line indices to stellar population synthesis models.
}\label{fig:lick}
\end{figure*}

\begin{figure*}
\centering
\includegraphics[scale=0.5]{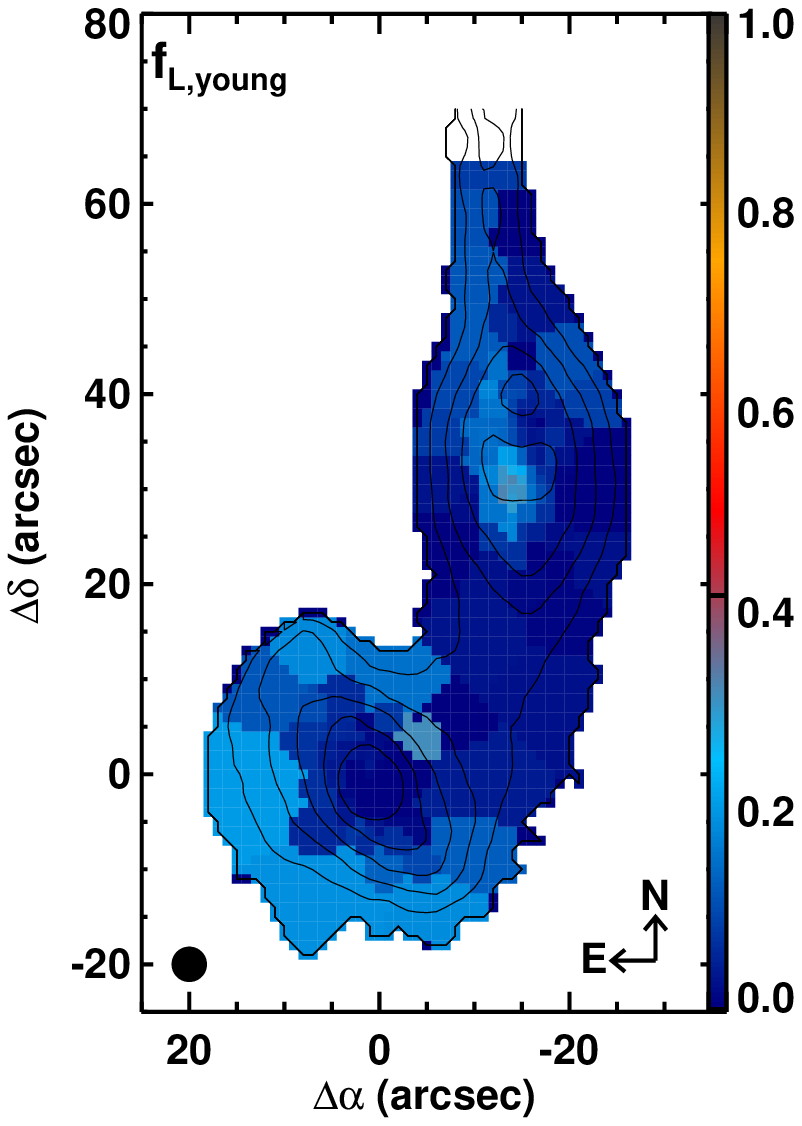}
\includegraphics[scale=0.5]{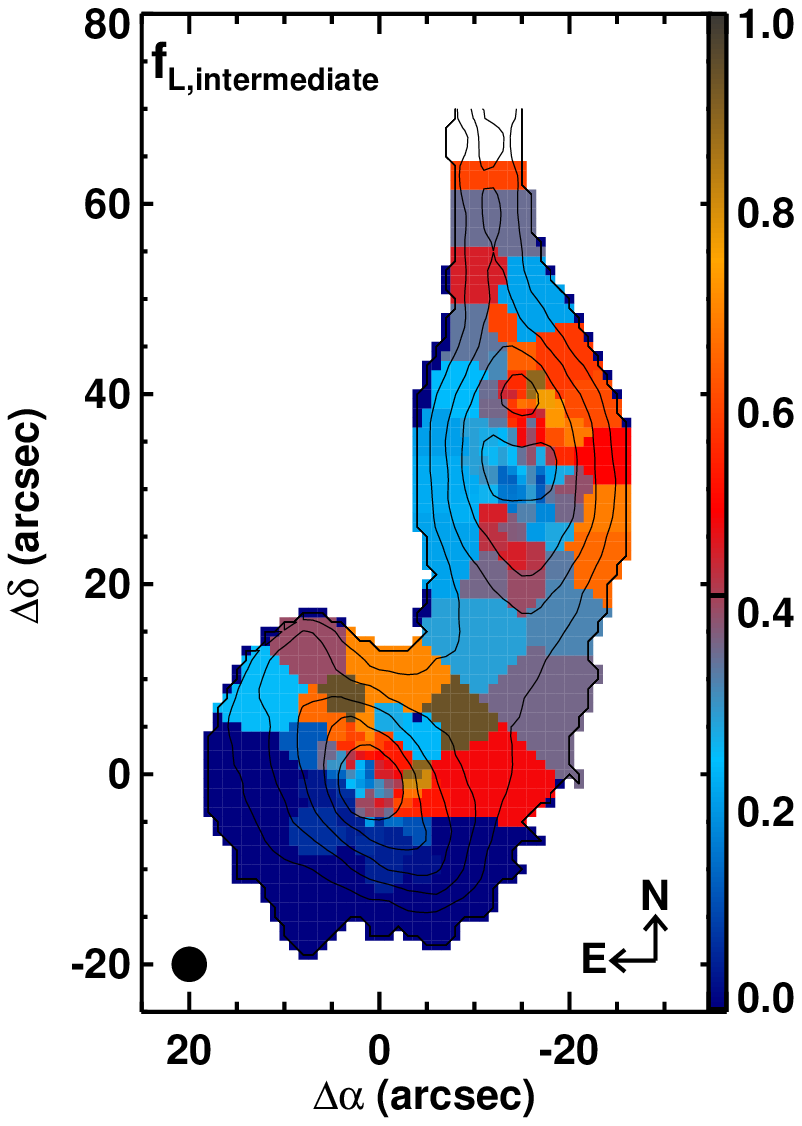}
\includegraphics[scale=0.5]{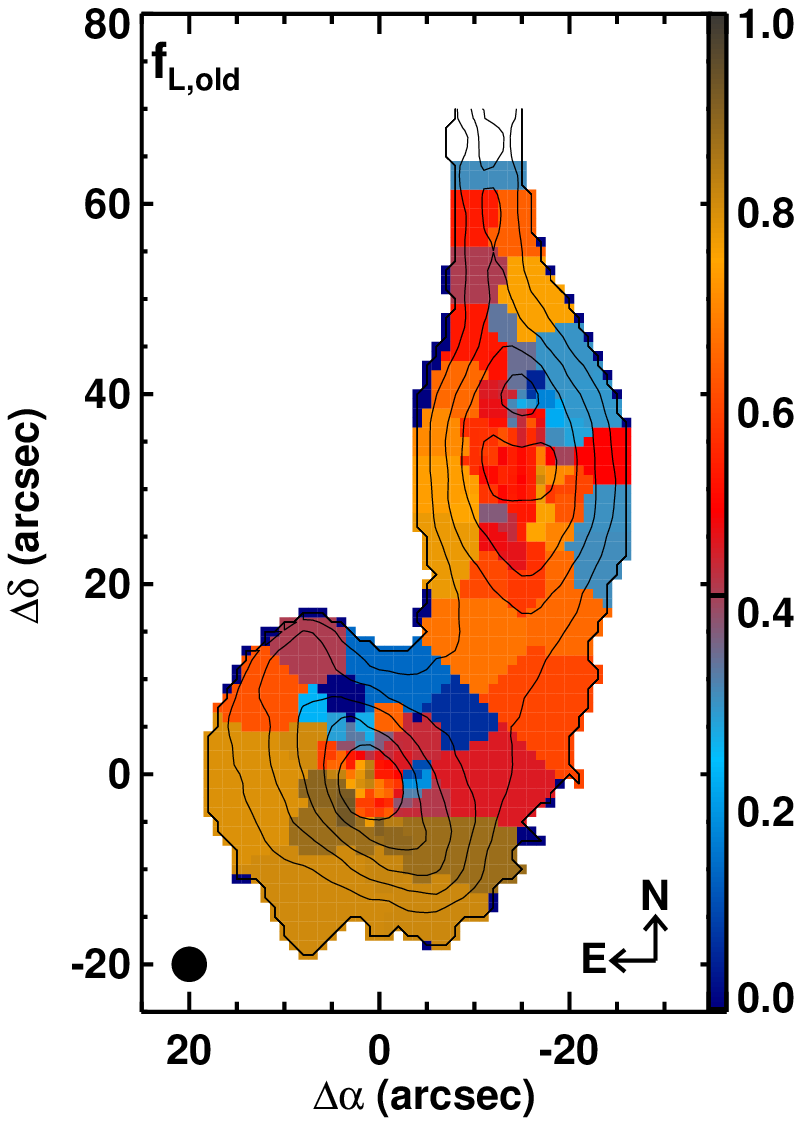}\\
\includegraphics[scale=0.5]{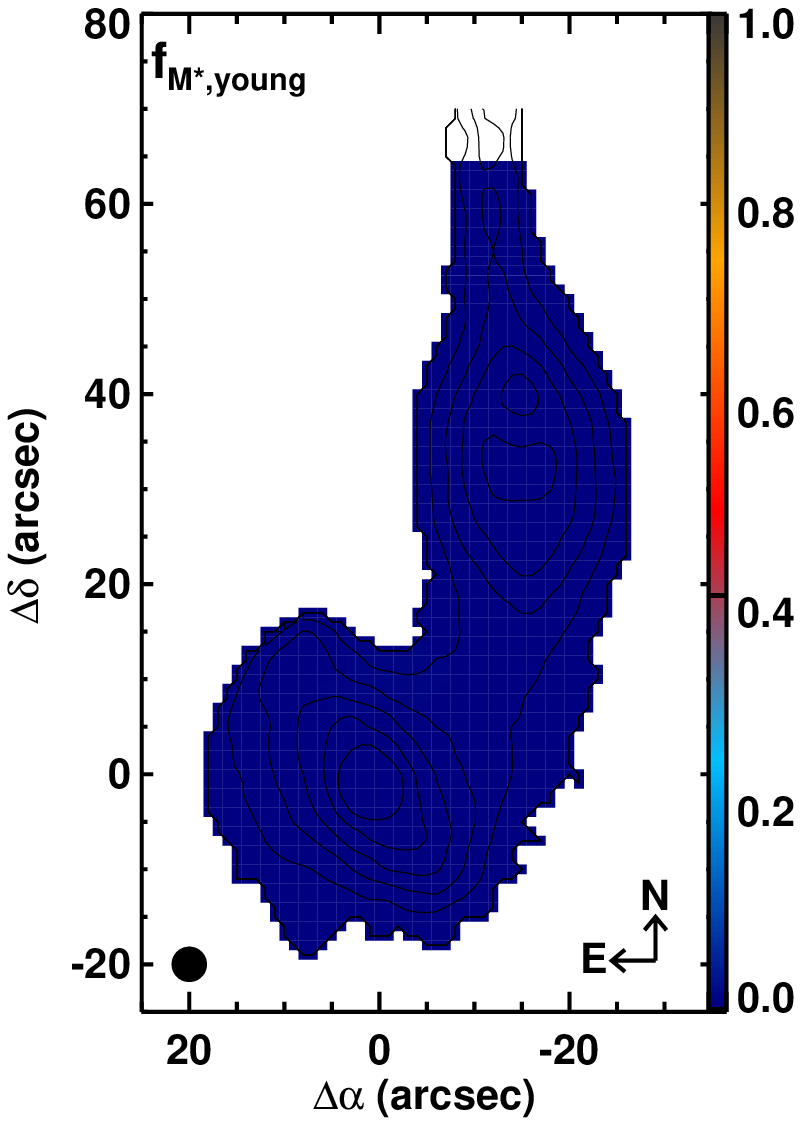}
\includegraphics[scale=0.5]{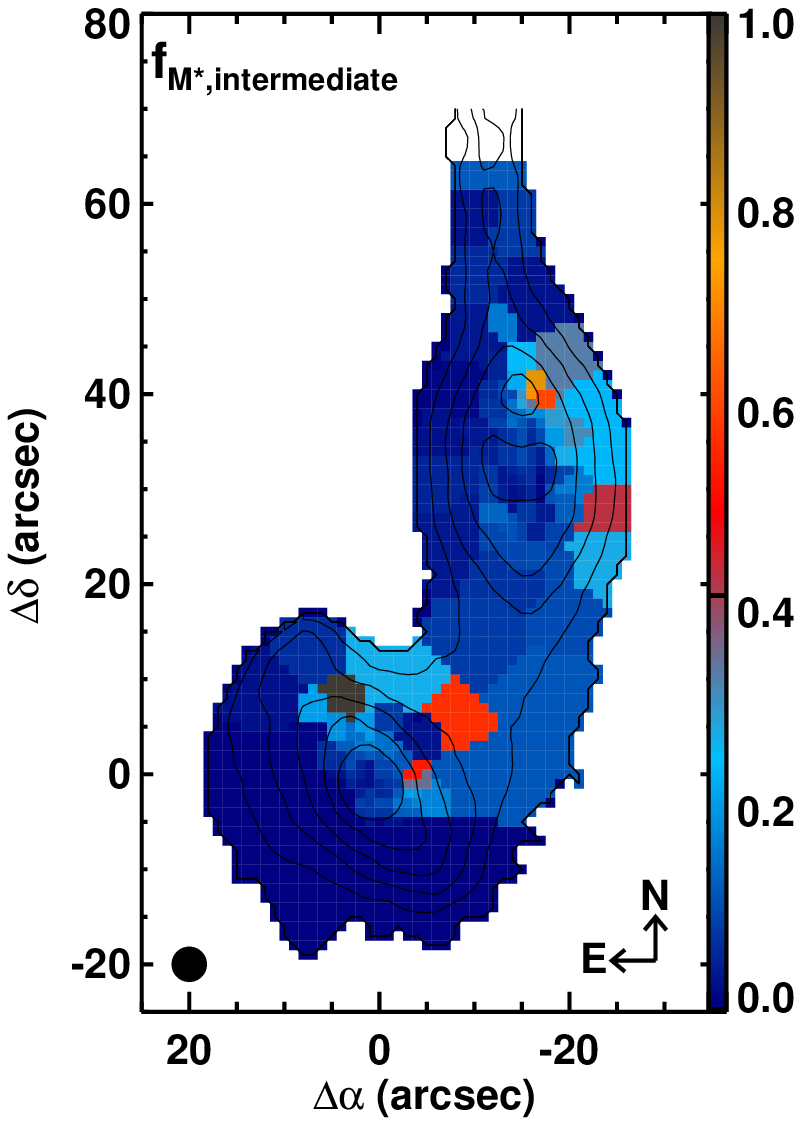}
\includegraphics[scale=0.5]{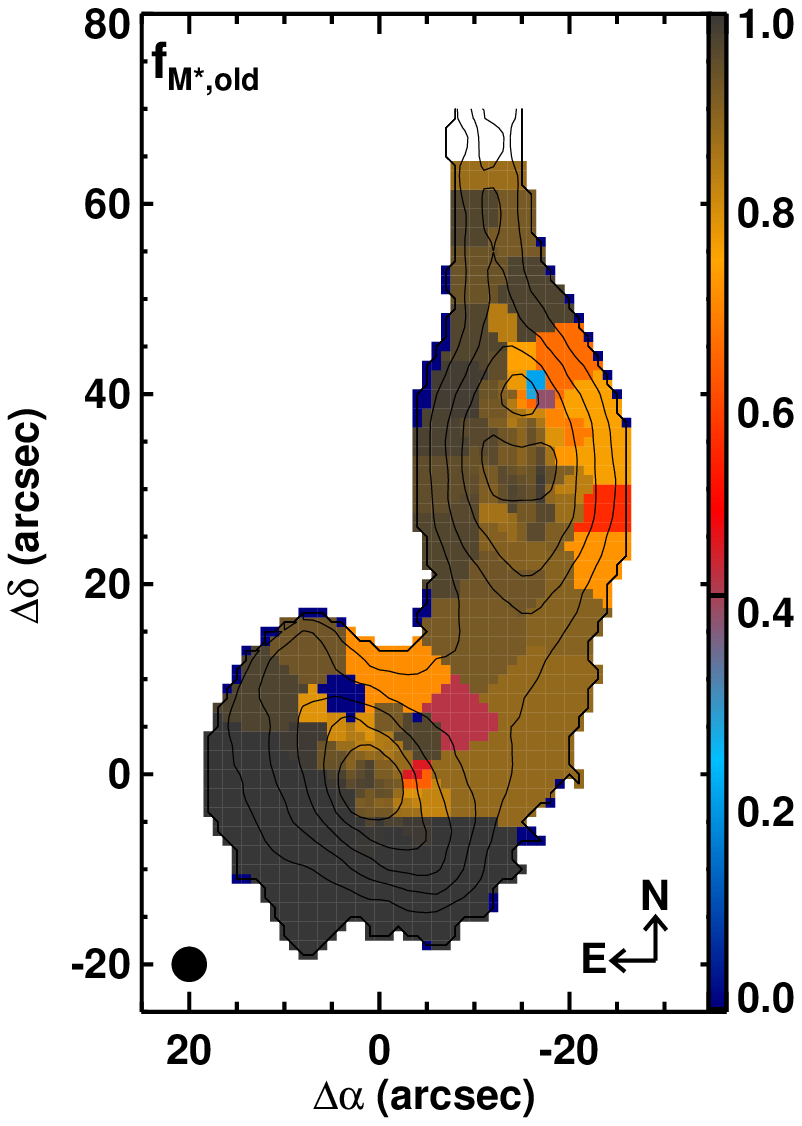}\\
\caption{The fraction of light at 5635\AA\ (top panels) and mass
  (lower panels) contributed by young ($t<$140\,Myr), intermediate
  (140\,Myr$<t<$1.4\,Gyr) and old ($t>$1.4\,Gyr)
  populations. }\label{fig:2dsfh}
\end{figure*}

\begin{figure*}
\centering
  \includegraphics[scale=0.6]{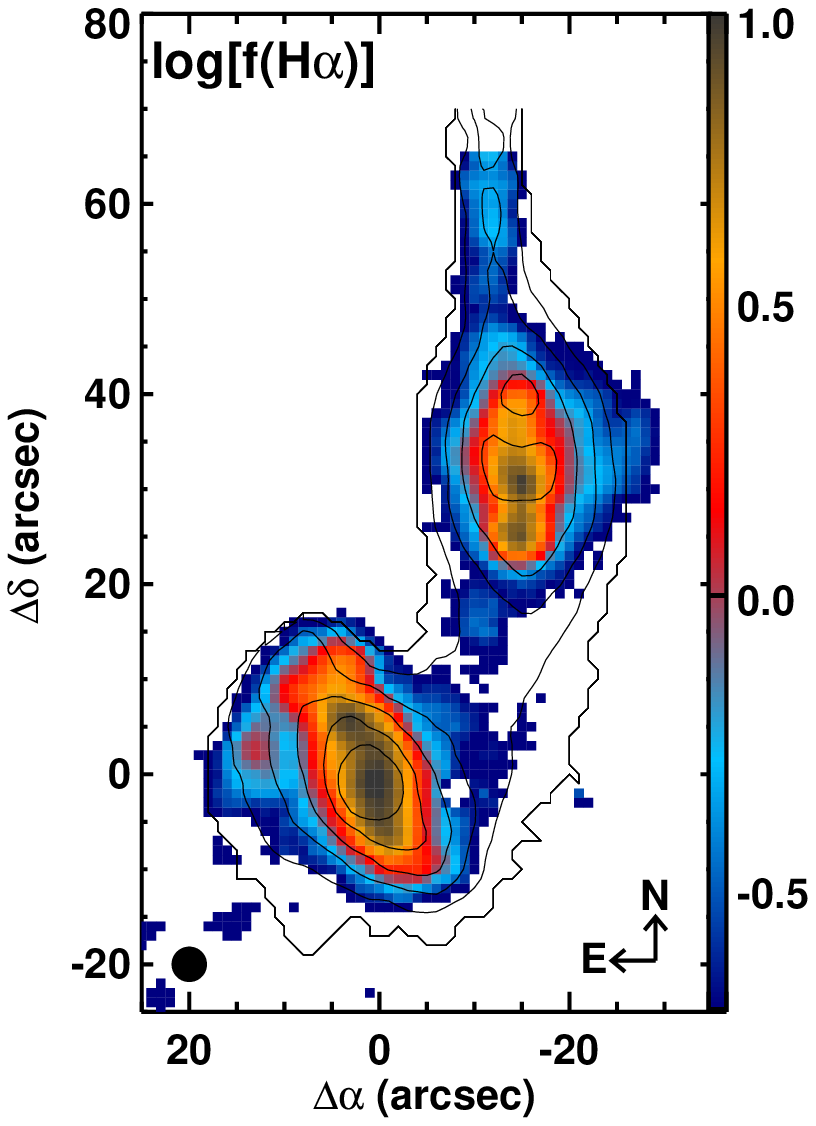}
  \includegraphics[scale=0.6]{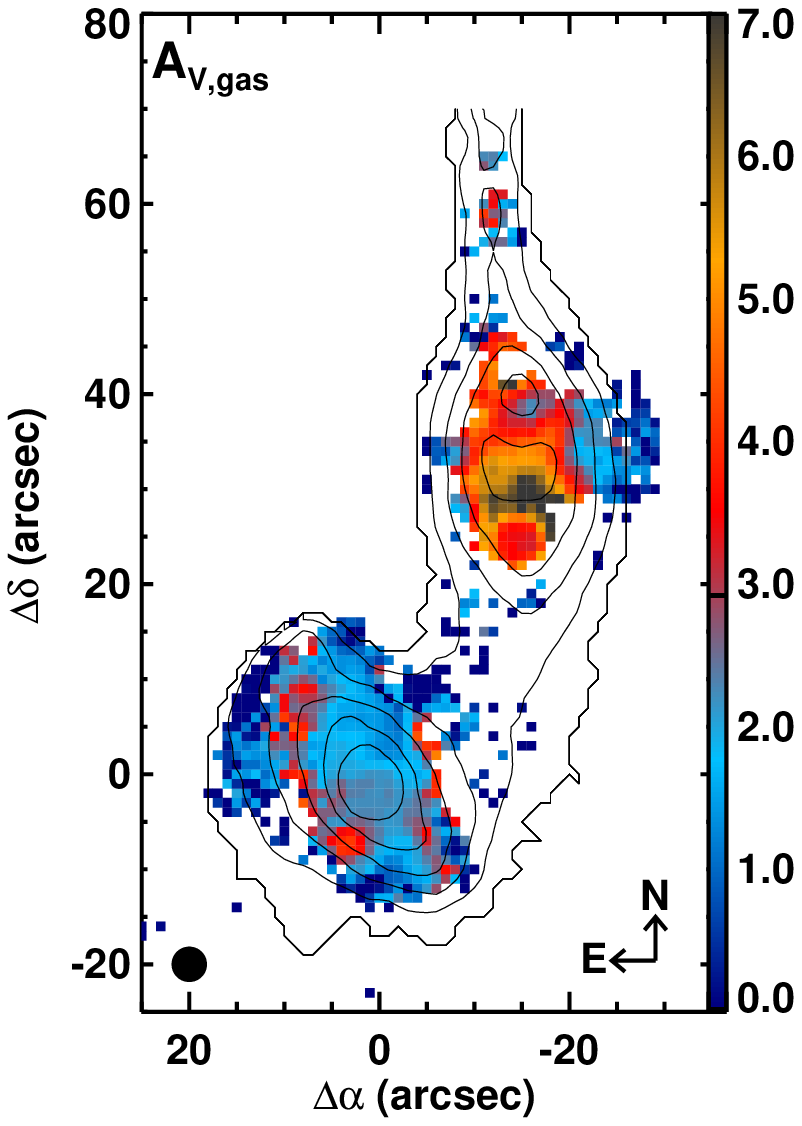}
  \includegraphics[scale=0.6]{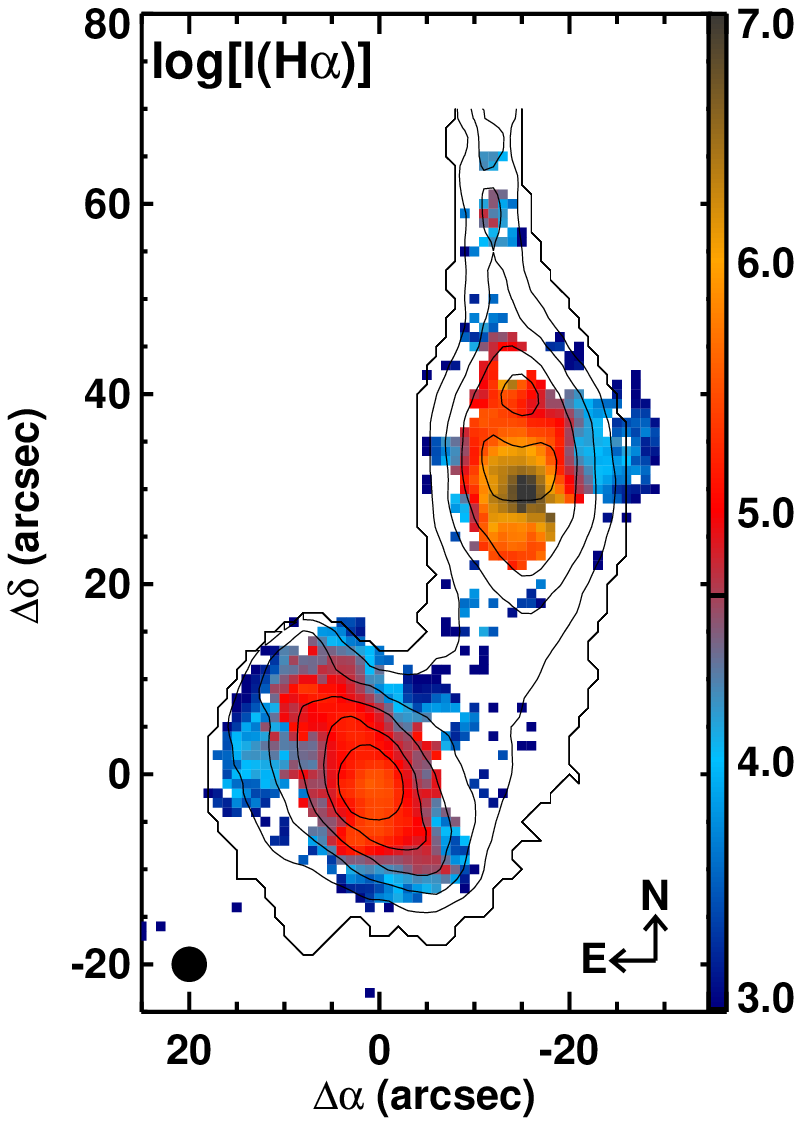}
  \caption{\emph{Left:} \ha\ emission line flux
    ($10^{-16}$\ergflux\,arcsec$^{-2}$). \emph{Centre:} Emission line
    attenuation at 5500\AA\ (magnitudes), calculated from the observed
    \ha/\hb\ emission line ratio and dust attenuation law suitable for
    emission lines in low redshift star-forming galaxies
    \citep{Wild:2011p6538}.  \emph{Right:} dust attenuation corrected
    \ha\ luminosity surface density ($L_\odot$\,kpc$^{-2}$).
  }\label{fig:em_flux}
\end{figure*}

In this section we use the stellar continuum shape and strength of
stellar absorption features to characterise the spatially resolved properties of the
stellar population and constrain the stellar masses
and star formation history of each galaxy. From the simulations of
\citet{Barnes:2004p6826} we expect first passage to have occurred about
170\,Myr ago, and optical spectra of star clusters indicate some star
formation occurred at that time \citep{Chien:2007p7317}. Here we
investigate how wide spread that star formation was and whether it has
continued to the time of observation. 

The stellar population analysis is performed on the combined V500 and
V1200 cube, using the same Voronoi binning as for the stellar
kinematics. While it is difficult to quantify the uncertainties
associated with the extraction of physical properties from stellar
population model fitting \citep[see
e.g.][]{Panter:2007p9207,Conroy:2009p9193},
\citet{CidFernandes:2013p9197} present an analysis of the errors
caused by use of different spectral synthesis models and
spectrophotometric calibration, applied to the CALIFA data. They find
that noise and shape-related errors at the level expected for CALIFA
lead to uncertainties of 0.10–0.15\,dex in stellar masses,
light-weighted mean ages and metallicities, with larger uncertainties
on star formation histories and therefore on mass-weighted
quantities. There are even larger uncertainties associated with the
choice of population synthesis model, at the level of
0.2-0.3\,dex. For this reason, we present raw line index maps
alongside derived quantities.

\subsection{Spectral indices}

We measure five standard stellar absorption line indices from
emission line subtracted spectra: the
4000\AA-break (D4000n), the Balmer absorption lines \hb\ and
\hda+\hga, and two composite metallicity-sensitive indices [Mg$_2$Fe]
and 
[MgFe]' as defined in \citet{2003MNRAS.344.1000B}
and \citet{Thomas:2003p8701}, respectively. Emission lines detected at greater than
$3\sigma$ significance are subtracted from the stellar continuum
using a Gaussian broadened template. Physical properties are then
derived through comparison of the line indices to a ``stochastic
burst'' library of stellar population synthesis models
\citep{2003MNRAS.344.1000B}, following the same Bayesian approach as
described in \citet{Gallazzi:2005p6450}. Typical errors on mean ages
are $\sim$0.12dex, varying between 0.1 and 0.2\,dex. 

Maps of three of the measured spectral indices are shown in the upper
panels of Figure \ref{fig:lick}. Derived properties of light-weighted
and mass-weighted mean stellar age and metallicity are shown in the
bottom panels.  The strength of the 4000\AA\ break
(top left) correlates to first-order with mean stellar age, averaged over
timescales of several Gyr, and to second-order with metallicity,
particularly in the older stellar populations. The strength of the
higher order Balmer absorption lines (top centre) measures mean stellar
age over a slightly shorter period of $\sim0.5$\,Gyr. Both the central
region of NGC~4676B (in the bulge), together with the east and west
flanks of NGC~4676A (above and below the disk), show strong break
strengths indicating older mean stellar ages than in the disks of the
two galaxies. The region with the youngest mean stellar age is the
north-east (NE) tidal arm of NGC~4676B. In star-forming galaxies the
strength of the Balmer lines and 4000\AA\ break strength are strongly
inversely correlated. It is therefore not surprising that (the inverse
of) the Balmer line index map is not dissimilar to that of D4000,
although with a slightly lower SNR\footnote{Note that the negative equivalent
widths of the Balmer lines do not indicate emission, measured values as low as
$-10$ are expected for old stellar populations \citep[see][for the expected values for
SDSS galaxies]{Gallazzi:2005p6450}.}. The right hand panel shows the
metallicity sensitive index, [Mg$_2$Fe]. Both the bulge of NGC~4676B
and regions above and below the disk plane of NGC~4676A show stronger
[Mg$_2$Fe] than the disks, indicating a more metal rich stellar
population. The index [MgFe]' shows similar results. 

Converting these observables into physical properties in the lower
panels of Figure \ref{fig:lick}, we find that the stars of both
galaxies are predominantly older than several Gyrs. Only the
northern tidal tail and the NE arm of NGC~4676B have distinctly
younger stellar populations, with light-weighted ages of
$\sim0.6$\,Gyr or mass-weighted ages of $\sim2.5$\,Gyr.  The bulge of
NGC~4676B and regions above and below the disk of NGC~4676A have the
oldest mean stellar ages of $\gtrsim4.5$\,Gyr. There is no evidence
for a significant intermediate age population that would be consistent
with having formed during first passage, therefore it is clear that
the merger has not yet had a significant impact on the stellar
populations.

\subsection{Young, intermediate age and old populations}\label{sec:sfh:yio}

To visualise the spatially resolved star formation history of
the galaxies, we turn to the full-spectrum fitting package {\sc
  starlight} \citep{2005MNRAS.358..363C,CidFernandes:2013p9201}. This
inverts the observed spectrum into stellar populations of different
ages, rather than fitting a library of models with analytic star
formation histories, as was done in the previous section. {\sc
  starlight} fits the full wavelength range with combinations of
simple stellar population (SSP) spectra from the population synthesis
models of \citet{Vazdekis:2010p7466} and
\citet{GonzalezDelgado:2005p7488}, using the Granada
\citep{Martins:2005p8406} and MILES \citep{miles} stellar libraries,
dust extinction following the \citet{1989ApJ...345..245C} law, a
Salpeter IMF, and stellar evolutionary tracks from
\citet{Girardi:2000p7718}. The SSP ages range from 1\,Myr to 14\,Gyr
and SSPs of four different metallicities are included ($Z=$0.0004,
0.008, 0.020, 0.033, where $Z_\odot\sim0.02$ for these
models). Emission lines are masked during the fit, and errors
propagated from the CALIFA error arrays. 

Figure \ref{fig:2dsfh} shows maps of the fraction of light at
5635\AA\ and mass arising from stars of different ages. We select age
bins that correspond to the main sequence lifetimes of stars with
distinctly different optical line and continuum features, namely young
($t<$140\,Myr), intermediate (140\,Myr$<t<$1.4\,Gyr) and old
($t>$1.4\,Gyr) populations, in order to maximise the robustness of the
spectral decomposition results.

Clearly the ongoing merger has thus far had little significant affect
on the stellar populations in terms of total stellar mass or global
star formation history. The fraction of stellar mass contributed by
stars younger than 140\,Myr is less that 5\% in all regions, although
$\sim$30\% of optical light in the nuclear regions of NGC~4676A and
tidal arm of NGC~4676B arises from stars younger than 140\,Myr. In
most regions $>90\%$ of the stellar mass is from stars older than
1.4\,Gyr. There is an excess of light from intermediate age stars seen
in the western half of NGC~4676A, inter-galaxy region, and NE bar of
NGC~4676B, compared to the disks of these galaxies. This is consistent
with the triggering of low level star formation in the gas flung out from
the galaxy disks at first passage $\sim$170\,Myr ago, perhaps through
dynamic instabilities or shocks \citep[see e.g.][and references
therein]{Boquien:2010p8247}.

\subsection{Stellar mass and star formation rate}\label{sec:mass}
From the integrated star formation history measured by {\sc
  starlight}, and correcting for recycling of matter back into the
interstellar medium (ISM), we obtain a total current stellar mass of $1.2\times10^{11}$ and
$1.5\times10^{11}$\Msol\ for NGC~4676A and B respectively, for a
Salpeter IMF. For NGC~4676A the {\sc CALIFA} derived stellar mass
agrees well with that derived from SDSS 5-band photometry
($1.6\times10^{11}$\msol, J. Brinchmann\footnote{The SDSS catalogue
  can be downloaded from here:
  http://home.strw.leidenuniv.nl/$\sim$jarle/SDSS/DR7/totlgm\_dr7\_v5\_2.fit.gz
  SDSS mass was increased by a factor of 1.8 to convert from a
  Chabrier to Salpeter IMF.}). For NGC~4676B the SDSS derived mass is
10 times lower, which is traceable to incorrect photometric
measurements in all bands in the SDSS catalogue, presumably caused by
the deblending algorithm which has identified three sources in this
galaxy.  The higher stellar masses than dynamical masses (Table
\ref{tab:basic}) may be due to the assumed Salpeter IMF or difficulty
in measuring a dynamical mass from a kinematically disturbed
system. The use of a Chabrier IMF would solve the discrepancy.

The decomposition of the stellar continuum can provide an approximate
estimate of the ongoing star formation rate in the galaxies,
independent of the ionised gas emission. The value obtained depends
sensitively on the width of the time bin over which the average is
calculated.  Varying the width of the time bin between 25 and 140\,Myr
yields a global SFR of 2.6-5\,\msolyr\ and 1-4\,\msolyr\ for NGC~4676A
and B respectively. While the errors on these values are difficult to
quantify, it is clear that neither galaxy is currently undergoing a
significant burst of star-formation, especially when their very large
stellar masses are taken into account. 

The young stellar population found in the centre of NGC~4676A implies
a nuclear SFR surface density of $\sim0.15$\sfrsd, averaged over 140\,Myr
in the inner $5\times5$\arcsec.  No young stellar population is found
in the central regions of NGC~4676B.

\section{Ionised gas emission}\label{sec:emission}

\begin{figure*}
\centering
  \includegraphics[scale=0.6]{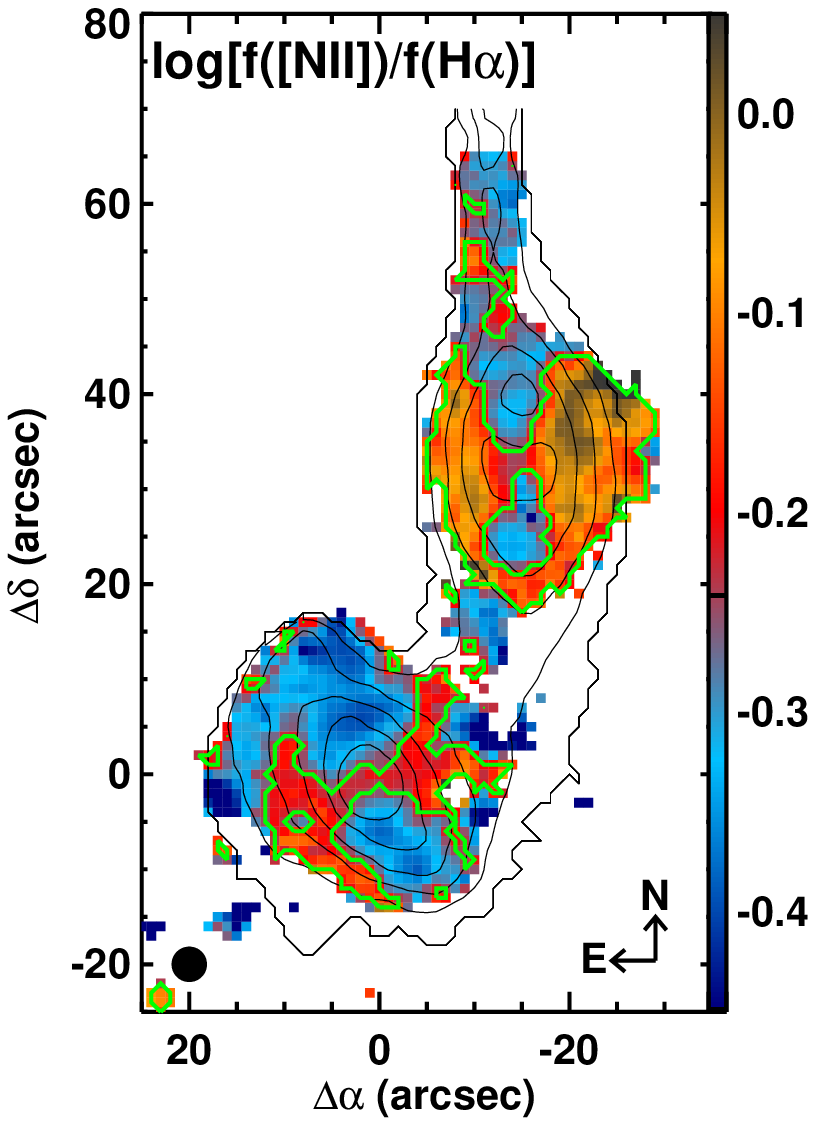}
  \includegraphics[scale=0.6]{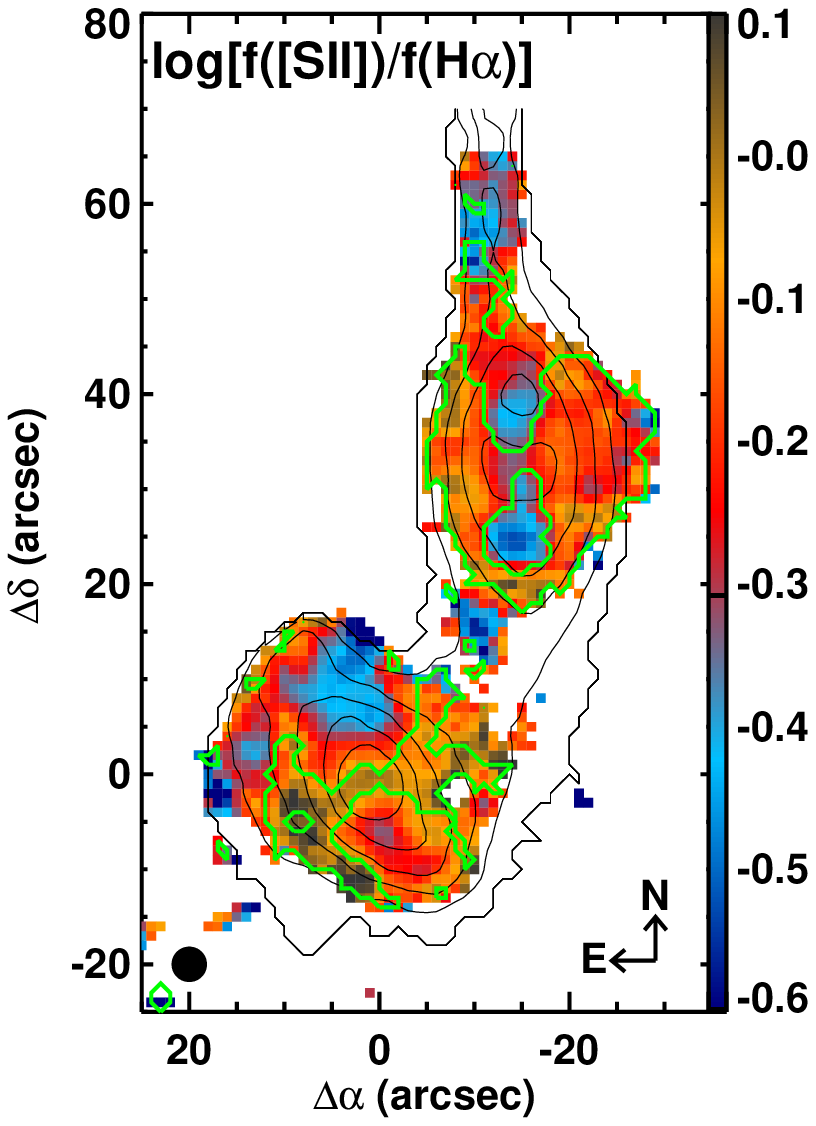}\\
  \includegraphics[scale=0.6]{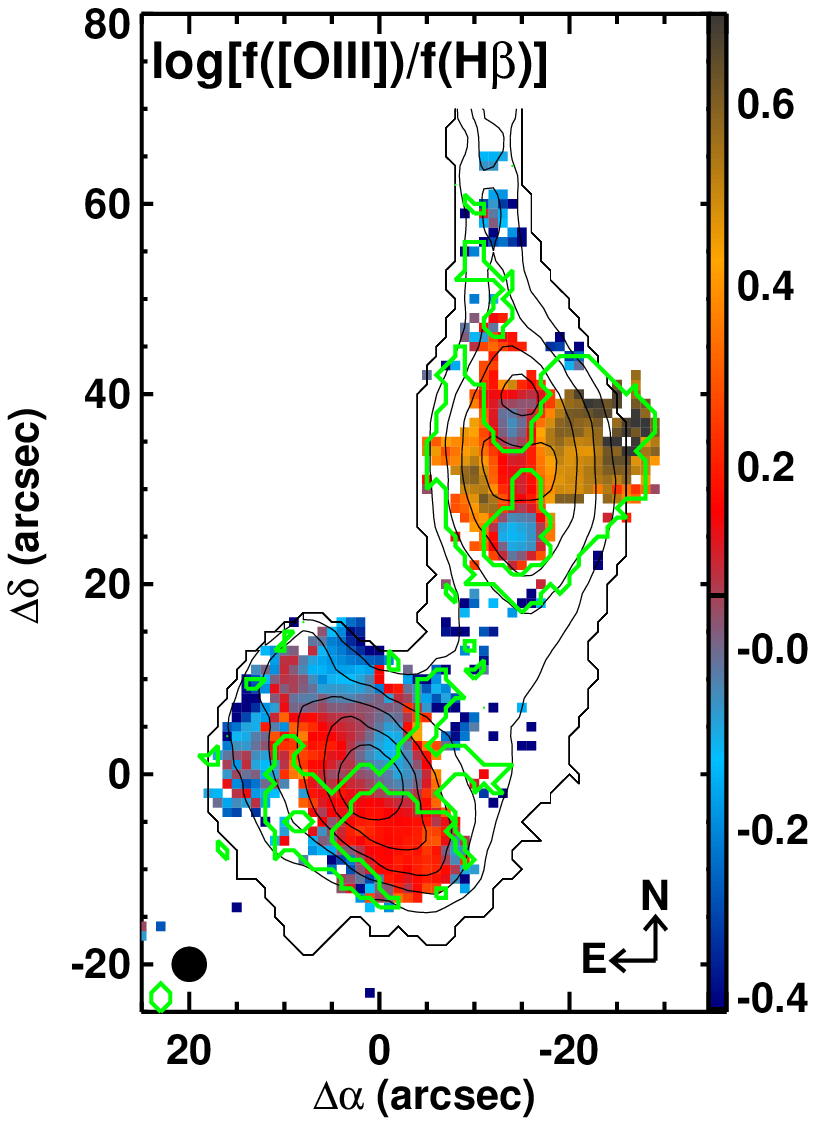}
  \includegraphics[scale=0.6]{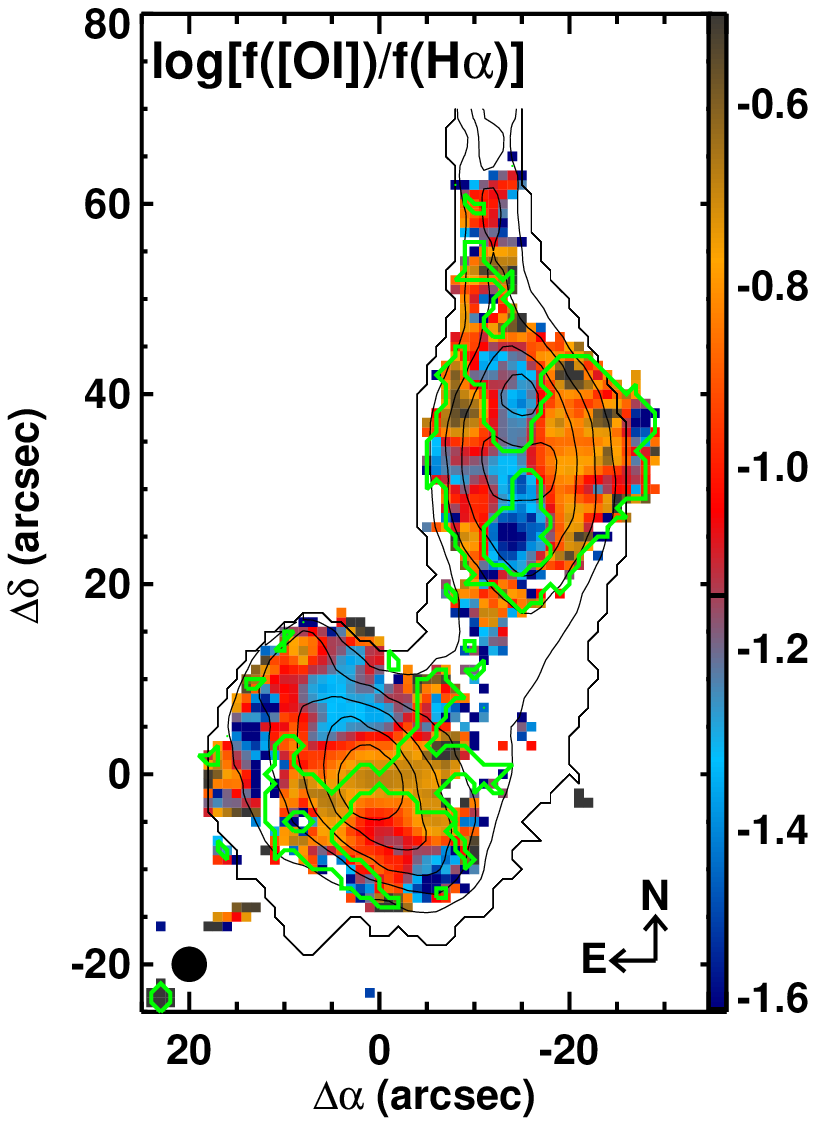}
  \caption{From top left to bottom right: emission line flux ratio maps for
    \nii$\lambda6583$/\ha, (\sii$\lambda6717$+\sii$\lambda6731$)/\ha,
    \oiii$\lambda5007$/\hb\ and \oi$\lambda6300$/\ha. To guide the eye
    in comparison between panels the black contours indicate $V$-band
    isophotes, and the thick green contours delineate those regions
    with $\log$(\nii/\ha)$>-0.25$. }\label{fig:em_ratio}
\end{figure*}

One of the highlights of the {\sc CALIFA} dataset is its wide
wavelength coverage, allowing measurements of all the strong emission
line species from \oii$\lambda3727$ to \sii$\lambda6731$. We measure
the total emission line fluxes in each spaxel by fitting Gaussian line
profiles to the stellar continuum subtracted spectra using the IFU
package fit3D \citep[see e.g.][]{Sanchez:2007p8294}. 

We use empirically derived flux thresholds to remove spaxels from the
analysis where the surface brightness is too low to obtain a reliable
line flux. For \ha\ and line ratios which include \ha\ we use a flux
threshold of $1.5\times10^{-17}$\ergflux, and for \hb\ and line ratios
which include \hb\ we use a flux threshold of
$5\times10^{-18}$\ergflux. Low SNR spectra at the outskirts of the
maps are visually checked to ensure that their line measurements are
reasonable. In order to measure emission line strengths in the fainter
outer regions, we performed the same analysis on a Voronoi-binned data
cube.  However, we found that no significant additional information
was gained from these maps, and some information was lost due to the
lack of connection between Voronoi bin boundaries and physical
components.

\subsection{\ha\ emission and dust attenuation}\label{sec:emfluxes}

The {\sc CALIFA} datacubes have sufficient continuum SNR, spectral
resolution and wavelength range to allow subtraction of the stellar
absorption from the Balmer emission lines. This allows us to measure
the dust attenuation affecting the nebular lines and thus estimate the
intrinsic distribution of ionised gas. The dust attenuation ($A_{\rm
  V,gas}$) map is constructed from the \ha/\hb\ ratio map using the
attenuation law from \citet{Wild:2011p6538} which is measured from,
and applicable to, emission lines in local star-forming galaxies
including local ULIRGs.


Figure \ref{fig:em_flux} shows the observed \ha\ line flux, the
emission line attenuation map ($A_{\rm V,gas}$) and the
dust-attenuation corrected (intrinsic) \ha\ line luminosity. The
three peaks in observed \ha\ flux along the disk of NGC~4676A have been
noted previously in the literature. The ionised gas bar of NGC~4676B
is clear, with the axis of the bar offset by about 25$^\circ$ in the
clockwise direction from the major axis of the continuum light profile
(contours). 

We measure a maximum effective attenuation of $\sim7$ magnitudes at
5500\AA\ close to the centre of NGC~4676A, which is consistent with
the dust lane visible in the HST images. For such large attenuations,
small variations in the assumed dust attenuation law or stellar
population models result in significant uncertainties in the
dust-corrected line fluxes. We show below that the dust attenuation
corrected total \ha\ flux is a factor of a few larger than expected
from multiwavelength observations (Section \ref{sec:MWsfr} and Table \ref{tab:sfr}). After correcting for dust
attenuation we find that ionised hydrogen emission
in NGC~4676A is concentrated in the central regions of the galaxy.

NGC~4676B has an average line extinction of a little over $\sim$1
magnitude, typical for an ordinary star-forming galaxy. The dust
content of the disk appears slightly higher than the bar. We note that
\ha\ emission is seen even in the region dominated by the old stellar
bulge. Either there are sufficient young stars are available
to ionise the gas, even though the continuum light is completely
dominated by old stars (Section \ref{sec:sfh}), or the gas
in the central regions is primarily ionised by the AGN. Alternatively,
the ionised gas is not coincident with the bulge, which would also be
consistent with the offset in velocities (Section \ref{sec:kin})

\subsection{Emission line ratios sensitive to ionisation source}\label{sec:emratios}

Standard line ratios sensitive to the shape of the ionising spectrum
have been calculated and the spatial distribution of a selection of
these are shown in Figure \ref{fig:em_ratio}. We focus on pairs of
lines that are close enough in wavelength space for their ratios to
not be strongly affected by dust attenuation. The most obvious feature
of these maps are the butterfly shaped bicones of higher ionisation
gas orientated along the minor axes of both galaxies. Comparing the
four line ratio maps, the bicones in the two galaxies are clearly
different: in NGC~4676A the bicones are visible in all maps,
in NGC~4676B the bicones have higher \sii/\ha, \nii/\ha, and \oi/\ha\ line ratios,
but are not visible in \oiii/\hb. High ionisation gas is 
also found following an arc to the south of NGC~4676A. 

The line ratios make it clear that a substantial fraction of the
line emission originates from sources other than photo-ionisation by
stars. To guide the eye, we overplot contours to delineate those
regions with $\log$(\nii$\lambda6584$/\ha)$>-0.25$, which is
approximately the maximum ratio observed in high-metallicity star
forming regions \citep{2003MNRAS.346.1055K}.  We will analyse 
the possible causes of these bicones in Sections \ref{sec:outflowA} and
\ref{sec:biconeB} below.

\subsection{Emission line ratios sensitive to gas density}\label{sec:sii}

\begin{figure}
\centering
  \includegraphics[scale=0.7]{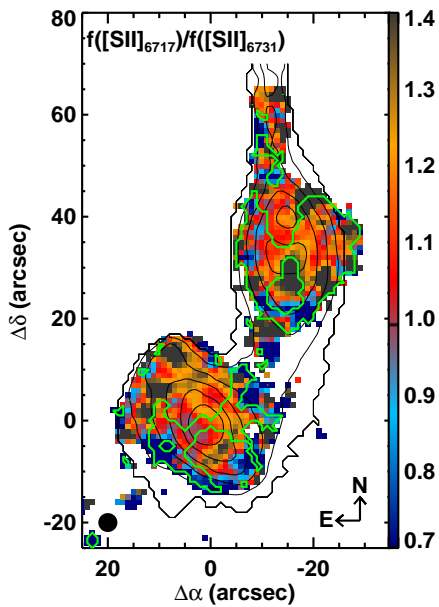}
  \caption{ Map of the \sii\ doublet ratio
    (\sii$\lambda6717$/\sii$\lambda6731$), a sensitive electron
    density estimator. Higher density gas has lower ratios.  The thin
    black contours show V-band continuum flux, and the thick green
    contours delineate those regions with $\log$(\nii/\ha)$>-0.25$.
  }\label{fig:em_ratio2}
\end{figure}

In Figure \ref{fig:em_ratio2} we show the
\sii$\lambda6717$/\sii$\lambda6731$ doublet ratio which is primarily
sensitive to the electron density of the emitting gas ($n_e$). This
ratio varies
between 0.4 and 1.4 for $n_e\gtrsim10^5{\rm cm}^{-3}$ and
$n_e\lesssim100{\rm cm}^{-3}$ \citep[at
$T_e=10^4$K,][]{Osterbrock:2006p4475}. Both galaxies show complex
structure in their gas density.  Regions with particularly low density
(high doublet ratio) are the NE tidal arm of NGC~4676B, and the edges
of the western bicone in NGC~4676A. The apparently very high density
in the outskirts of NGC~4676B and the region to the south of NGC~4676A
might be due to shocks caused by the interaction and gas inflows, but
higher quality data should be obtained to confirm these measurements.

The median doublet ratio in the high ionisation bicones of NGC~4676A
is 1.1, which implies a typical electron density of
$n_e\sim400{\rm cm}^{-3}$. In the very outer regions of the western
cone, the median \sii\ ratio increases slightly to 1.2, equivalent to
an electron density of $n_e\sim200{\rm cm}^{-3}$. The central electron
density
is no higher than in the gas throughout the cones.

\subsection{Gas-phase metallicities and star formation rates}\label{sec:metals}

We calculated metallicities from the \oiii\ and \nii\ emission lines
\citep{Pettini:2004} in regions where the line ratios indicate the gas
is primarily photoionised by hot stars, i.e. in the disks of the
galaxies and in the northern tidal tail. The metallicity of the gas
throughout both galaxies is approximately solar, consistent with star
clusters located in the tidal tails of the two galaxies \citep{Chien:2007p7317}.

We additionally note that the \nii/\ha\ ratio
map is noticeably flat in the disk and bar of NGC~4676B. Given that
this galaxy is massive, with a prominent central bulge, a metallicity
gradient is expected
\citep{Pilyugin:2004p8581,Sanchez:2012p8557}. Flat gradients can be
caused by mergers, but such a flat gradient is expected to only occur
in the later stages of a merger
\citep{Rupke:2010p7180,Rupke:2010p4468,Kewley:2010p6114}. The presence
of a strong bar complicates the picture, and simulations by
\citet{Martel:2013p8702} show that mixing of gas from the outer
regions to the centre through a bar may occur even before star formation is
induced.

For the regions of the galaxies where the gas is photoionised by light
from stars, we can measure the SFR from the dust attenuation corrected
\ha\ line luminosity, using the standard conversion to star formation
rate \citep{1998ARA&A..36..189K}.  For the remaining regions we obtain
only an upper limit on the SFR, because a fraction of the \ha\ flux
must arise from processes unrelated to star formation. For NGC~4676A we
measure a total \ha\ luminosity of $4.4\times10^{42}$\ergs, of which
$2.2\times10^{42}$\ergs\ (50\%) arises from regions with line ratios
consistent with stellar photoionisation. These lower and upper limits
lead to a SFR for NGC~4676A of 17-35\,\msolyr.  In the same manner,
for NGC~4676B we measure a total dust-attenuation corrected \ha\
luminosity of $6.1\times10^{41}$\ergs\ of which 56\% arises from
regions with line ratios consistent with photoionisation by young
stars. This gives a SFR for NGC~4676B of 3-5\,\msolyr.

We can use the dust attenuation corrected \ha\ luminosity to estimate
the central SFR surface density in the inner $5\times5$\arcsec\ of
each galaxy. For NGC~4676A we find $\Sigma_{\rm SFR}<4$\sfrsd, and for
NGC~4676B we find $\Sigma_{\rm SFR}<0.2$\sfrsd.  The upper limit
estimated from the dust corrected \ha\ luminosity is similar to that
measured from the spectral decomposition for NGC~4676B
($\sim0$\sfrsd), but much higher for NGC~4676A ($\sim0.15$\sfrsd). As
previously, we note that the global dust attenuation corrected \ha\
luminosity appears to overestimate the global SFR compared to other
multiwavelength methods (Section \ref{sec:MWsfr} and Table
\ref{tab:sfr}). In the nuclear regions, where the dust attenuation is
large, it is possible that the dust attenuation correction is leading
to incorrect emission line strength estimates.

\section{Multiwavelength analysis of global star formation rates}\label{sec:MWsfr}

\begin{table}
    \caption{The star formation rates of NGC~4676A and NGC~4676B
      estimated from CALIFA and multiwavelength
      observations. \label{tab:sfr} }
\centering
\begin{tabular}{ccc}\hline\hline
  Method & SFR$_A$/\msolyr & SFR$_B$/\msolyr\\\hline
  (\ha,\hb)\tablefootmark{a} &17-35&3-5\\
  (\ha,mid-IR)\tablefootmark{b} &4.3-6.9&1.3-2.1\\
  stellar continuum\tablefootmark{c}  &2.6-5&1-4\\
  (FUV,mid-IR)\tablefootmark{d} &6.2&2.3\\
  \neii \tablefootmark{e} &$<$10.1&$<$1.4\\
  (FIR, radio) \tablefootmark{f} & 14 & 4 \\
  \hline 
\end{tabular}
\tablefoot{
\tablefoottext{a}{Dust attenuation corrected \ha\ luminosity using the Balmer decrement
(Section \ref{sec:emfluxes}). The errors on the dust attenuation
correction in the nucleus of NGC~4676A are likely to be 
significant, due to the very high dust content.}
\tablefoottext{b}{Combination of \ha\ and 24\mum\ luminosities \citep{Smith:2007p8201}.}
\tablefoottext{c}{Decomposition of the stellar continuum using {\sc starlight}
(Section \ref{sec:mass}).}
\tablefoottext{d}{Combination of  24\mum\ and FUV luminosities
  \citep{Smith:2010p8200}.}
\tablefoottext{e}{Mid-IR \neii\ emission line \citep{Haan:2011p7283}.}
\tablefoottext{f}{Combined FIR and radio continuum luminosities \citep{Yun:2001p7305}.}
}
\end{table}

In the previous sections we have estimated the ongoing star formation
rate of the Mice galaxies from both the ionised gas recombination
lines and the stellar continuum in the CALIFA datacube, however, both
estimates are uncertain. Although the higher SFR for NGC~4676A
estimated from the emission lines compared to the stellar continuum
may plausibly indicate a recent increase in SFR, we urge caution in
this interpretation.  A significant fraction of the line emission may
arise from gas that has not been photoionised by stars, and the very
large dust attenuation causes significant uncertainty in the central
region of NGC~4676A. The SFR obtained from decomposition of the
stellar continuum depends sensitively on the star formation history
fitted by the {\sc starlight} code. While both methods suggest that
only moderate levels of star formation is ongoing in both galaxies,
there are several multiwavelength observations that can be used to
verify this result. These are summarised in Table \ref{tab:sfr} and
discussed in more detail below.

The Mice have been observed in the mid-IR with the MIPS instrument on
board Spitzer \citep{Smith:2007p8201} and the NUV and FUV with GALEX
\citep{Smith:2010p8200}. This provides us with two alternative SFR
estimates which account for optically thick dust obscuration by direct
detection of the thermal dust emission.  Firstly, we combine the
24\mum\ luminosity with the total \ha\ line luminosity, using both the \citet{Calzetti:2007} coefficient
of 0.031 derived from \HIIr\ regions and the
\citet{Kennicutt:2009p4470} coefficient of 0.02 which includes a
correction for diffuse emission.  We note that this calibration
depends on metallicity \citep[e.g.][]{Relano:2007p8582}, but is
appropriate for the solar metallicity that we measure for the Mice
galaxies in
Section \ref{sec:emratios}. This method results in a SFR of 
4.3-6.9\,\msolyr\ for NGC~4676A and 1.3-2.1\,\msolyr\ for NGC~4676B,
where the ranges account for the different coefficients and the upper
and lower limits on the \ha\ line flux that arises from \HIIr\
regions. 

We can combine the 24\mum\ with the FUV luminosities to obtain a SFR
independent of the {\sc CALIFA} data, along with an attenuation in the
FUV ($A_{\rm FUV}$), following the prescription in
\citet{IglesiasParamo:2006p8585}\footnote{FUV luminosities are
  converted into SFRs using SB99 \citep{Starburst99} and a Salpeter
  IMF with $M_{\rm low}=0.1$\,\msol\ and $M_{\rm
    up}=100$\,\msol. 24\mum\ luminosities are converted to total IR
  luminosities (from 8\mum\ to 1000\mum) using the models by
  \citet{Chary:2001p8627}.}. This results in a SFR of 6.2\,\msolyr,
with an attenuation of $A_{\rm FUV} = 3.3$\,mag for NGC~4676A, and
SFR of 2.3\,\msolyr\ with an attenuation of $A_{\rm FUV} = 1.3$\,mag
for NGC~4676B.

We can use the \neii12.81\mum\ flux to provide a nebular line estimate
of SFR that is unaffected by dust obscuration. Taking the \neii\ flux
reported in \citet{Haan:2011p7283} and the calibration of
\citet{DiamondStanic:2012} for galaxies with $L_{IR}<10^{11}$\lsol,
gives a SFR of $<$10.1 and $<$1.4\,\msolyr, for NGC~4676A and B
respectively. The upper limits arise because, like \ha, \neii\ can be
emitted by processes other than star-formation.

Finally, \citet{Yun:2001p7305} combine FIR IRAS observations with
radio continuum flux to estimate the FIR luminosity of each galaxy,
even though they are not spatially resolved in IRAS. Assuming a factor
of 2.5 conversion from \citet{Helou:1988p1394} to total infrared flux,
and \citet{1998ARA&A..36..189K} conversion to SFR we obtain a SFR of
14 and 4\msolyr, for NGC~4676A and B respectively. Combining \ha\
luminosity with our own estimate of IRAS total infrared flux from
template fitting between 8-1000\mum\ and the conversion given by
\citet{Calzetti:2013p9396} results in a lower total SFR for the two
galaxies, but consistent within the scatter of the other measurements.

Given the complexity of the galaxies, it is perhaps surprising that
the many different estimates of SFR are so close. Excluding the \neii\
based estimate, for which we only have an upper limit, we obtain
median values of 6.2 and 2.3\,\msolyr, for NGC~4676A and B respectively.

Dividing the total SFR of the galaxies by their stellar mass (Section
\ref{sec:mass}) gives a specific SFR (sSFR) of
$5\times10^{-11}$yr$^{-1}$ and $1.3\times10^{-11}$yr$^{-1}$ or
log(sSFR/yr$^{-1}$) of $-$10.3 and $-$10.9 for NGC~4676A and B
respectively.

\section{The bicone in NGC~4676A}\label{sec:outflowA}

\begin{figure*}
\centering
\includegraphics[scale=0.36]{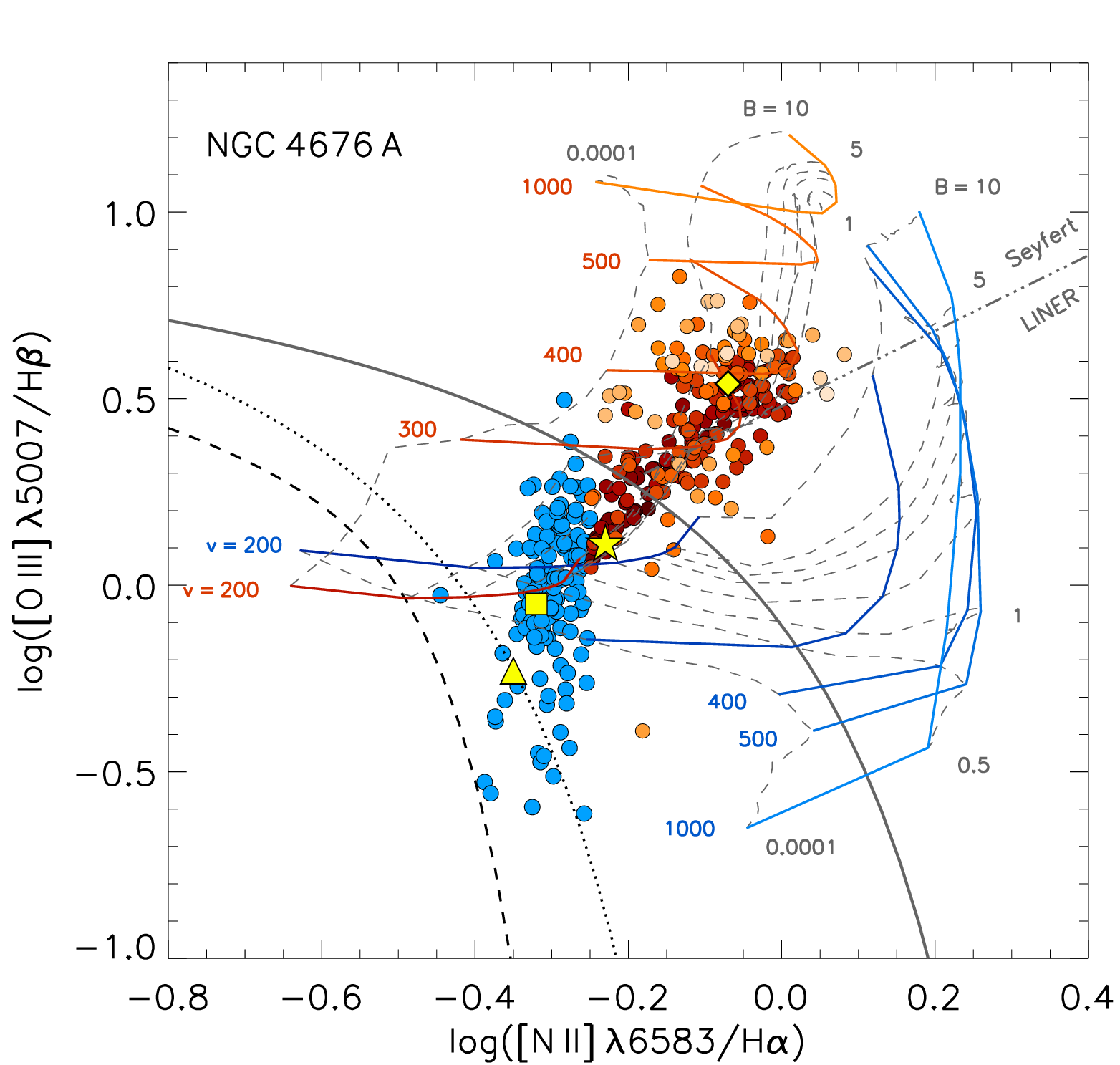}
\includegraphics[scale=0.36]{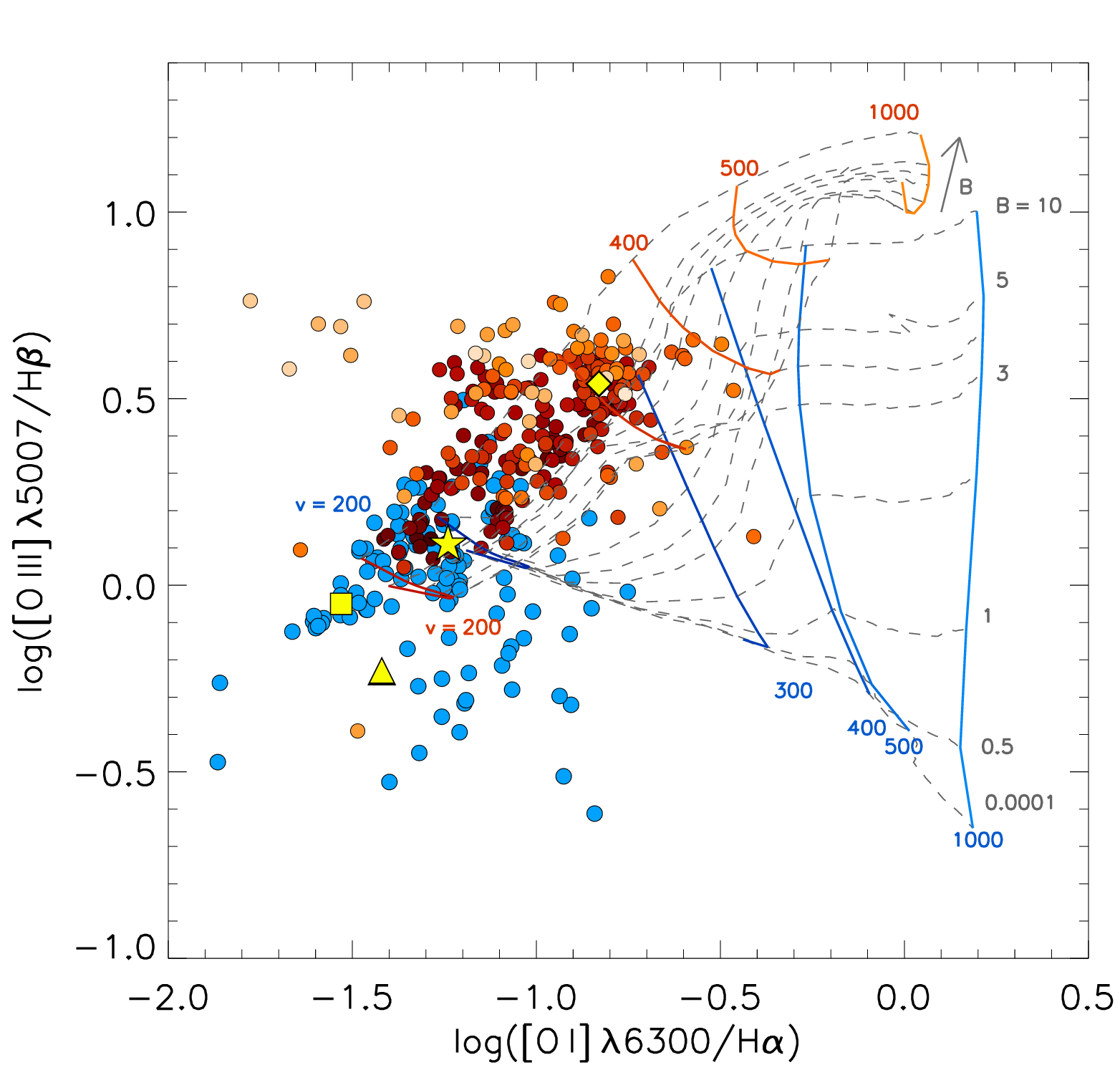}
\includegraphics[scale=0.42]{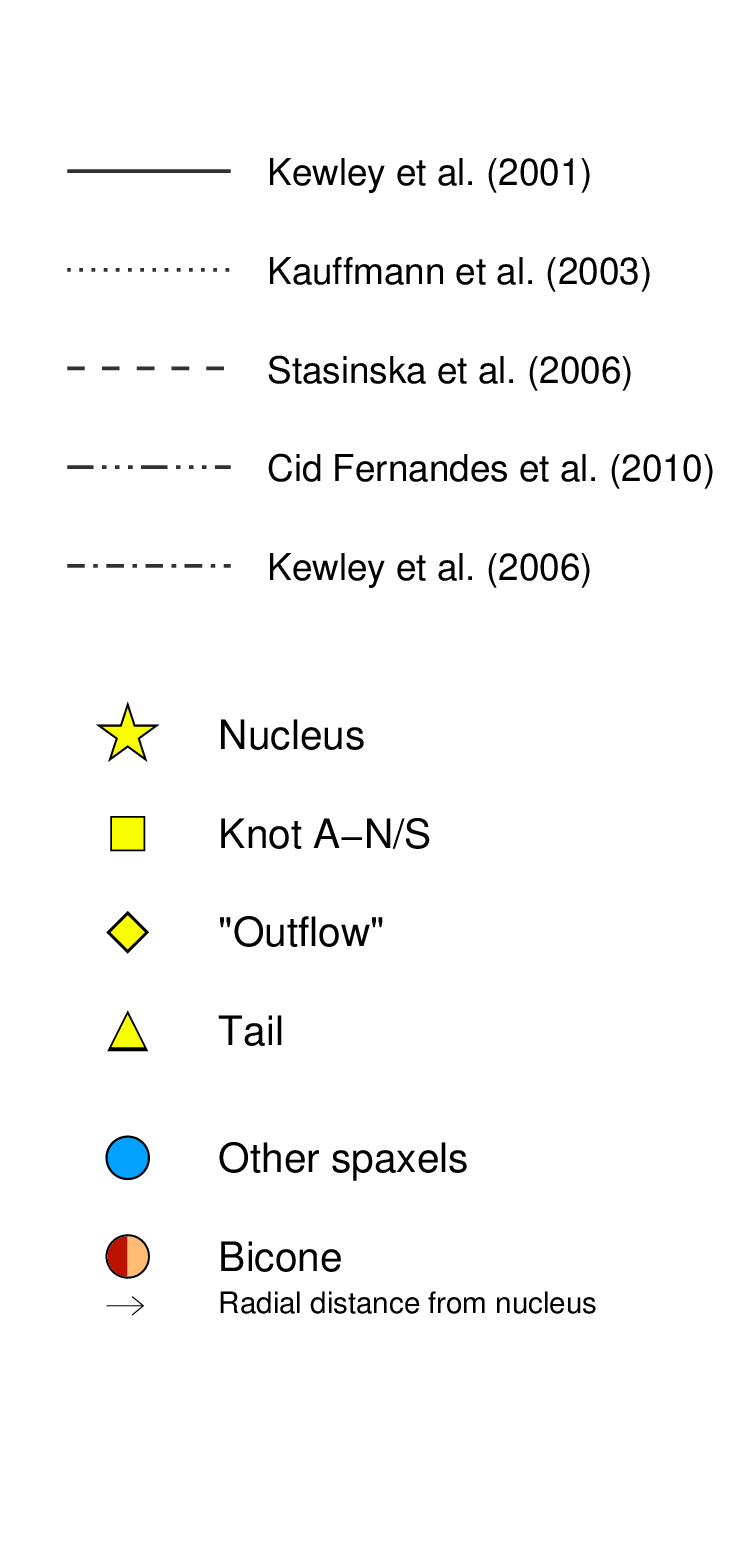}
\caption{Line ratio diagnostic diagrams, showing line ratios for
  independent spaxels in NGC~4676A. Those spaxels which lie in the
  bicones are coloured orange, shaded according to radial distance
  from the nucleus, the spaxels lying in the non-shocked regions are included as blue
  circles. Line ratios in some key regions of interest (see
  Fig. \ref{fig:regions}) are plotted as
  yellow symbols. Overplotted as black lines are empirically and
  theoretically derived separations between LINERs/Seyferts and \HIIr\
  regions. Overplotted as coloured lines are line ratios predicted for
  the photoionisation of gas by fast shocks from Allen et al. (2008), as described in the
  text. The blue model tracks show predicted line ratios when
  ionisation comes from the shock front alone, orange lines also
  include pre-ionisation by a precursor.
}\label{fig:bpt_a}
\end{figure*}

In Figure \ref{fig:bpt_a} we show two emission line diagnostic
diagrams for individual spaxels in NGC~4676A.  Spaxels in the disk are
coloured blue and those in the bicones are coloured orange, with
lightness of tone increasing with distance from the nucleus. While gas
in the northern tidal tail and north and south disk shows line ratios
similar to high metallicity star-forming galaxies in the local
Universe \citep{2003MNRAS.346.1055K,Stasinska:2006p7671}, the bicones
have line ratios which indicate increasing hardness of the ionisation
field with height above the mid-plane of the disk. At the outer extent
the line ratios lie primarily in the region commonly occupied by
Seyfert galaxies
\citep{2006MNRAS.372..961K,CidFernandes:2010p4744}. Increasing
hardness of the radiation field with distance from the mid-plane, and
extended soft X-ray emission \citep{Read:2003p7304} is inconsistent
with an AGN being the primary source of ionisation, but instead
suggests that shocks driven by a superwind are ionising the gas
\citep[][hereafter HAM90]{HAM1990}.

Overplotted on Figure \ref{fig:bpt_a} are predictions from the
fast-shock models of \citet{Allen:2008p6993}\footnote{These were
  calculated using the ITERA package: \citep{Groves:2010p4895}.}, for
a range of shock velocities ($v_s$) and magnetic field strengths
(B). We use solar metallicity as measured from line ratios in the disk
(Section \ref{sec:metals}), the abundance set of \citet{Grevesse:2010p8449} and a
pre-shock density of 1\cmcube.  In photoionising
shocks, the flux of ionising radiation emitted by the shock increases
proportional to $v_s^3$, leading to a strong increase in
pre-ionisation level of the gas as shock velocity increases. At the
highest velocities, the ionisation front can expand ahead of the shock
front, leading to an ``ionised precursor'' which can contribute
significantly to the optical emission of the shock. Magnetic fields
can limit compression across the shock, thereby allowing the
ionisation front to proceed into the post-shocked gas more quickly
than the case with no  magnetic fields, again leading to an increase in
pre-ionisation level of the gas. We overplot predicted line ratios
from models both with and without inclusion of the additional
photoionisation caused by the precursor (orange and blue lines). The
observed line ratios in the cone of NGC~4676A are consistent with a
fast shock, including an ionised precursor, with velocity increasing
from $\sim$200\,\kms\ in the centre of the galaxy to $\sim$350\,\kms\
at the outer edge of the bicone. We note that this apparent increase in shock
velocity could also arise from a line-mixing scenario, where the
emission from \hii\ regions falls off more rapidly with height above
the disk than emission from shocks. Further data would be needed to
confirm whether or not the shock was increasing in velocity from the
mid plane.  

Recent low velocity shock models described in
\citet{Rich:2010,Rich:2011} and \citet{Farage:2010} are appropriate
for shocks with velocities $\lesssim200$\,\kms. As these models are
not available, we compare to the figures published in
\citet{Rich:2011} for NGC~3256, which has a similar metallicity to
NGC~4676A. We find that the large observed \oiii/\hb\ line ratio in
NGC~4676A is inconsistent with the slow shock models, and the closest
models have velocities of $\lesssim100$\,\kms, well below the measured
velocity dispersion of the lines in the bicones.

\subsection{Outflow kinematics}
Shocks heat the gas through which they pass, and lead to increased
linewidths of the post-shock emitting gas from thermal motions. Line
splitting of absorption and/or emission lines is sometimes observed in
galactic outflows, due to the bulk motions of gas towards and away
from the observer.  An increase in \ha\ emission line width is
observed in radial fingers extending along the minor axis of NGC~4676A
(Figure \ref{fig:kin}), with a maximum velocity dispersion of
$\sigma\sim200$\,\kms. Unfortunately, with the limited spectral
resolution of the CALIFA data we are unable to resolve the different
kinematic components, thus we can only use the measured line widths to
place limits on the bulk kinetic motions of the gas. If the increased
linewidth is caused by bulk outflow, then the line-of-sight component
of the outflow velocity is estimated from half of the
full-width-half-maximum (FWHM) of the line to be
$\sim$235\kms. Including a significant tangental component, as might
be expected from an outflowing wind, this is consistent with the
estimate of the shock velocity estimated from line ratios above.

\subsection{Energetics of the superwind fluid}

Spatially resolved, optical line emission observations can constrain
the rates at which the fast moving wind fluid in NGC~4676A carries
mass, momentum and energy out of the galaxy. Within the context of the
superwind model of HAM90, the wind in NGC~4676A has expanded beyond
the initial ``hot bubble'' phase, into the ``blow out'' or free
expansion phase, during which the wind propagates at approximately constant velocity
into the intergalactic medium. The optical emission lines arise from
clouds and wind shell fragments that are shock heated by the
outflowing wind fluid.

The density of the medium into which the wind is propagating ($n_1$)
can be determined from the electron density ($n_e$) measured from the
\sii\ line ratios (Section \ref{sec:sii}), combined with knowledge of
the type of shock causing the optical line emission (see Appendix A
for details):
\begin{equation}
  n_1 [{\rm cm}^{-3}] = 0.12 \left(\frac{n_e[{\rm cm^{-3}}]}{100}\right)  \left(\frac{350}{v_s[{\rm km\,s^{-1}}]}\right)^{2} 
\end{equation}
where $v_s$ is the shock velocity.  From this, the thermal pressure of
the clouds where the \sii\ emission arises is given by:
\begin{equation}
P_{\rm cloud} = n_1 m_p \mu\, v_s^2 = 6.6\times10^{-11}{\rm Nm}^{-2}
\end{equation}
where $m_p$ is the proton mass, $\mu=1.36$ accounts for an assumed
10\% Helium number fraction and we have set $n_e=200$cm$^{-3}$ from the
observed \sii\ line ratio at the outer extent of the bicones. We note that, for
fast shocks, the thermal pressure of the cloud is independent of the
shock velocity. This is equivalent to a pressure of
$6.6\times10^{-10}$dynes\,cm$^{-2}$ or $P_{\rm cloud}/k =
4.8\times10^{6}$K\,cm$^{-3}$. The pressure at the outer extent of the
wind in NGC~4676A is a little larger than the range measured by
HAM90 for 6 far-infrared galaxies (FIRGs) of
$2.5-5\times10^{-10}$dynes\,cm$^{-2}$.

In the ``blow-out'' phase of a superwind, the pressure source in the
outer regions of the wind is the ram pressure of the wind fluid itself
\footnote{A lower limit on the ram pressure of the wind fluid can also
  be estimated from the X-ray data. Taking the density and temperature
  from \citet{Read:2003p7304} we find $P/k >
  1\times10^{5}$K\,cm$^{-3}$, consistent with the CALIFA
  results. However, we note that this value is highly uncertain: the
  lack of counts in the X-ray data have deterred other authors from
  fitting the X-ray spectrum to obtain a temperature
  \citep{GonzalezMartin:2009p6907}. Improved X-ray data would be
  required to estimate a filling factor for the X-ray emitting gas.}:
$P_{\rm wind} \sim P_{\rm cloud}$. This allows us to use the
information derived from the shock heated gas to determine the
momentum flux of the wind flowing into a solid angle
$\Omega$. Following HAM90:
\begin{equation}
\dot{p}_{\rm wind} =  P_{\rm wind}(r) r^2 \frac{\Omega}{4\pi}= 1.3\times10^{30} {\rm N} 
\end{equation}
where $P_{\rm wind}(r)$ is the wind pressure at radius $r$ and we have
estimated $r=6.6\,$kpc and $\Omega/4\pi\sim0.37$ from the {\sc CALIFA}
maps.  This is equivalent to $1.3\times10^{35}$dynes. This compares
to values ranging from $0.3-12\times10^{35}$dynes for the winds in
HAM90, although we note that these will be upper limits, as
without IFS data the authors assumed $\Omega/4\pi=1$.

Both the energy and mass outflow rates can be estimated from the
momentum flux, however, they depend upon the unknown velocity of the
wind fluid. For typical values for superwinds of
$v_{w}=1000-3000$\,\kms\  \citep[][HAM90]{Seaquist:1985p9395,Hopkins:2013p8621} 
we estimate a total energy flux of
\begin{equation}\label{eqn:Edotwind}
\dot{E}_{\rm wind}=0.5 P_{\rm wind} r^2 v_w \frac{\Omega}{4\pi}=
[6.4-20.4]\times10^{42}\,{\rm erg\,s^{-1}}
\end{equation}
and a mass outflow rate of 
\begin{equation}\label{eqn:Mdotwind}
\dot{M}_{\rm wind} = \frac{P_{\rm  wind}r^2}{v_{ w}} \frac{\Omega}{4\pi} = [8-20]{\rm M_\odot yr}^{-1}.
\end{equation}
This mass outflow rate is a factor of 1.5-3 larger than the global
star formation rate of the galaxy ($\dot{M}^*\sim6$\,\msolyr, Section
\ref{sec:MWsfr}), which is typical for galaxy outflows
\citep[e.g.][]{1999ApJ...513..156M,Veilleux:2005p7318}.

\begin{figure*}
\centering
\includegraphics[scale=0.35]{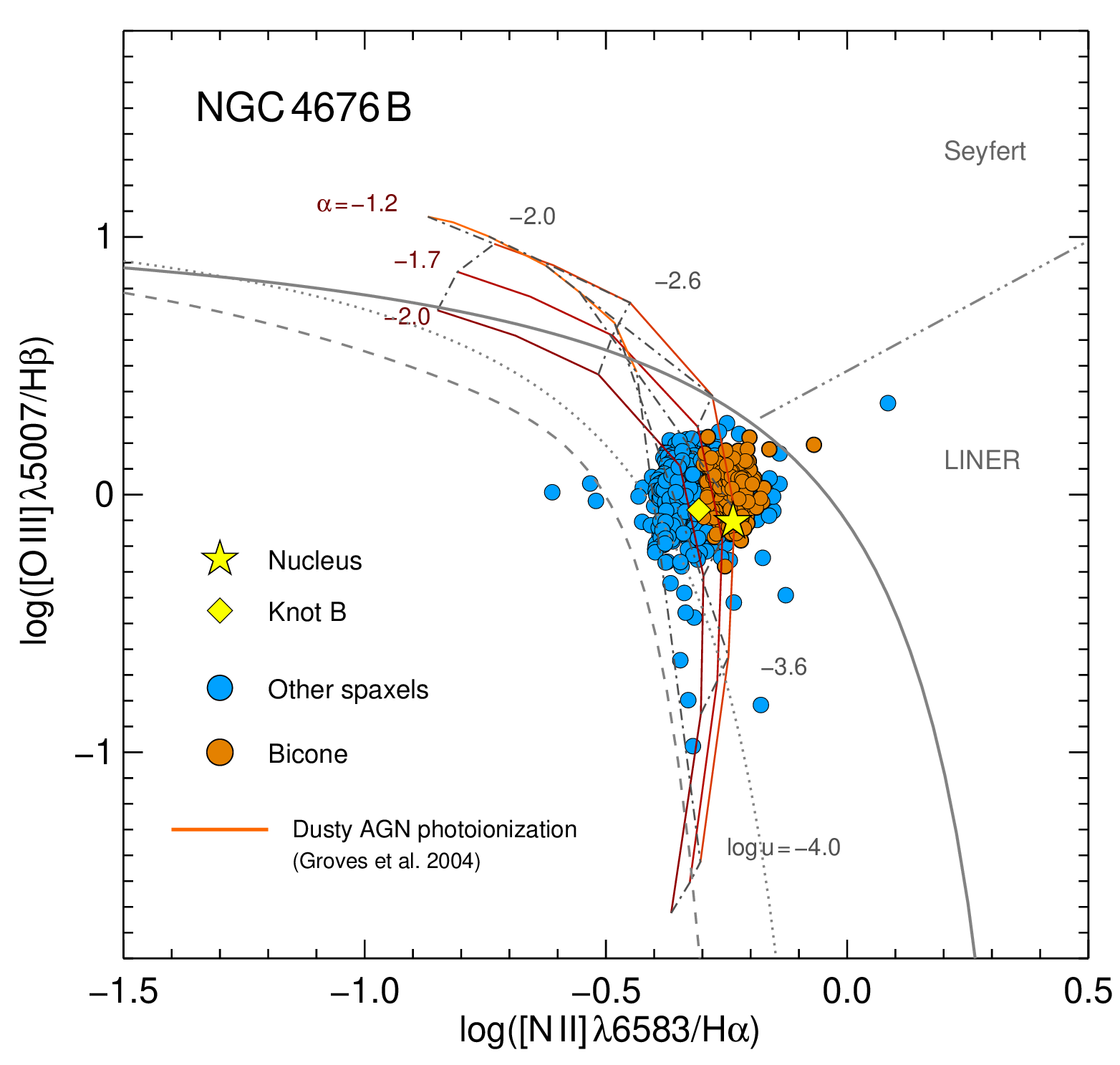}
\includegraphics[scale=0.35]{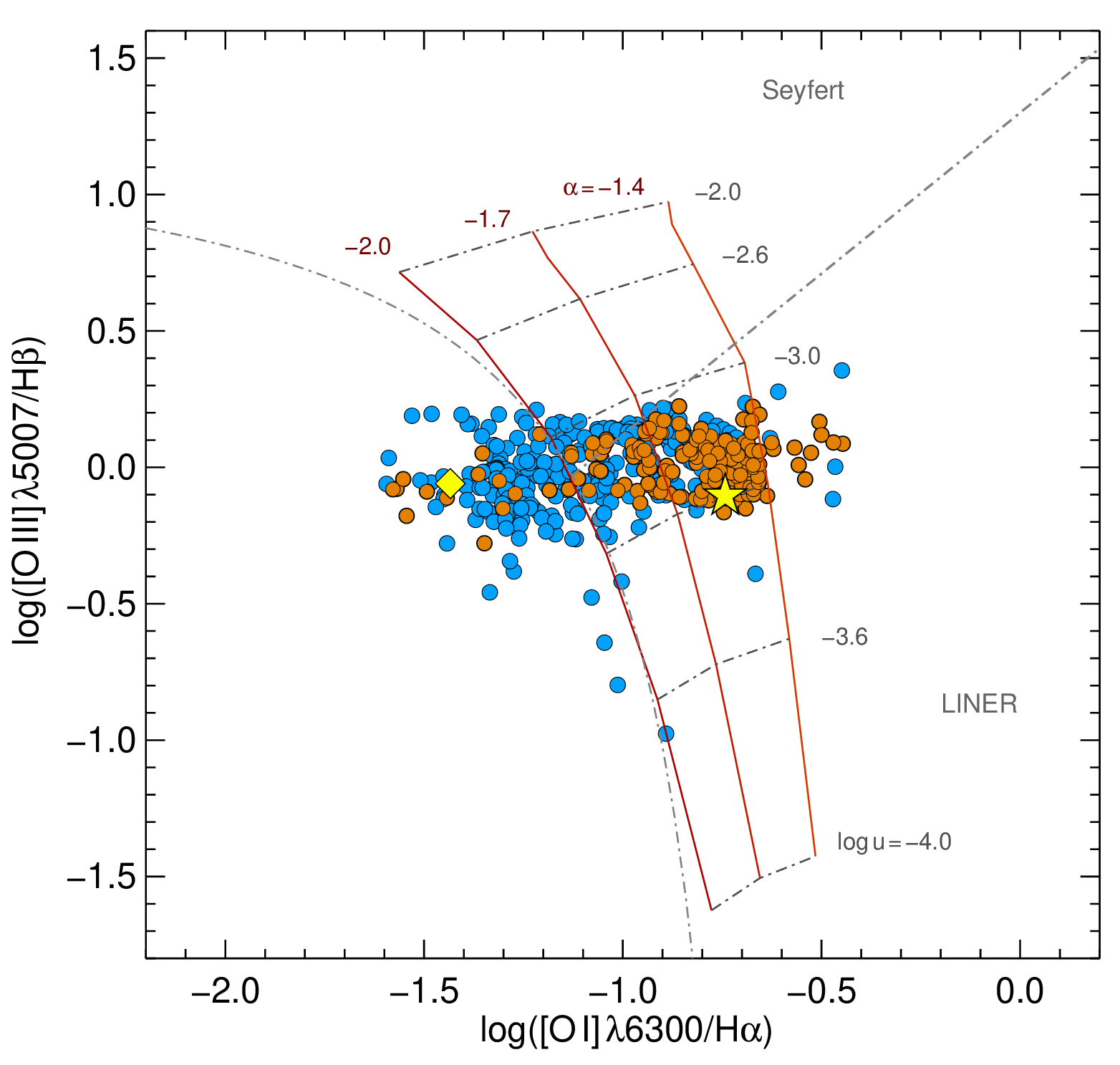}
\caption{Line ratio diagnostic diagrams, showing line ratios for
  independent spaxels in NGC~4676B. Those spaxels which lie in the
  bicones are coloured orange, the remainder of the spaxels are
  included as blue circles. Line ratios in some key regions of
  interest are shown as yellow symbols (see
  Fig. \ref{fig:regions}). Overplotted as black lines are empirically
  and theoretically derived separations between LINERs/Seyferts and
  \HIIr\ regions, as given in Figure \ref{fig:bpt_a}. Overplotted as
  coloured lines are AGN photoionisation model predictions by
  \citet{Groves:2004p4703}. }\label{fig:bpt_b}
\end{figure*}

\subsection{The optical emission nebulae}

Given the LINER- and Seyfert-like line ratios observed in the bicones,
the optical line emission seen at large distances from the galaxy must
largely be a direct result of the fast, radiative shocks which are
being driven into the ambient medium by the superwind.

To obtain the total dust attenuation corrected \ha\
luminosity\footnote{Where \hb\ is not measured we use uncorrected \ha\
  line luminosities, although this is only the case in the fainter
  outer regions and makes no difference to the final results.}  in the
bicones we sum over regions with $\log$(\nii/\ha)$>-0.25$ and
conservatively exclude emission within 1.3\,kpc (3\arcsec) of the
nucleus, where the \nii/\ha\ line ratio lies in the composite region
of the BPT diagram and we might expect a significant fraction of line
emission to instead arise from photoionisation by stars in the
starburst. We also exclude the highly ionised gas to the north and
south of the nucleus, as this may have a different origin. We find the
total \ha\ luminosity of the nebulae is $L_{neb} =
7\times10^{41}$\ergs, leading to a bolometric energy loss rate of
$\dot{E}_{\rm neb}\sim5.5\times10^{43}$\ergs, where we have assumed a
bolometric conversion factor of 80, appropriate for ionisation by
shocks with $v_s\gtrsim140$\,\kms\ \citep{Rich:2010}. This is a factor
of 2-6 larger than the energy flux in the wind fluid estimated from
the geometry and cloud pressure ($\dot{E}_{\rm wind}$, Equation
\ref{eqn:Edotwind}). Given the uncertainties in the many assumptions
made to obtain these two numbers, we consider them to be in good
agreement, implying a near 100\% efficiency in converting wind energy
into radiation.  However, it is possible that an additional source of
ionisation is contributing to the optical nebulae, or our zeroth order
calculations have underestimated the energy flux of the wind by a
factor of a few. In comparison, HAM90 find the total dust
attenuation corrected line luminosity in FIRGs to range from
$1.5\times10^{42} - 1.5\times10^{44}$, assuming a bolometric
correction of 30. They also note that $\dot{E}_{\rm wind} \approx L_{\rm
  neb}$ implies a high efficiency for converting the wind energy into
emission lines.

From the optical emission flux we can also estimate the total mass
of ionised gas currently decelerating at the shock fronts:
 \begin{equation}
 M_{ion} = \frac{\mu m_p L_{H\alpha,0}}{\gamma_{H\alpha}(T) n_e} 
 \end{equation}
 where, for Case B recombination and purely photoionised gas of
 electron temperature $T$, the effective volume emissivity is
 $\gamma_{H\alpha}(T)=3.56\times10^{-25}T_4^{-0.91}$\,erg\,cm$^{-3}$\,s$^{-1}$
 ($T_4=T/10^4K$). Taking $T=10^4$ and the median electron density
 throughout the bicones of $n_e=400$cm$^{-3}$ we find a
 total ionised gas mass of $\sim6\times10^6$\msol.  

 For a shock front travelling at 350\,\kms\ the crossing time to reach
 the outer edge of the bicones is $\sim$18\,Myr. Assuming the shocks
 cause a bulk motion of the entrained gas at the velocity of the
 shock, this gives an ionised gas mass outflow rate of
 0.3\,\msolyr. We see that the mass of outflowing ionised gas is
 negligible compared to the total mass in the wind
 (Eqn. \ref{eqn:Mdotwind}). 

\subsection{Comparison to the energy injection rate by SNe}

The mechanical energy injection rate into the ISM by supernovae (SNe) and stellar
winds can be estimated from evolutionary synthesis models of
populations of massive stars \citep{Starburst99,Veilleux:2005p7318}:
\begin{equation}
\dot{E}_*=7\times10^{41} \left(\frac{\rm SFR}{\rm M_\odot
    yr^{-1}}\right) = 4.3\times10^{42}\,{\rm erg\,s}^{-1}
\end{equation}
for SFR$=6.2$\msolyr (Section \ref{sec:MWsfr}), and we have taken the
limiting assumption that the mechanical energy from all of the stars
being formed throughout the entire galaxy is available to drive the
wind. Even under this assumption, this barely provides sufficient
energy to drive the lowest velocity wind assumed above, and is a
factor of 10 too little to power the emission line nebulae.  The
geometry of the outflow as seen in the CALIFA observations suggests
that only the central star formation is driving the wind, increasing
the discrepancy. This implies that other forms of energy injection
into the biconical nebulae are required beyond simple mechanical
energy, such as photo-heating from the young stars and radiation
pressure (see e.g. Hopkins et al. 2013 for recent simulations
including these effects).

\section{The bicone in NGC~4676B}\label{sec:biconeB}

A bicone is also evident in the line ratio maps of NGC~4676B (Figure
\ref{fig:em_ratio}). The different pattern of line ratios to those
seen in NGC~4676A suggest that different physical mechanisms are
responsible.  In Figure \ref{fig:bpt_b} we plot two emission line
diagnostic diagrams for individual spaxels in NGC~4676B.  As seen in
the map, the bicone does not have distinct \oiii/\hb\ line ratios
compared to the disk.  The nucleus has line ratios in the
``composite'' region in \nii/\ha, and in the LINER region in \oi/\ha.

The hard X-ray detection, compact distribution of soft X-rays and
unusually high ratio of mid-IR excited H$_2$ emission to PAH emission
suggest an AGN may be present in NGC~4676B \citep[see Section
\ref{sec:intro}]{Read:2003p7304,GonzalezMartin:2009p6907,Masegosa:2011p6966}.
If so, the AGN is weak, with a hard X-ray luminosity of
$L_{\rm X,2-10keV}=1.48\times10^{40}$\ergs.  Using a bolometric conversion
factor of 50 suitable for LINERs \citep{Eracleous:2010p7428}, results
in a bolometric luminosity of $7.4\times10^{41}$\ergs\ or
$\log(L/L_\odot)= 8.3$, typical for LINER galaxies. From the {\sc
  CALIFA} nuclear spectrum we measure an \oiii\ luminosity of
$\log(L_{\rm [OIII]}/L_\odot)= 5.9$. Using a bolometric conversion factor
of 600 \citep{heckman04}, leads to a bolometric luminosity of
$1.7\times10^{42}$\ergs\ or $\log(L/L_\odot)= 8.7$. Given the errors
inherent in these conversions, these values are consistent. 

Predicted line ratios from the AGN photoionisation models of
\citet{Groves:2004p4703} are overplotted in Figure \ref{fig:bpt_b},
for a range of dimensionless ionisation parameter ($u\equiv q$/c) and
spectral index ($\alpha$). The line ratios in the nucleus are
consistent with an AGN with $\log u\sim-3.3$ and $\alpha\sim-1.5$,
with line mixing from \hii\ regions in the host galaxy disk causing
the variation across the galaxy.  It is also worth considering the
potential of post-AGB stars in the underlying old stellar population
to contribute to the ionising photon field, particularly in the
massive bulge of NGC~4676B \citep[see
e.g.][]{Eracleous:2010p8632}. However, the observed \ha\ equivalent width in the
nucleus of NGC~4676B is $\sim25$\AA, which firmly rules out any
significant contribution to the line emission from post-AGB stars
\citep{CidFernandes:2011p8588}. Finally, the nuclear line ratios are
also consistent with slow-shock models including some line mixing from
low level star formation
\citep[e.g.][]{Rich:2011}. Ultimately, understanding fully the source of the
high ionisation lines in NGC~4676B will require observations with higher
SNR, spatial and spectral resolution.

\section{Comparison to simulations}\label{sec:simn}

\begin{figure}
  \includegraphics[scale=0.35]{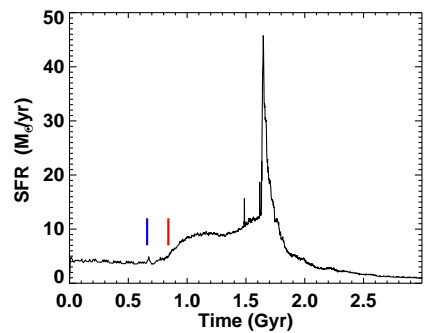}
   \caption{The combined SFR of the mock Mice over the duration of the
    merger simulation. The blue and red vertical marks indicate the time of first
    passage and time of observation at 180\,Myr after first
    passage.}\label{fig:simn3}
\end{figure}

\begin{figure*}
\centering
\includegraphics[scale=0.6]{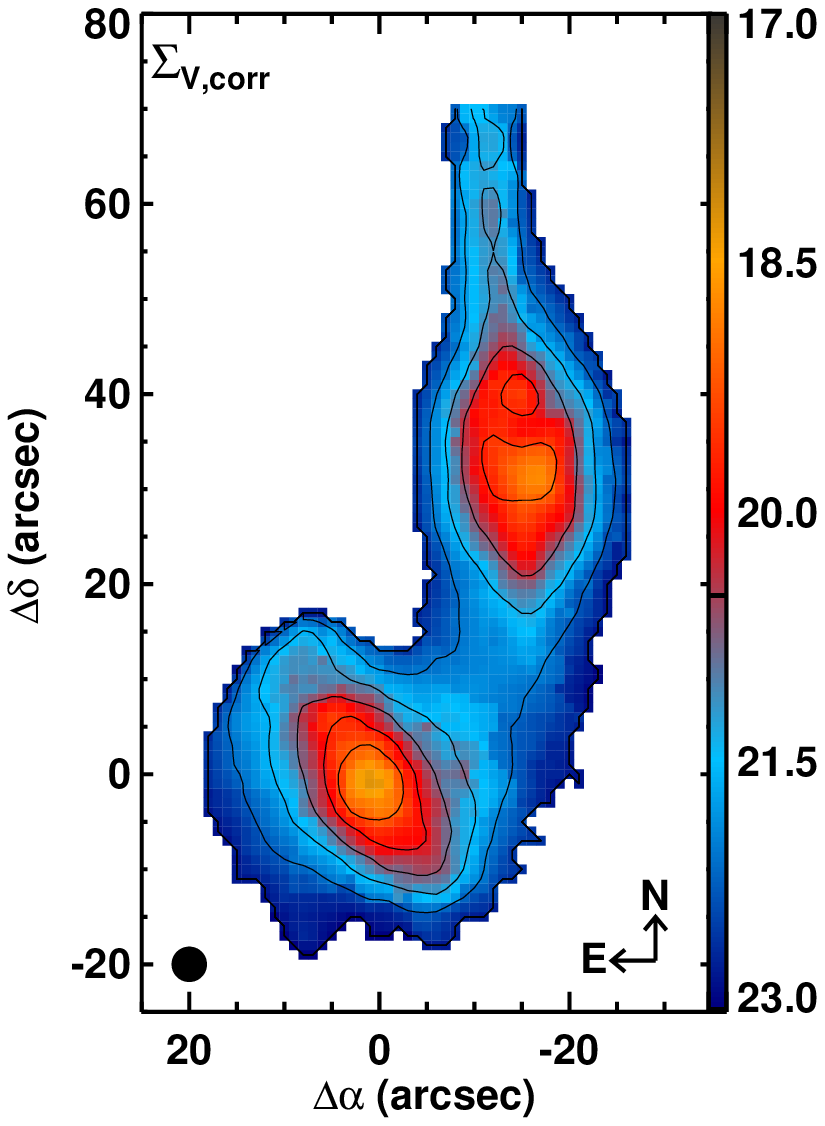}
\hspace*{0.5cm}
\includegraphics[scale=0.6]{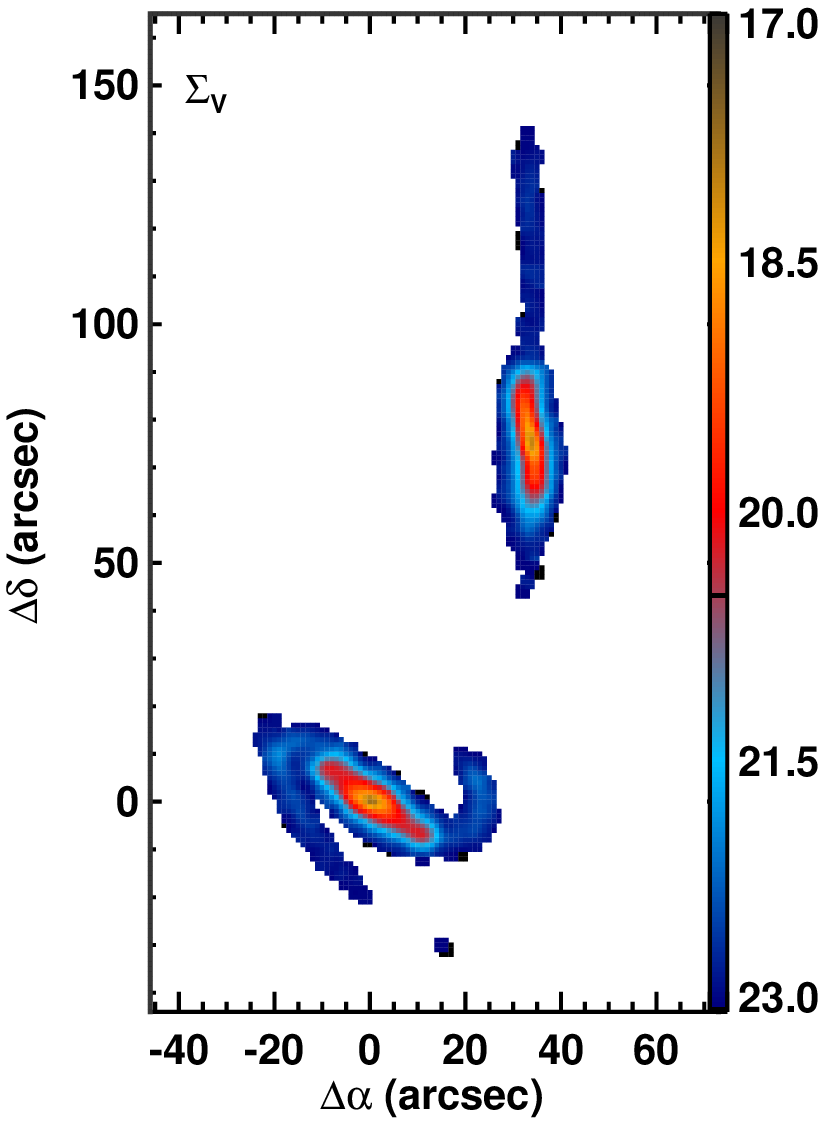}

\includegraphics[scale=0.6]{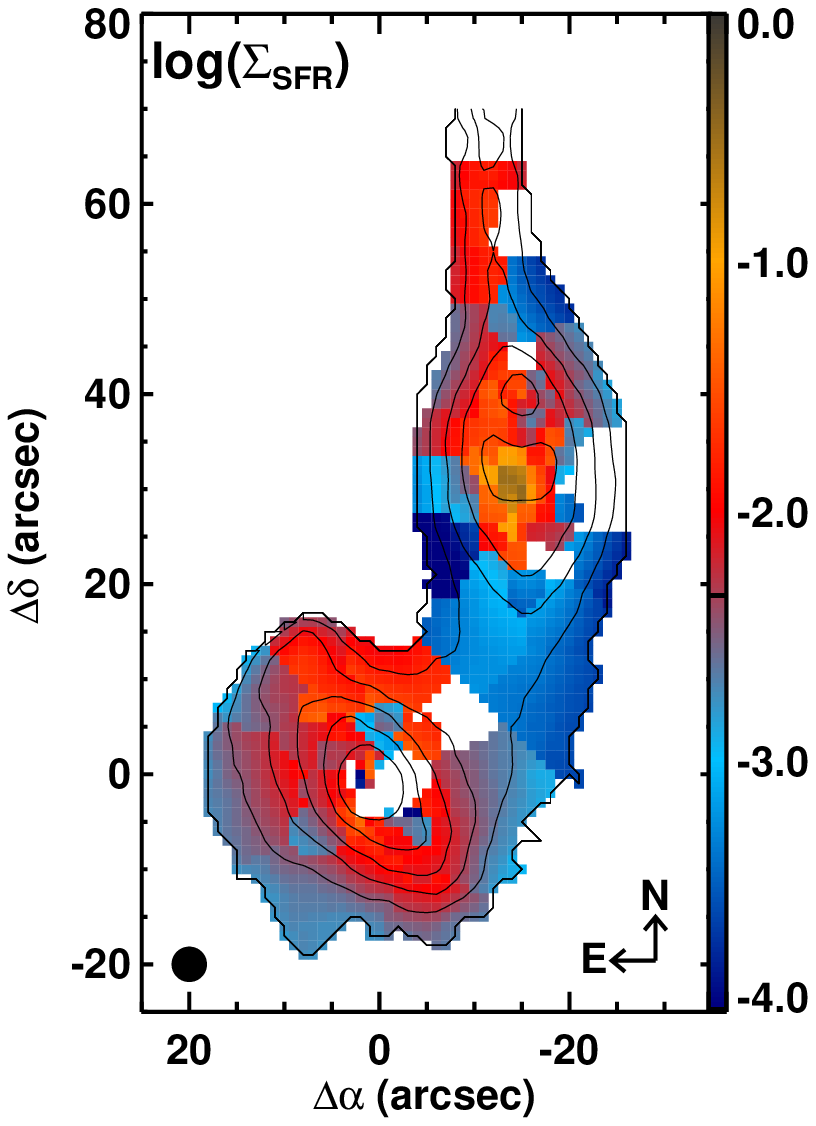}
\hspace*{0.5cm}
\includegraphics[scale=0.6]{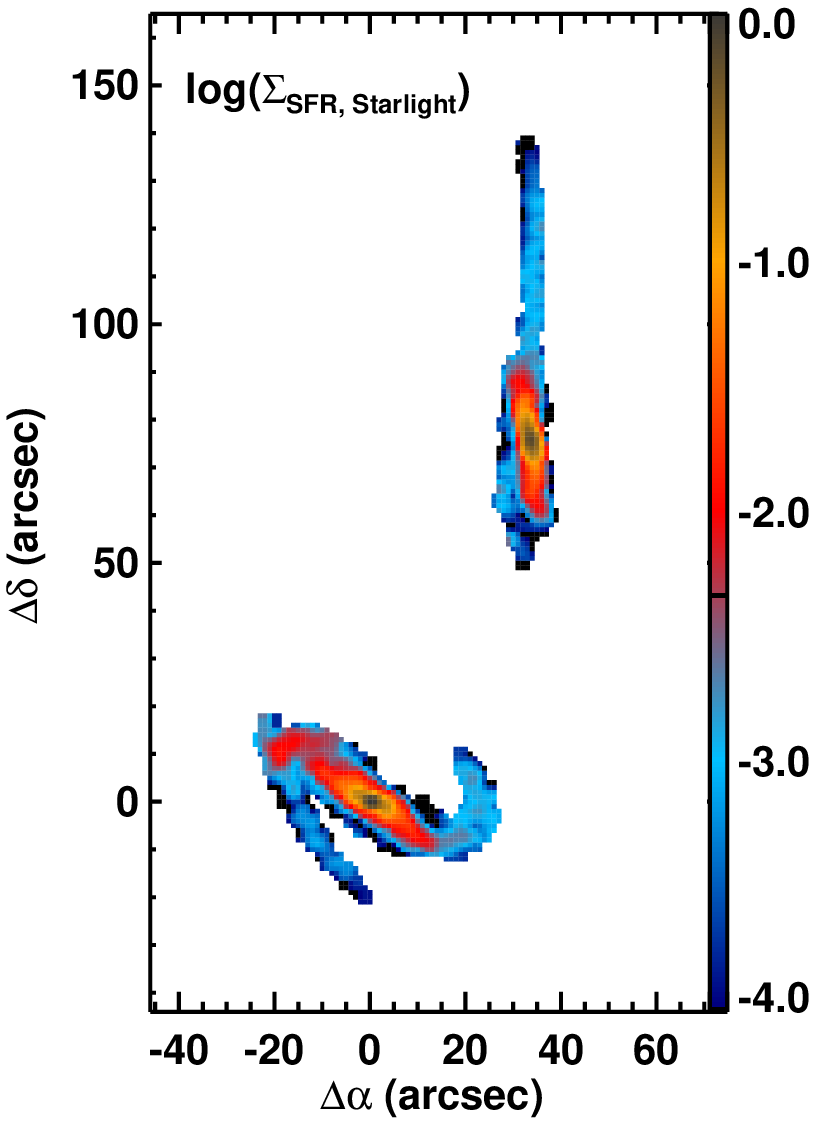}

\caption{\emph{Top:} The V-band surface brightness
  (mag/$\square\arcsec$) of the real (\emph{left}) and mock
  (\emph{right}) Mice. The surface brightness of the real Mice has
  been corrected for dust attenuation using the analysis from the {\sc
    starlight} code (Section \ref{sec:sfh}). The surface brightness of
  the mock Mice has been calculated by post-processing an SPH
  simulation to obtain a mock IFU datacube with the same spectral
  range and resolution as the CALIFA data. This has been analysed
  using the same pipelines as the real data. \emph{Bottom:} The star
  formation rate surface density (\sfrsd) measured from the stellar continuum
  using the {\sc starlight} code, averaged over the last 140\,Myr, for
  the real (\emph{left}) and mock (\emph{right}) Mice. White
  indicates regions with no measurable star formation.
} \label{fig:simn1}
\end{figure*}

\begin{figure*}
\centering
\includegraphics[scale=0.6]{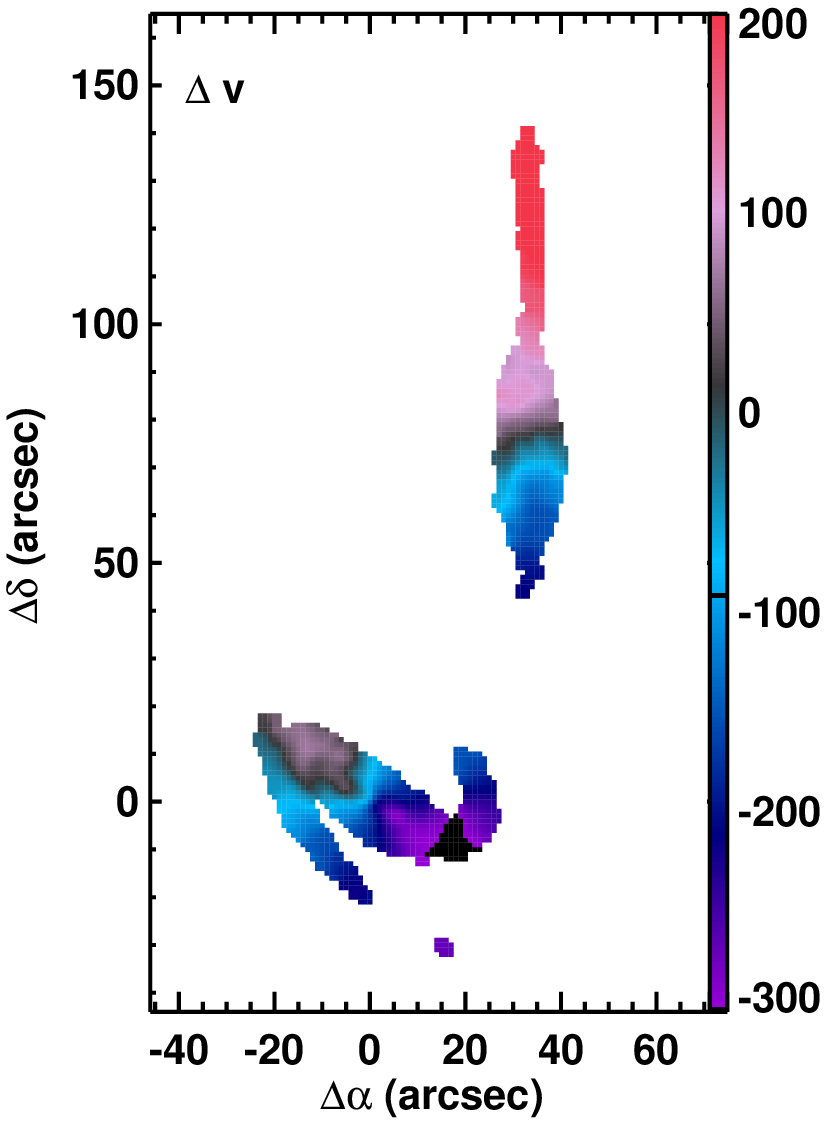}
\hspace*{0.5cm}
\includegraphics[scale=0.6]{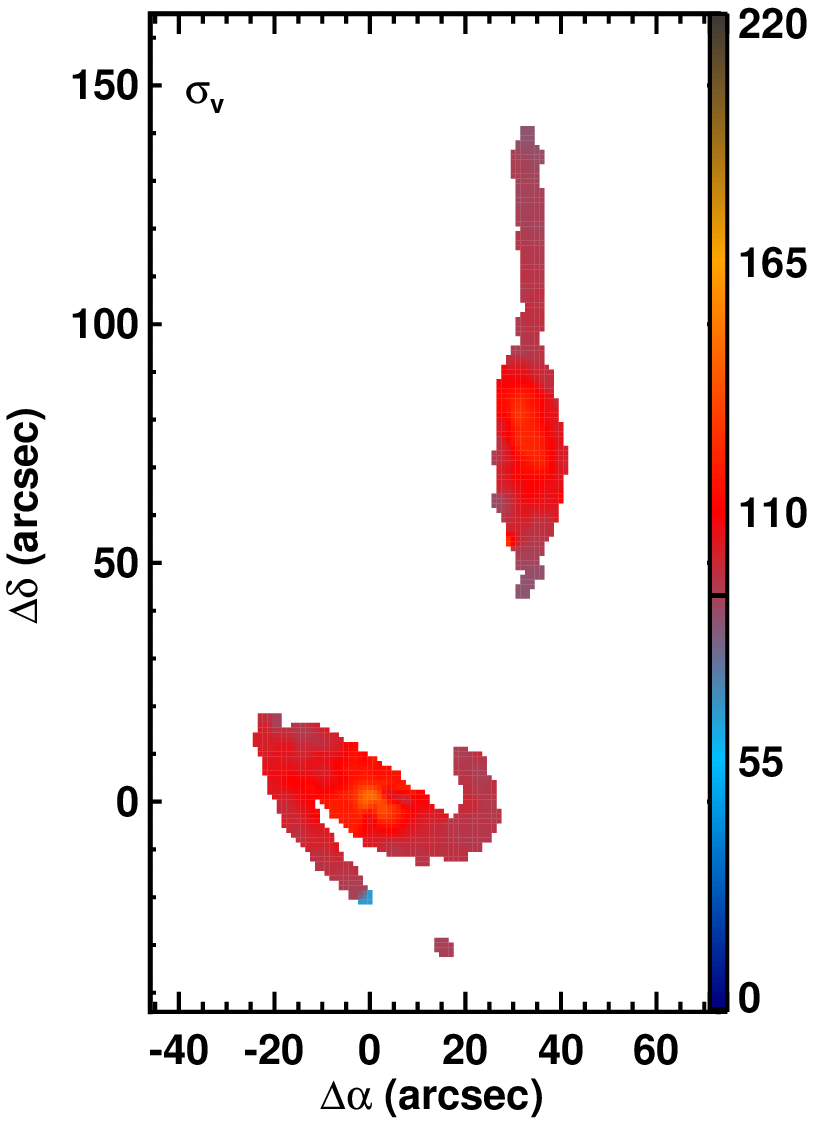}
\caption{The stellar velocity field and dispersion of the mock
  Mice. See Figure \ref{fig:kin} to compare with the data.} \label{fig:simn2}
\end{figure*}

Since the initial galaxy merger simulations by
\citet{1972ApJ...178..623T}, a large research effort has focussed upon
trying to reproduce the spatial distribution of stars, gas, star
formation and stellar cluster ages in mergers
\citep[e.g.][]{Karl:2010p7179,Lanz:2014p9226}.  With their tidal tails
constraining the orbital parameters reasonably well, the Mice have been an
obvious target for such studies. For example, \citet{Mihos:1993p7313}
found the SFR of NGC~4676A to be higher in models than data, which
could result from limitations of the data available at the time
(e.g. inability to correct for dust attenuation) or limitations of the
model.  \citet{Barnes:2004p6826} used the Mice as a case study to
advocate a shock-induced star formation law, rather than the standard
gas density driven star formation implemented in hydrodynamic
simulations \citep[see also][]{Teyssier:2010p7166}. Recently,
\citet{Privon:2013p9225} presented a new solution for the orbital
parameters of the Mice from purely N-body simulations, which differ
from those of  \citet{Barnes:2004p6826}, largely due to the use of a
different mass model for the galaxies. The degeneracy between mass
model and orbital parameters are discussed in detail in
\citet{Privon:2013p9225}  and inhibit easy comparison between
simulations and observations. 

With observations advancing towards an era in which spatially resolved
star formation histories, kinematics, ionisation mechanisms and AGN
strengths are well constrained for single objects, simulations are
facing more stringent observational constraints than ever before
\citep[e.g.][]{Kronberger:2007p8642}. Given the primarily
observational nature of this paper, we present here an illustrative,
rather than exhaustive, comparison of the CALIFA results with a
standard smoothed particle hydrodynamic (SPH) merger simulation. It is
simulations such as these that have driven many of the currently
favoured models of galaxy evolution over the last decade, and it is
worth investigating how well they match the latest generation of
observations. For conciseness, we focus on the stellar kinematics, SFR
surface densities and star formation histories of the galaxies,
leaving a more complete study for further work.

 The simulation represents the collision of two galaxies with a
  baryonic mass ratio of 1.3: a pure exponential disk galaxy and a
  galaxy with a bulge-to-disk ratio of 0.27. These parameters are very
  close to the observed mass ratio and morphologies of the Mice
  galaxies. We take the orbital parameters provided in
  \citet{Barnes:2004p6826}, i.e. a close to prograde-prograde orbit
  with a closest approach of $\sim$10\,kpc occurring $\sim$170\,Myr
  ago.  Star formation and the associated supernova feedback is
  implemented using the sub-resolution multiphase model developed by
  \citet{2003MNRAS.339..289S}, in which cold gas forms stars when its
  density reaches above a certain density threshold (${\rm
    n_H}=0.128\,{\rm cm}^{-3}$, appropriate for the resolution of the
  simulation). For the purposes of this CALIFA comparison, black hole
  feedback is not included.

Spectral energy distributions of each of the star particles
were calculated from their star formation history, using the models of
\citet{2003MNRAS.344.1000B}. In order to make a fair comparison
between observations and simulations, we convert the simulation into
an IFS data cube with the same spectral and
spatial resolution as the data, and analyse this cube using the same
codes as employed on the real data.  Full details of the simulations
and post-processing applied to produce the mock observations are given
in Appendix \ref{app:simn}.

Figure \ref{fig:simn3} presents the global SFR of both galaxies in the
simulation, over the full 3\,Gyr run-time. The time of first passage
and time of observation are marked as vertical lines. The small burst
of star formation caused by first passage is evident. At final
coalescence the simulations predict a 40-50\msolyr\ starburst;
however, more accurate gas mass fractions of the real Mice would be
required to improve the accuracy of this prediction. 

The upper panels of Figure \ref{fig:simn1} show maps of the $V$-band
surface brightness of the real and mock Mice.  The data has been
corrected for dust extinction using the $V$-band stellar continuum
attenuation map estimated by the {\sc starlight} code. On altering the
viewing angle of the simulation a strong bar becomes visible in
NGC~4676A. This lends support to our argument that the boxy shape of
NGC~4676A is caused by an edge-on bar rather than a classical bulge
(Section \ref{sec:galfit}). While the sizes and overall surface
brightnesses of the real and mock galaxies are largely comparable, the
mock galaxies are separated by a distance on the sky that is a factor
of 2 larger than observed. This may result from incorrect initial
orientations, orbits, mass profiles, or shapes of the dark matter
halo. The new orbital parameters presented in \citet{Privon:2013p9225}
differ in several aspects to those used here. However, because these
authors assume very different mass models for the progenitor galaxies,
their orbital parameters are no more likely to provide a better
match. Simply slowing the initial velocities of the simulated galaxies
resulted in too short tidal tails, and further investigations into the
origin of the discrepancy are beyond the scope of this primarily
observational paper.

The lower panels of Figure \ref{fig:simn1} show the SFR surface
density averaged over the last 140\,Myr, calculated for both the real
and mock Mice from the decomposition of the stellar continuum by the
{\sc starlight} code (Section \ref{sec:sfh}). The SFR surface density
reconstructed from the mock IFS data cube using {\sc starlight}
closely matches the instantaneous SFR surface density taken directly
from the simulation. The global SFR of each of the mock galaxies is
$\sim2.5$\msolyr, in agreement with the observed SFR for NGC~4676B
(2\msolyr) and a little lower than observed for NGC~4676A
(6\msolyr). 

The SFR surface densities in the central $5\times5$\arcsec\ (2.2\,kpc)
are $\sim0.3$\sfrsd\ for both mock galaxies. The value averaged over
140\,Myr using the {\sc starlight} spectral decomposition is close to
the instantaneous value obtained directly from the simulation. This
compares well to a central SFR density of 0.15\sfrsd\ for NGC~4676A
and is not substantially different from the $\sim0$\sfrsd\ measured
for NGC~4676B from the {\sc starlight} spectral decomposition (Section
\ref{sec:sfh}). However, the observed decrease in SFR surface density
towards the central regions of NGC~4676B is not reproduced by the
simulations.

Figure \ref{fig:simn2} shows the stellar velocity field and velocity
dispersion maps of the mock Mice.  The velocity map of NGC~4676A is a
good match to the data, and including a bulge results in a
significantly worse match. The velocity map of NGC~4676B is dominated
by a rotating disk, with the bulge causing the pinched effect at the
centre and small rise in velocity dispersion. The unusual twisted
stellar disk velocity field seen in the observations, with the
rotation axis offset from the minor axis, is not observed at this
stage of the simulation. During the full run time of the simulation,
the high surface brightness of the tidal tails in the Mice is clearly
a transient feature. There is also evidence for a slight twist in the
velocity field of NGC~4676B at some epochs of the simulations, similar
to that seen in the observations. Precisely reproducing these
transient morphological and kinematic features may make it possible to
set tighter constraints on the orbital parameters of the Mice merger
and mass distribution of the galaxies in the future.

\section{Discussion and Summary}\label{sec:disc}

\begin{table}
  \caption{A summary of the main properties of the Mice
    galaxies measured in this paper. For a summary of properties
    available from the literature see Table
    \ref{tab:basic}. }\label{tab:summary}
  \centering
   \begin{tabular}{ccc}\hline\hline
     Parameter      & NGC~4676A (NW) & NGC~4676B (SE) \\\hline
      B/T\tablefootmark{a}        & 0 (SBd)     & 0.5 (S0/a) \\
      v$_{\rm nucleus, gas}$ (km/s)\tablefootmark{b}    & 6585 & 6581 \\
      v$_{\rm nucleus, stars}$ (km/s)\tablefootmark{b}  & 6652 & 6493\\
      M$^*$ ($10^{11}$M$_\odot$)\tablefootmark{c}  & 1.2  & 1.5  \\
      SFR (M$_{\odot}$/yr)\tablefootmark{d} & $\sim$6 & $\sim$2 \\
      P/k$_{\rm wind}$ ($10^{6}$K\,cm$^{-3}$)\tablefootmark{e} & 4.8 & ... \\
      $\dot{\rm M}_{\rm out}$ (\msolyr)\tablefootmark{e} & 8-20 & ... \\
      Morphological PA\tablefootmark{f} & 2.8 & 33.2 \\
      Kinematic PA$_{\rm stars}$\tablefootmark{f}  & 33.9, 27.6 & 165.4, 150.4   \\
      Kinematic PA$_{\rm gas}$\tablefootmark{f}  & 31.9, 24.5 & 154.4, 165.0 \\
\hline
    \end{tabular}
    \tablefoot{
      \tablefoottext{a}{ Bulge-to-total flux ratio in the $I$-band, from
          image decomposition of HST-ACS image
          (Section \ref{sec:galfit}).}
      \tablefoottext{b}{Recessional velocity, from kinematic fits to the
        central spaxel with the highest $V$-band luminosity 
        (Section \ref{sec:kin}).}
      \tablefoottext{c}{Total stellar mass, from full spectrum decomposition assuming a Salpeter IMF (Section
        \ref{sec:sfh}).}
      \tablefoottext{d}{Global SFR taken from the median of CALIFA and
        multi-wavelength measurements (Section \ref{sec:MWsfr}).}
      \tablefoottext{e}{The ram pressure and mass outflow rate of the bi-conical wind
        from NGC~4676A (Section
        \ref{sec:outflowA}).}
      \tablefoottext{f}{ Position angles (approaching
          and receding for kinematic PAs), measured
          within 10\arcsec\ (Barrera-Ballesteros et al. in prep., Section \ref{sec:kin}).}
    }
\end{table}

Here we summarise the impact of the close encounter on the
morphology, kinematics, star formation rate and history, and ionised
gas of the merging galaxies. Table \ref{tab:summary}
collates some of the key properties measured in this paper. 

\begin{enumerate}
\item{\bf Merger induced bars in NGC~4676B and NGC~4676A?}  The strong
  bar in NGC~4676B is clear, both in the image and kinematics maps:
  the characteristic Z-shaped iso-velocity contours are visible in
  both the ionised gas and, unusually, in the stellar velocity
  field. While NGC~4676A has previously been classified as an S0, the
  CALIFA maps of the stellar kinematics show no evidence for a
  classical bulge, with low nuclear velocity dispersion and constant
  rotation above with height above the major axis. A young, thin, disk
  is visible in the stellar population maps, with older populations
  extending into the ``boxy'' shaped bulge. The high dust and gas
  content of the galaxy is also inconsistent with its classification
  as an S0.  It is possible that the boxy morphology evident in the
  imaging (See Figure \ref{fig:hst}) comes from a strong bar
  \citep[e.g.][]{Kuijken:1995,Bureau:1999p9430,Bureau:2005p8704,
    MartinezValpuesta:2006p7585,Williams:2011}; further detailed
  investigation of higher order moments in the velocity map may help
  to confirm this.  The first passage of a major merger is seen to
  induce strong bars in simulations of disky galaxies
  \citep[e.g.][]{Barnes:1991p7419,Lang:2014p9453}, including the mock Mice simulation
  presented here. While we cannot rule out the pre-existence of bars
  in the merging galaxies, the presence of such strong bars in both
  galaxies, combined with predictions from simulations, suggests that
  these have been induced by the recent close encounter.

\item{\bf Z-shaped stellar velocity field in NGC~4676B.} The CALIFA
  data reveals that the dominant rotation of both the ionised gas and
  stars in NGC~4676B is  close to being around the major axis of
    the galaxy (offset between morphological and kinematic PAs of
    50-60 degrees). While these Z-shaped (or S-shaped) isovelocity
  contours are a common feature in the gas velocity fields of strongly
  barred galaxies \citep[e.g.][]{Contopoulos:1980p8403}, and weak
  warps can be seen in the outer regions of stellar disks
  \citep{Saha:2009p9231}, such strong disturbance of the inner stellar
  velocity field is unusual \citep[for one other example see NGC
  4064, ][]{Cortes:2006p9398}.  \citet{BarreraBallesteros:2014p9452}
  have confirmed that bars have no significant effect on stellar
  velocity fields in the full CALIFA sample. Studies of larger samples
  are needed to see whether such strong twists are common in merging
  galaxies (see Barrera-Ballesteros et al, in prep., for such a study
  within CALIFA).

\item{\bf No substantial increase in global star formation rates.}
  We can compare the sSFR of the Mice galaxies to the distribution of
  aperture corrected sSFR as a function of stellar mass for galaxies
  in the SDSS survey \citep[Figure 24 of][]{2004MNRAS.351.1151B}. We
  find that NGC~4676A lies at the upper end of stellar masses measured
  for star-forming galaxies in the local Universe and has exactly
  typical sSFR for its stellar mass. NGC~4676B has a slightly low sSFR
  for its stellar mass, but still lies well within the range observed
  in the general population. Both galaxies have SFR surface densities
  typical of star-forming disk galaxies. We conclude that there is no
  evidence that either galaxy is currently undergoing a substantial
  galaxy-wide burst of star formation. These results are broadly
  consistent with the mock Mice simulations, where the galaxies have
  only undergone a small increase in global SFR since first passage.
  Circumstantial evidence for increased star formation in close pairs
  has existed for decades
  \citep[e.g.][]{Bernloehr:1993p8522}. Statistically significant
  evidence for the enhancement in star formation due to interactions
  and mergers has come from large samples of spectroscopically
  observed galaxies in the last decade
  \citep{Barton:2000p8514,Lambas:2003p8630,Patton:2011p7164,FreedmanWoods:2010p4650,Patton:2013p9428}.
  Estimates of the occurrence, strength and duration of SFR
  enhancement vary. For example, \citet{FreedmanWoods:2010p4650}
  find that SFR increases by a modest factor of 2-3 averaged over the
  duration of the enhancement, which is around a few hundred million
  years. Such enhancements occur in about 35\% of close pairs. These
  observational results are consistent with hydrodynamic galaxy merger
  models \citep{DiMatteo:2008p8552} and show that the lack of SFR
  enhancement in the Mice galaxies is not unexpected. 

\item{\bf No substantial post-starburst population.} First passage of
  the Mice merger occurred roughly 170\,Myr ago
  \citep{Barnes:2004p6826,Chien:2007p7317} and star formation induced
  at this epoch should be identifiable through decomposition of the
  stellar continuum to measure the fraction of light from A and F
  stars. Referring to the fraction of light emitted by stars formed
  between 140\,Myr and 1.4\,Gyr ago in Figure \ref{fig:2dsfh}, we see
  that during this time star formation is located primarily in the
  disks of both galaxies, as expected for continuous star formation
  histories. There is some enhancement in the intermediate
  age population in the outer extents of the galaxies. Our results
  show that, while some low level star formation at large radii may
  have been triggered by the first passage, the fraction of total
  stellar mass formed was not significan.

\item{\bf Weak nuclear starburst in NGC~4676A.} Due to the attenuation
  by dust in the centre of NGC~4676A, and poor spatial resolution of
  far infrared observations, our estimates of the nuclear SFR are
  uncertain. The SFR surface density in the central $\sim$5\arcsec\
  estimated from the {\sc starlight} spectral decomposition is
  $\sim$0.15\sfrsd, placing it at the upper end of the range for local
  spiral galaxies \citep{1998ApJ...498..541K}. This is in good
  agreement with the mock Mice simulation.

\item{\bf No ongoing star formation in the centre of NGC~4676B.} Our
  stellar continuum decomposition of NGC~4676B shows a 
  complete absence of young stars within $\sim$1\,kpc of the nucleus
  (Figure \ref{fig:2dsfh}). The upper limit on the SFR surface density
  from the dust attenuation corrected \ha\ luminosity is
  $\Sigma_{SFR}<0.2$\sfrsd\ within the central
  5\arcsec. It is clear that NGC~4676B is not
  currently undergoing or has not recently undergone a nuclear
  starburst. The strong bar has not yet driven significant gas into
  the nucleus. This decrease in SFR surface density in the central
  regions of the galaxy is not reproduced by the mock Mice simulation. 

\item{\bf Spectacular bicones driven by fast shocks in NGC~4676A.}
  The extreme line ratios seen at the outer edge of the bicones are
  consistent with being caused by fast shocks ($v_s\sim350$\,\kms)
  driven by a superwind. The emission of NGC~4676A in soft X-rays is
  also found to be elongated along the minor axis of the galaxy,
  coincident with the ionised gas bicones seen here
  \citep{Read:2003p7304}, again implying a strong galactic outflow. In
  the nearby Universe observational studies of galactic superwinds have
  focussed primarily on LIRGs and ULIRGs, where a clear driving source
  is present \citep[see][for a review]{Veilleux:2005p7318}.  As
  NGC~4676A is less luminous than a LIRG, and has a moderate star formation
  rate despite its very large mass, it would not have been an obvious
  candidate for superwind investigations.  Stacking of rest-frame
  optical and UV spectra finds evidence for winds in a large fraction
  of star forming galaxies at all redshifts
  \citep[e.g.][]{Weiner:2009p5902,Chen:2010p7414}, implying that our
  census of superwinds is still far from complete and NGC~4676A may
  not be so unusual. Many more outflows may be found in ongoing and
  upcoming IFU galaxy surveys \citep[e.g.][in commissioning data for
  the SAMI survey]{Fogarty:2012p8646}.

\item{\bf What drives the outflow in NGC~4676A?} Observational results
  suggest a minimum SFR surface density of
  $\sim\Sigma_{SFR}=0.1$\sfrsd\ is required for superwinds to be
  launched from galaxies in the local and distant Universe (Heckman
  2002, and see the recent compilation by Diamond-Stanic et
  al. 2012)\nocite{Heckman:2002p7644, DiamondStanic:2012}. Thus,
  although the nuclear starburst in NGC~4676A is not significant in
  terms of global SFR and is weak with respect to local starbursts, it
  is powerful enough to launch a wind in principle. The mass outflow
  rate from NGC~4676A is about 1.5-3 times the global SFR of the
  galaxy, which is typical for galactic outflows in the local Universe
  \citep{1999ApJ...513..156M}. On the limiting assumption that the
  star formation in the entire galaxy is driving the superwind, we
  show that the mechanical energy available from SNe and stellar winds
  is a factor of 10 too low to explain the optical line emission in
  the bicones above 1.3\,kpc from the plane of the galaxy, and only
  sufficient to explain the energy outflow rate if the velocity of the
  wind fluid is very low ($<1000\,$\kms). If only the nuclear star
  formation is driving the wind, as suggested by the CALIFA maps, the
  discrepancy increases. Additional sources of energy input could be
  from ionising photons and radiation feedback
  \citep{Murray:2011p8622,Agertz:2012p8618,Hopkins:2013p8621}. Alternatively,
  the additional energy could have been provided by a source that has
  since switched-off, such as an AGN or a more intense starburst at
  first passage.

\item{\bf Extended narrow line region (ENLR) in NGC~4676B.} The
  kiloparsec scale bicones in NGC~4676B are not accompanied by higher
  velocity dispersion gas, and the line ratios are consistent with
  ionisation by an AGN. The detection of hard X-rays in the nucleus
  and point source like distribution of soft X-rays suggests the
  presence of a weak AGN. We conclude that there is no evidence for a
  galactic outflow from NGC~4676B \citep[as seen in e.g. Cen A,
  ][]{Sharp:2010p8649}, and the bicones are more likely caused by the
  excitation of off-planer gas by an AGN \citep[as seen in e.g. NGC
  5252, ][]{Morse:1998p9224}. Kiloparsec scale ENLRs are not unusual
  in local AGN \citep{Gerssen:2012p9429}.

\item{\bf Areas for improvement in the simulations.} The new
  constraints afforded by the CALIFA data of the Mice galaxies
  indicate new directions where improvements can be made in the
  simulations. Firstly, there is some tension between the geometry of
  the tidal tails and the relative position of the two Mice, with the
  simulation Mice flying twice as far apart to obtain the same length
  and surface brightness of the tails. Further useful constraints on
  the orbital parameters of the merger and mass profiles of the
  galaxies may be obtained by detailed matching of transient
  morphological and kinematic features, such as the twist in the disk
  of NGC~4676B, the orientation of the two bars and the surface
  brightness of the tidal tails. Secondly, observations of the gas
  mass fraction of the galaxies could be used to further test the
  ability of the models to reproduce the SFR surface density maps, and
  in particular understand the very low star formation rate observed
  in the nucleus of NGC~4676B.

\end{enumerate}

\section{The future is a LIRG... and then?}

While the combination of CALIFA and multiwavelength data has
conclusively shown that first passage has not triggered substantial
star formation, the considerable gas content of the two galaxies
makes it likely that final coalescence will cause a substantial
starburst. The two galaxies have a combined molecular gas mass of
$7.2\times10^9$\msol\ \citep{Yun:2001p7305}, which on conversion to
stars will lead to a SFR of order 100\,\msolyr\ over a timescale of
$10^7$years and efficiency of 10\% (or timescale of $10^8$years and
efficiency of 100\%), values typical for starburst galaxies
\citep{1998ApJ...498..541K}. This would result in an 8-1000\mum\
luminosity of $L_{IR}\sim5\times10^{11}$\lsol, i.e. a LIRG. The
simulation presented in Section \ref{sec:simn} predicts a similar peak
SFR of $\sim$50\msolyr.

On coalescence the galaxy will have a stellar mass of
$\sim3\times10^{11}$\msol\ (Salpeter IMF), assuming that each galaxy
increases its mass by 10\% during the starburst as
found observationally \citep{Wild:2009p2609,
  Robaina:2009p8635}. Converting to a Chabrier IMF ($1.8\times10^{11}$\msol)
and comparing to the galaxy mass function of \citet{Baldry:2012p8634}
we see that this places the merger remnant in the very highest mass galaxy
population in the local Universe. Galaxies of this mass and above have
a number density of a few $\times10^{-4}$\,Mpc$^{-3}$.

Extending the predictions for the properties of the system beyond
coalescence of the galaxies requires input from simulations. By the
end of the hydrodynamic merger simulation presented in Section
\ref{sec:simn}, in about 2.2\,Gyr time, the merger remnant will have a
SFR of 1\msolyr, with the decay in star formation occurring as a
natural result of gas consumption. This would lead to an elliptical
galaxy with a specific star formation rate (SFR/M$^*$) of
$5\times10^{-12}$\,yr$^{-1}$, within the range measured for galaxies
of this mass in the local Universe \citep{Schiminovich:2007p9431}.

\section*{Acknowledgements}
This study makes use of the data provided by the Calar Alto Legacy
Integral Field Area ({\sc CALIFA}) survey (http://www.califa.caha.es).
Based on observations collected at the Centro Astron\`{o}mico Hispano
Alem\'{a}n (CAHA) at Calar Alto, operated jointly by the
Max-Planck-Institut f\"{u}r Astronomie and the Instituto de
Astrofisica de Andalucia (CSIC). {\sc CALIFA} is the first legacy
survey being performed at Calar Alto. The {\sc CALIFA} collaboration
would like to thank the IAA-CSIC and MPIA-MPG as major partners of the
observatory, and CAHA itself, for the unique access to telescope time
and support in manpower and infrastructures.  The {\sc CALIFA}
collaboration thanks also the CAHA staff for the dedication to this
project.

The authors would like to thank the anonymous referee for comments
that significantly improved the paper; Dimitri Gadotti for providing
extensive help and advice with the image decomposition; Mike Dopita
and Tim Heckman for their patience explaining the effects of shocks in
superwinds; Daria Dubinovska for creating undistorted ACS PSF images
from the Tiny Tim images and Carolin Villforth for further help with
the ACS PSF; Lia Athanassoula for help interpeting the
kinematic maps; Daniel Pomarede for help getting SDvision running;
Jeremy Sanders and Roderik Johnstone for help interpreting the X-ray
observations; Eva Schinnerer for pointing out the effect of beam
smearing; all other interested researchers who have contributed with
questions and comments following discussions and presentations of this
work over the last 2 years.

The numerical simulations were performed on facilities hosted by the
CSC -IT  Center for Science in Espoo, Finland, which are financed by the
Finnish ministry of education.

Funding and financial support acknowledgements: 
V.~W. from the European Research Council Starting Grant (P.I. Wild SEDmorph), European Research Council Advanced Grant
(P.I. J.~Dunlop) and Marie Curie Career Reintegration Grant (P.I. Wild
Phiz-ev);  
J.~M.~A. from the European Research Council Starting Grant (P.I. Wild
SEDmorph); 
F.~F.~R.~O. from the Mexican National Council for Science and Technology (CONACYT);
A.~G. from the European Union Seventh Framework Programme (FP7/2007-2013) under grant agreement n. 267251. The Dark Cosmology Centre is funded by the Danish National Research Foundation.
P.~H.~J. from the Research Funds of the University of Helsinki; 
J.~F.~B. from the Ram\'on y Cajal Program, grants AYA2010-21322-C03-02 and AIB-2010-DE-00227 from the Spanish Ministry of Economy and Competitiveness (MINECO), as well as from the FP7 Marie Curie Actions of the European Commission, via the Initial Training Network DAGAL under REA grant agreement no. 289313;
R.~G.~D. and R.~G.~B. from the Spanish project AYA2010-15081;
A.~M.-I. from the Agence Nationale de la Recherche through the STILISM
project (ANR-12-BS05-0016-02) and from BMBF through the Erasmus-F project (grant number 05 A12BA1).
R.~A.~M. from the Spanish programme of International Campus of Excellence Moncloa (CEI); 
K.~J. from the Emmy Noether-Programme of the German Science Foundation
(DFG) under grant Ja 1114/3-2; 
P.~P. from a Ciencia 2008 contract, funded by FCT/MCTES (Portugal) and POPH/FSE (EC); 
I.~M.~P. from Spanish grant AYA2010-15169 and the Junta de Andalucia through TIC-114 and the Excellence Project P08-TIC-03531; 
J.~M.~G. from grant SFRH/B PD/66958/2009 from FCT (Portugal).
C.~J.~W.  from the Marie Curie Career Integration Grant 303912;
J.~I.~P. and J.~V.~M. from the Spanish MINECO under grant AYA2010-21887-C04-01, and from Junta de Andaluc\'{\i}a Excellence
Project PEX2011-FQM7058;
E.~M.~Q. from the European Research Council via the award of a
Consolidator Grant (PI McLure);
M.~P. from the Marie Curie Career Reintegration Grant (P.I. Wild Phiz-ev).\\
This work was supported in part by the National Science Foundation under Grant No. PHYS-1066293 and the hospitality of the Aspen Center for Physics.

\bibliography{refs_mice_v7}

\appendix

\section{Pre- and post-shock gas density}



Assuming pressure equilibrium between the post-shock gas and the gas
that has cooled sufficiently to emit \sii, the {\sc CALIFA} observations can be
used to measure the pre-shock density ($n_1$) in the following way. The
post-shock temperature ($T_2$) is given by
\begin{equation}
T_2 [K] = 1.38\times10^5\left(\frac{v_s [{\rm km/s}]}{100}\right)^2
\end{equation} 
for a fully ionised precursor \citep{Dopita:2003}. The temperature of
the gas emitting \sii\ is $T_3\sim$8000\,K, and the density of this gas is
measured from the\sii\ line ratio ($n_3=n_e$). For a fast shock the
post-shock density is four times the pre-shock density
($n_2=4n_1$). The pre-shock gas density is then given by:
\begin{equation}
n_1 [{\rm cm}^{-3}] =\frac{n_3 T_3}{4 T_2} = 0.12 \frac{n_e[{\rm cm^{-3}}]}{100}  \left(\frac{350}{v_s[{\rm km/s}]}\right)^{2} 
\end{equation}
From this, we can estimate the thermal pressure of the clouds where the \sii\ lines
are produced: 
\begin{equation}
P_{\rm cloud} = \rho_1 v_s^2 = n_1 m_p \mu v_s^2
\end{equation}
where $\rho_1$ is the mass density of the medium into which the shock
is propagating, $v_s$ is the shock velocity, $m_p$ is the proton mass,
$\mu=1.36$ accounts for an assumed 10\% Helium fraction. This gives a
relation between thermal pressure and electron density
\begin{equation}
P_{\rm cloud}=3.3\times10^{-11} \frac{n_e[\rm{cm^{-3}}]}{100}{\rm Nm^{-2}}
= 3.3\times10^{-12} n_e{\rm dynes\,cm^{-2}}
\end{equation}
close to the value given by HAM90 of
$4\times10^{-12}n_e{\rm dynes\,cm}^{-2}$ based on shock
models by Shull \& McKay (1985).


The bolometric luminosity per unit area of a shock can be
independently estimated from $\mathcal{L}_{\rm shock}=0.5\rho v_s^3$,
and this provides a consistency check on the shock speed estimated
from the model fit to the observed line ratios.  Summing over the full
area of the two bicones, which we estimate from the {\sc CALIFA} maps
to cover 1/4 of a sphere, we find a bolometric energy loss rate of
$\dot{E}_{\rm shock} =3.2\times10^{43}$\ergs\ for
$v_S=350$\,\kms. This is only a factor of $\sim$2 smaller than the
estimated bolometric luminosity of the bicones, which lies well within
the errors of these calculations, and supports the argument that
350\kms\ fast shocks are the most likely candidate for causing the
ionisation of the gas in the cones.

\section{Simulation methodology}\label{app:simn}


The details of the simulation are presented in \citet{Johansson:2009}
and \citet{Johansson:2009p7314}, here we provide a brief summary of
the relevant details. The simulations were performed using the entropy
conserving TreeSPH-code GADGET-2 \citep{2005MNRAS.364.1105S} which
includes radiative cooling for a primordial mixture of hydrogen and
helium together with a spatially uniform time-independent local UV
background.

 The assumed dark matter profile has a significant impact on the
  evolution of the merger. Here we use an ``NFW-like'' profile,
  i.e. an analytical \citet{1990ApJ...356..359H} dark matter profile,
  with a concentration parameter $c=9$, matched to the empirical
  \citet{Navarro:1996p9444} profile as described in
  \citet{2005MNRAS.361..776S}.  Disks have exponential profiles, with
a total baryonic disk mass of $M_{d} = m_d M_{vir}$, where $m_d =
0.041$ and $M_{vir}$ is the total virial mass. The disks are composed
of stars and gas, with a fractional gas content of $f_{gas} =
M_{gas,d}/(M_{gas,d}+M_{*,d}) = 0.2$. The stellar bulge(s) have
profiles closely approximating a de Vaucouleurs law
\citep{1990ApJ...356..359H} and a stellar mass of $M_{*,b} =
\frac{1}{3}M_d = 0.27M_{*,d}$ i.e. close to the observed bulge-to-disk
ratio for NGC~4676B of $M_{*,b}/M_{*,d}\sim$0.25. All stellar and gas
particles are embedded in a dark matter halo.

The galaxies contain 240,000 disk particles, 60,000 gas particles
(i.e. disk gas fraction of 20\%), and 400,000 dark matter
particles. NGC~4676B additionally has 100,000 bulge particles, the
bulge is omitted from NGC~4676A to match the observations.  This
leads to a baryonic mass ratio of 1.3, close to that observed in the
real Mice galaxies. The baryonic and dark matter particles have masses
of $1.8\times10^{5}$\msol\ and $3.2\times10^{6}$\msol\ respectively.

The initial orbital parameters and time of observation (180\,Myr
following first passage\footnote{Snapshots are extracted every 20\,Myr.})
were taken from \citet{Barnes:2004p6826}. The viewing angle was chosen
to provide the best match by eye to the CALIFA spatial and velocity
maps, using the package SDvision
\citep{Pomarede:2008p8081}\footnote{http://irfu.cea.fr/Projets/COAST/visu.htm}.

We follow the method of \citet{Wild:2009p2609} to assign initial
smooth star formation histories to the disk and bulge stars. For
bulge stars, we assume a single formation epoch at a lookback time of
13.4\,Gyr (the age of the Universe at the redshift of the Mice). The
disk stars have close to constant star formation rates, with a slight exponential
increase to earlier times to ensure consistency between the current
mass and previous star formation history. The
simulation then provides the star formation history up until the point
of observation, and we extract the spectral energy distribution of
each particle using the spectral synthesis models of
\citet{2003MNRAS.344.1000B}. Yields are not tracked in these
simulations, so metallicity is kept fixed at solar in agreement with
the measured metallicity of the galaxies (Section
\ref{sec:emratios}). Similarly, attenuation of the starlight by dust
is not included due to the many assumptions required, and we compare
to observed quantities that have been dust attenuation corrected.

\end{document}